\documentclass[12pt]{article}
\usepackage{amssymb,slashed, color}
\usepackage{amsmath,amsfonts,euscript,mathrsfs,multirow}
\usepackage[all]{xy}
\usepackage{epsf}
\usepackage{graphicx}
\csname @addtoreset\endcsname{equation}{section}
\newcommand{\SU}{\mathop{\rm SU}}

\newcommand{\U}{\mathop{\rm {}U}}

\newcommand{\Sl}{\mathop{\rm {}SL} }

\newcommand{\su}{\mathfrak{su}}

\newcommand{\IIVo}{
\put(0,0){\circle{4}}
\put(2.2,0){\line(1,0){2}}
\put(6.4,0){\circle{4}}
\put(8.6,0){\line(1,0){2}}
\put(12.6,0){\circle{4}}
\put(14.6,0){\line(1,0){2}}
\put(18.8,0){\circle{4}}
\put(-.5,-.5){\tiny{$1$}}
\put(5.9,-.5){\tiny{$2$}}
\put(12.1,-.5){\tiny{$4$}}
\put(18.2,-.5){\tiny{$2$}}
}
\newcommand{\IIVob}{
\put(0,0){\circle{4}}
\put(2.2,0){\line(1,0){2}}
\put(6.4,0){\circle{4}}
\put(8.6,0){\line(1,0){2}}
\put(12.6,0){\circle{4}}
\put(14.6,0){\line(1,0){2}}
\put(18.8,0){\circle{4}}
\put(-.5,-.5){\tiny{$1$}}
\put(5.9,-.5){\tiny{$2$}}
\put(12.1,-.5){\tiny{$3$}}
\put(18.2,-.5){\tiny{$2$}}
}

\newtheorem{theorem}{Theorem}[section]

\newtheorem{prop}[theorem]{Proposition}

\newtheorem{problem}{Problem}

\DeclareFontFamily{U}{rsf}{} \DeclareFontShape{U}{rsf}{m}{n}{ <5> <6> rsfs5 <7> <8> <9> rsfs7 <10-> rsfs10}{}
\DeclareMathAlphabet\Scr{U}{rsf}{m}{n}
\definecolor{pink}{rgb}{1,0,1}

  \usepackage[pdftex]{hyperref}
\begin{document}

 %%%%%%%%%%%%%%%%%%%%%%%%%%%%%%%%%%%%%%%%%%%%%%%%%%%%%%%%%%%
\begin{titlepage}
\begin{center}
\baselineskip=14pt
{\LARGE 
Small resolutions of    $\SU(5)$-models  in F-theory.\\
}
\vspace{.6 cm}
{\large  Mboyo Esole$^{\spadesuit,\clubsuit}$  and Shing-Tung  Yau$^{\spadesuit}$
  } \\
\vspace{.3 cm}
${}^\spadesuit$ Department of Mathematics,  
 Harvard University, 
  Cambridge, MA 02138, U.S.A.\\

$^\clubsuit$Jefferson Physical Laboratory, Harvard University, Cambridge, MA 02138, U.S.A.\\

\end{center}

\vspace{1cm}
\begin{center}

{\bf Absract}
\vspace{.3 cm}
\end{center}
{\small

We provide an explicit desingularization and study the resulting fiber geometry of  elliptically fibered fourfolds defined by  Weierstrass models  admitting a split $\tilde{A}_4$ singularity over a divisor of the discriminant locus. 
Such varieties  are used to geometrically engineer  $\SU(5)$ Grand Unified   Theories  in   F-theory. The desingularization is given by a small resolution of singularities. 
The $\tilde{A}_4$ fiber naturally appears after resolving the singularities in codimension-one in the base.  The   remaining  higher codimension singularities  are then  beautifully described by a four dimensional affine binomial variety which leads to    six different  small  resolutions of  the elliptically fibered fourfold. These six small resolutions define  distinct fourfolds  connected to each other by a network of flop transitions forming  a    dihedral  group. 
The location of these exotic fibers in the base is mapped to conifold points of the  threefolds  that defines  the type IIB orientifold limit of the F-theory. 
The full resolution has interesting properties, specially for fibers in codimension three: the rank of the singular fiber does not necessary increase and the fibers are not necessary in the list of  Kodaira and some are not even (extended) Dynkin diagrams.
\vfill
$^\spadesuit$Email:{\tt    esole at  math.harvard.edu,  yau  at math.harvard.edu}
}

\end{titlepage}
\addtocounter{page}{1}
 \tableofcontents{}
\newpage

\newpage

\section{Introduction }

The theory of  elliptic curves is  an elegant and  vast subject in mathematics that can be traced back to  ancient Greece and beyond. An elliptic curve is a nonsingular irreducible curve of genus one with a  distinguished point   on it. 
 It is well known that elliptic curves  play an important   role in number theory and  are  instrumental  in   cryptography and  geometric modeling. Elliptic curves have also invaded many branches of theoretical physics through their modular properties. They are also familiar in several string theory dualities. 
The ultimate example of applications of elliptic curves to string theory is probably F-theory.

F-theory was introduced by Vafa \cite{Vafa:1996xn}  as a non-perturbative formulation of  type IIB string theory   providing a geometrization of its non-perturbative $\Sl(2,\mathbb{Z})$ symmetry  known as {\em S-duality}.
F-theory  provides a geometric formulation of type IIB string theory  compactification on a complex variety $B$  with a varying  axio-dilaton field $\tau$. The variation of the axio-dilaton field is geometrically realized in F-theory by an elliptic fibration $\pi:Y\rightarrow B$ where the variety $B$ is the base of the fibration: 
$$
\xymatrix{ T^2 \ar[r] &  Y\ar[d]^{\pi } \\& B \ar[u]_{\iota} } 
$$
The  period modulus of the elliptic fiber is interpreted as the axio-dilaton field:
$$T^2\simeq \mathbb{C}/(\mathbb{Z}+\tau\mathbb{Z}), \quad \tau=C_0+i e^{-\phi},\quad e^{\phi}=g_s.$$
 The  S-duality group is then geometrically realized as the $\Sl(2,\mathbb{Z})$ modular group  of the elliptic  fiber. The axio-dilaton field can be used to probe the presence of $(p,q)$-seven-branes   since  seven-branes and D-instantons are magnetic duals. 
In F-theory, the elliptic fibration is a Calabi-Yau space and the base of the fibration is usually  a complex curve, surface or threefold  leading respectively (after compactification) to a theory in $8$, $6$  or $4$ real spacetime dimensions. There is another road to F-theory by starting from  M-theory, a theory defined in an eleven dimensional spacetime. 
The duality between F-theory and M-theory is understood via   a chain of dualities. 
Type IIB string theory compactified on a circle is T-dual to type IIA on a dual circle, which is then dual to  M-theory compactified on a 2-torus\cite{Schwarz:1995jq}:
$$\text{M-theory} \ \ \overset{\text{\footnotesize{ S-duality}}}{\longleftrightarrow}  \ \    \text{IIA}  \ \  \overset{\text{\footnotesize{ T-duality}}}{\longleftrightarrow}   \ \  \text{IIB }\  \  \longleftrightarrow \  \  \text{F-theory }
$$
In F-theory, one usually starts with a singular elliptic fibration given by a singular Weierstrass model and the smooth elliptic fibration is obtained by a {\em resolution} of the singularities of the elliptic fibration. 
Starting form a smooth elliptic fibration $Y$, the type IIB limit of F-theory is taken in two steps: first, shrink to zero area all fiber components not meeting the chosen section, then shrink the  remaining components of each fiber to zero area.
For singular fibers over codimension one loci in the base, the  structure of the singular fibers can be deduced without performing an explicit resolution thanks to Tate's algorithm. But this is not true for singular fibers that  project to codimension-two or higher in the base. 
A resolution of singularity only changes the singular part of the variety while it is an isomorphism on the smooth part. Even so, a resolution is in general not unique and the singular part of the variety is usually replaced by a locus of higher dimension, typically a divisor. Additional requirements can be imposed on a resolution. One can for example ask the resolution to be {\em small} so that a singular locus of codimension $k$ is replaced by a locus which is smaller than a divisor. A more restricted definition of small resolution requires that a  singular locus of codimension $k$  is replaced  by a locus of codimension  less than $k/2$. Small resolutions are not always possible. 
One can also ask the resolution to be {\em crepant}. This is very common in string theory as it ensures that the Calabi-Yau condition is preserved. 
In this paper, we will analyze a particular model where  F-theory is defined from a singular geometry for which  the desingularization has been conjectured to have a remarkable structure but has never been analyzed systematically. We will use this specific example as  an appetizer to attack a more general questions also relevant for mathematics.

We will consider an elliptic fibration endowed with  a smooth section  $\iota: B\rightarrow Y$.
Fiberwise, the section selects a point on every fiber so that the fiber is a bona fide elliptic curve. In F-theory, the section provides a concrete embedding  of the base  $B$ (on which type IIB is compactified)  inside the elliptic fibration $Y$.
 An elliptic fibration with a section has the nice property of being birationally equivalent to a  (singular) Weierstrass model. 
Our model of interest will be given explicitly by a Weierstrass model in Tate's form:
$$
\mathscr{E}: zy^2+a_1 x y z + a_3 y z^2=x^3+ a_2 x^2 z + a_4 x z^2 + a_6 z^3,
$$
where $[x:y:z]$ are projective coordinates of a $\mathbb{P}^2$ bundle over the base $B$ and the coefficients $a_i$ are sections of appropriate line  bundles. We will come back to the definition in section \ref{ellipticfibrations}. 
In this paper, the base of the fibration is a complex variety $B$ of complex dimension three (a threefold) so that after the compactification we end up with a theory in 4 spacetime dimensions. 

The elliptic fibration we will be interested in, is given by a singular Weierstrass model with a specific type of singularities (a  $\tilde{A}_4$ singularity) over a divisor of the base. This singularity is  motivated by the geometric engineering of $\SU(5)$ Grand Unified Theories in F-theory. We will resolve the singularities and study carefully the resulting fiber geometry. Over the divisor over which the singularity is defined, we will get a singular fiber of type $I_5$ in Kodaira's notation and the singular fiber will degenerate further over certain curves and points in the base. More precisely, the fiber is a split $I_5$, meaning that each components of the singular fiber  is defined without requiring a field extension. We assume that after the resolution, we have a smooth elliptic fibration with a smooth section and  the fibration is required to be {\em flat} (all fibers have the same dimension) and the resolution to be crepant. All the results will be independent of the Calabi-Yau condition.  The structure of the resolved geometry is conjectured in the physics literature to naturally realize a  cascade of  specializations to other types of singular fibers over certain loci of the base  in order to  geometrically engineer the matter content and the Yukawa couplings of the $\SU(5)$ Grand Unified Theory. 

In this paper, our point of view will be to use the physics as a motivation for the  geometry we are studying. But all the results of this paper rely solely on a  mathematical analysis. We will be able not only to resolve the geometry but we also present certain transitions between different resolutions of the singularities not anticipated by physicists.  In the rest of the introduction, we will discuss the fiber geometry in F-theory, present the problem that we want to attack and its physical origin and summarize our results. The rest of the paper will be a systematic derivation of these results. 

\subsection{Fiber degenerations and F-theory}
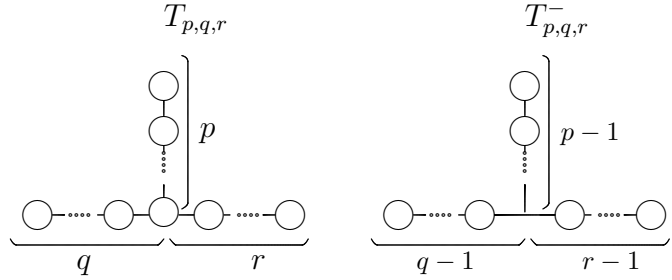
\begin{figure}
\begin{center}
\setlength{\unitlength}{.4 mm}
\begin{picture}(100,50)(0,-50)
%\multiput(0,0)(0,15){3}{\circle{10}}
    \put(0,20){$T_{p,q,r}$}
\put(0,0){\circle{10}}
\qbezier(0,-5)(0,-7)(0,-10)
\put(0,-15){\circle{10}}
\qbezier(0,-20)(0,-21)(0,-22)
\multiput(0,-22.5)(0,-2){4}{\circle{1}}
\qbezier(0,-33)(0,-34)(0,-37)
\put(0,-42){\circle{10}}
%\qbezier(0,-47)(0,-49)(0,-52)
\put(-7,3){\qbezier(15,5)(15,-19)(15,-42)\qbezier(13,7)(14.9,6.5)(15,5)\qbezier(13,-44)(14.9,-43)(15,-42) \put(19,-20){$p$}}
%%%%
\put(0,-43){
\put(-15,0){\circle{10}}
\qbezier(-20,0)(-21,0)(-22,0)
\multiput(-25,0)(-2,0){4}{\circle{1}}
\qbezier(-33,0)(-34,0)(-37,0)
\put(-42,0){\circle{10}}
\qbezier(-5,0)(-8,0)(-10,0)
\put(-7,-25){\qbezier(5,15)(-19,15)(-42,15)\qbezier(5,15)(5.5,15.1)(7,17)\qbezier(-44,17)(-43,15.1)(-42,15) \put(-22,7){$q$}}
}
\put(0,-43){
\put(15,0){\circle{10}}
\qbezier(20,0)(21,0)(22,0)
\multiput(25,0)(2,0){4}{\circle{1}}
\qbezier(33,0)(34,0)(37,0)
\put(42,0){\circle{10}}
\qbezier(5,0)(8,0)(10,0)
\put(9,-25){\qbezier(-5,15)(19,15)(37,15)\qbezier(-5,15)(-5.5,15.1)(-7,17)\qbezier(39,17)(38,15.1)(37,15) \put(20,7){$r$}}
}

\put(120,0){
    \put(0,20){$T^-_{p,q,r}$}
\put(0,0){\circle{10}}
\qbezier(0,-5)(0,-7)(0,-10)
\put(0,-15){\circle{10}}
\qbezier(0,-20)(0,-21)(0,-22)
\multiput(0,-22.5)(0,-2){4}{\circle{1}}
\qbezier(0,-33)(0,-34)(0,-42)
%\put(0,-42){\circle{10}}
%\qbezier(0,-47)(0,-49)(0,-52)
\put(-7,3){\qbezier(15,5)(15,-19)(15,-42)\qbezier(13,7)(14.9,6.5)(15,5)\qbezier(13,-44)(14.9,-43)(15,-42) \put(19,-21){\footnotesize{$p-1$}}}
%%%%
\put(0,-43){
\put(-15,0){\circle{10}}
\qbezier(-20,0)(-21,0)(-22,0)
\multiput(-25,0)(-2,0){4}{\circle{1}}
\qbezier(-33,0)(-34,0)(-37,0)
\put(-42,0){\circle{10}}
\qbezier(10,0)(-8,0)(-10,0)
\put(-7,-25){\qbezier(5,15)(-19,15)(-42,15)\qbezier(5,15)(5.5,15.1)(7,17)\qbezier(-44,17)(-43,15.1)(-42,15) \put(-29,7){\footnotesize{$q-1$}}}
}
\put(0,-43){
\put(15,0){\circle{10}}
\qbezier(20,0)(21,0)(22,0)
\multiput(25,0)(2,0){4}{\circle{1}}
\qbezier(33,0)(34,0)(37,0)
\put(42,0){\circle{10}}
\qbezier(5,0)(8,0)(10,0)
\put(9,-25){\qbezier(-5,15)(19,15)(37,15)\qbezier(-5,15)(-5.5,15.1)(-7,17)\qbezier(39,17)(38,15.1)(37,15) \put(10,7){\footnotesize{$r-1$}}}
}
}
\end{picture}
\end{center}
\caption{Fiber of type $T_{p,q,r}$ and $T^-_{p,q,r}$ with $1\leq p\leq q\leq r$. The fiber of type $T^-_{p,q,r}$ is obtained from $T_{p,q,r}$ by replacing the central node by a point common to the three branches. A fiber of type $T_{p,q,r}$ has rank $p+q+r-2$, while a fiber of type $T^-_{p,q,r}$ has rank $p+q+r-3$. \label{Tpqr}}
\end{figure}

As it is usually the  case for fibrations, many interesting  properties of an elliptic  fibration are encoded in the structure of its singular fibers. 
In F-theory, the structure of the singular fibers is a central piece in the dictionary between geometry and physics. 
In the presence of a $(p,q)$-seven-brane, the axio-dilaton field undergoes   $\Sl(2,\mathbb{Z})$-monodromies that determine  the type of $(p,q)$-seven-branes.  From the point of view of the elliptic fibration, such monodromies require the elliptic fiber to be singular over certain divisors of the base $B$. It follows that   singular fibers geometrically detect the locations (in the base $B$) and types of $(p,q)$ seven-branes. The locus of points of the base over which the  elliptic fiber is singular is called the {\em discriminant locus}.  For an elliptic fibration given by a Weierstrass model, the discriminant locus  is a divisor in the base $B$ defined by the equation $\Delta=0$ ,where  $\Delta$ is defined  in equation \eqref{Deligne} or equivalently in equation \eqref{discr}.  
The discriminant locus can be   composed of  several irreducible components of different multiplicities $\Delta=\prod_i \Delta_i^{m_i}$ and can admit  a  sophisticated structure of singularities.  The generic  fiber over a component of the discriminant locus can  degenerate further at the intersection with another component  or more generally at a  singularity of the discriminant locus. Such degenerations  occur in codimension-two or higher in the base and are  called  {\em collisions of singular fibers}. 
Here, we will restrict ourself to {\em flat fibrations} so that all the fibers have the same dimension. In particular, a singular fiber  is  composed of a finite number of  intersecting irreducible curves.   The  singular fibers of a smooth Weierstrass model are nodal and cuspidal irreducible curves. When the Weierstrass model is singular one can get  in addition a rich spectrum of reducible singular fibers after resolving the singularities. To each reducible singular fiber, it is convenient to  associate a {\em dual graph} representing the intersection structure  of the irreducible rational curves composing the singular fiber. The dual graph can take several shapes. It is a beautiful aspect of the theory of elliptic fibrations that the dual graph of the  singular fiber over a generic point of a  component of the discriminant locus is an  ADE Dynkin diagram plus an extra node. We will denote projective ADE Dynkin diagrams by the corresponding Lie algebra  notation  $A_n$, $D_n$ and $E_n$ where the index $n$ refers to the number of nodes of  the diagram and therefore to the rank of the  Lie algebra. The  associated  affine ADE Dynkin diagrams will be denoted with a tilde: $\tilde{A}_n$, $\tilde{D}_n$, $\tilde{E}_n$. In the case of affine Dynkin diagrams, the number of nodes is $(n+1)$. To cover all the type of diagrams that will occur in this paper,  it is useful to introduce  the graphs  of type $T_{p,q,r}$ and $T_{p,q,r}^-$. They are presented in figure \ref{Tpqr}. They generalize diagrams of type $\tilde{E}_n$, $E_n$ , $D_n$ and $A_n$.

The common wisdom on the physical meaning of singular fibers in F-theory can be summarized in three points:

(1) Each irreducible component of the discriminant locus corresponds to  a divisor  wrapped  by  $(p,q)$-seven-branes. The gauge group  associated to  branes wrapping   a divisor of the discriminant locus is  determined by the type of singular fibers above that component.  In the spirit of  Mckay's correspondence, the type of singular fibers over a component of the discriminant locus  identifies  the non-Abelian gauge group living on the brane wrapping that component\footnote{If the fiber is not split, there is monodromy that exchange certain components  of the singular fiber and by identifying such components, we can also get  non-simply laced gauge groups.}.

(2) Singular fibers in codimension-two  in the base are related to charged  chiral matter following  Katz-Vafa's  approach and the  matter content can be read off following the branching rules of  representation theory \cite{FTheoryTate,Katz:1996xe,Vafa:2009se}.  When the base is a threefold, these codimension-two loci are called {\em matter curves}. This follows from a familiar picture in D-brane modeling under which chiral matter can be localized at D-branes intersections. In F-theory, this point has been supported by  direct comparisons with the heterotic string in the case of  elliptic threefolds leading to theories in six spacetime dimensions.

(3) Codimension-three singularities are related to Yukawa couplings.

 \begin{table}[hbt]$$\xymatrix{*+[F]+{\text{Gauge group}} \ar[d]&*+[F]+{ \text{Matter fields}}\ar[d] & *+[F]+{\text{Yukawa Couplings} }\ar[d]  \\*+[F]+{ \text{Codim-1 sing.}}  \ar[r]  &*+[F]+{ \text{Codim-2 sing.} } \ar[r] &*+[F]+{  \text{Codim-3 sing.}}  } $$ \caption{F-theory description of gauge theories. }\end{table}

When the singularities are more involved, already the second step (identification of the matter representation) can become delicate to handle\cite{morrisonwati}. 
Another approach based on the field theory on the seven-branes world volume has been developed recently \cite{Donagi:2008ca,Donagi:2009ra,Beasley:2008dc,Tbranes}.
However, the relation between the details of the field theory on the world volume of the seven branes and the singular fibers of the elliptic fibration is not always clear.

\subsection{  $\SU(5)$ Unification in F-theory }

The Standard Model of particle physics describes with great accuracy (up to an energy scale of few hundred GeVs) 
 all the known non-gravitational  fundamental interactions of Nature; namely 
 the electromagnetism force, the weak and the strong force . Mathematically, the Standard Model  is described by a  gauge theory based on the semi-simple gauge group 
$G_{SM}=\SU(3)\times \SU(2)\times\U(1)$.  Fundamental particles are described by irreducible representations of the gauge group.
The Standard Model interactions are mediated by gauge bosons transforming in the adjoint representation of $G_{SM}$ and   fermionic particles  (quarks and leptons) transforming under representations  $G_{SM}$ that  seem at first look rather complicate and arbitrary. The Standard Model has also  several unexplained properties, like for example the quantization of the electric charge, which  beg for a more fundamental explanation. 

 In the 1970s, theoretical physicists pursued an ambitious program called ``Grand Unification'' with the goal of finding a  simple and  elegant  reformulation of the Standard Model of particle physics that would  make sense of its  puzzling structure while  providing at the same time a description of physics above      its   energy scale. 
The Standard Model being  a gauge theory, representations of Lie algebras and  Lie groups  were the main mathematical tools for unification. This is reviewed for example in   \cite{Georgi} or for a mathematical audience in \cite{Baez}. 
Grand unified theories  ( GUTs)   are based on the idea that the  different interactions of the Standard Model are all unified at high energy  and described by a  unique gauge theory with a  Lie group $G_{GUT}$ containing $G_{SM}$ as a subgroup. 
The Grand Unified gauge group  $G_{GUT}$, is usually  required to be a  simple group  so that it depends on a  unique coupling constant.   A compact unified  gauge group would also automatically explain the quantization of the electric charge, one of the most famous unexplained aspect of the Standard Model.  
A Grand unified theory  naturally reorganizes the Standard Model particles into  fewer but larger representations. This usually implies subtle relations between apparently independent  parameters of the Standard Model. It also introduces new interactions leading to prediction of new physical processes,  like for example the decay of the proton (that has not yet been observed).

In the quest of the ultimate theory, string theory provides an even more ambitious program of unification since it includes a quantum description of gravity. The geometry of spacetime itself, through its extra dimensions, can geometrically generate the gauge group of a Grand Unified Theory. The idea of geometric engineering of  physical models is the source of  rich  dialogues between theoretical  physics and mathematics.    The first example of a Grand Unified Theory was the  $\SU(5)$ model proposed by  Georgi and Glashow in 1974 \cite{GG}.   $\SU(5)$ is the smallest simple group  containing the Standard Model gauge group $G_{SM}$ and able to contain the Standard Model representation for a given generation of particles. Supersymmetric versions of the Georgi-Glashow $\SU(5)$ model have been the center of important efforts to embed Grand Unified Theories in string theory.  F-theory, one of the most geometric corner of string theory, has recently played a leading role  in  providing phenomenological models of  $\SU(5)$ unification.  
Describing Grand Unified Theories by F-theory models is not only  a geometrically  elegant construction but is also believed to be a powerful approach to phenomenology\cite{Vafa:2009se}. 
$\SU(5)$ Unification was also studied from the point of view of M-theory on manifolds with $G_2$ holonomies by Friedman and Witten \cite{tamar}. 

\subsection{The conjectured fiber geometry of $\SU(5)$ models}

The first step to implement an   $\SU(5)$ model  in F-theory  is to have a component of the discriminant locus  over which the elliptic fiber admits a dual graph of type   {\em split} $\tilde{A}_4$.
The ``{split}'' condition  ensures that the different curves that compose the singular fiber are individually well defined without having to perform a field extension.   
The second step is to have the proper representations for the Standard Model fermions.
It is a remarkable fact that one generation of Standard Model fermionic particles can be elegantly summarized by only two  representations of  $\SU(5)$, namely  the complex conjugate of the fundamental representation   and the  second exterior power of the fundamental representation.  We will denote them respectively as
  $\overline{\mathbf{5}}$ and  $\mathbf{10}$ following the tradition in particle  physics where representations are denoted by their dimensions  and complex conjugation is denoted by an overall bar. 
 In the spirit of Katz-Vafa proposal,  these two representations  can be respectively realized by   the enhancement  $\tilde{A}_4\rightarrow\tilde{A}_5$ and   $\tilde{A}_4\rightarrow \tilde{D}_5$ over  two distinct curves in the component of the discriminant over which we have the  generic $\tilde{A}_4$ fiber. Such curves have to be singularities of the discriminant locus. One way they could occur is as intersection with a different component of the discriminant locus. Interestingly, imposing a  $\tilde{A}_4$ fiber ensures that the support of the discriminant contains at least two components. 
Finally, the third step is to accommodate the Yukawa couplings of a (supersymmetic) $\SU(5)$ model: it requires   further enhancements to $\tilde{D}_{6}$  and $\tilde{E}_6$ singular fibers at the points of intersections of the two matter curves. The first one ($\tilde{D}_{6}$) gives the Yukawa couplings of the down-type quarks and the second one ($\tilde{E}_{6}$) gives the Yukawa of the  up-type quarks. There are additional important  requirements that should be imposed to have a realistic model that takes care of the Higgs field, the number of generations, the  breaking of the $GUT$ group, etc. We will not consider them here. There are mostly specialisation of the geometry that we will analyze. Since we are looking at the fiber geometry resulting from the condition of having a split $\tilde{A}_4$ structure, we will refer to the elliptic fibration simply as the $\SU(5)$ model.

\begin{figure}[tbh]
\begin{center}
\setlength{\unitlength}{.40 mm}
\begin{picture}(190,225)(20,-140)
\put(0,-10){ 
\put(-10,10){$\tilde{A}_4$}
\put(0,0){\circle{10}}
\put(-20,-20){\circle{10}}
\put(20,-20){\circle{10}}
\put(15,-40){\circle{10}}
\put(-15,-40){\circle{10}}
\qbezier(-3,-3)(-5,-5)(-17,-17)
\qbezier(3,-3)(5,-5)(17,-17)
\qbezier(-19,-26)(-18,-30.5)(-17,-35)
\qbezier(19,-26)(18,-30.5)(17,-35)
\qbezier(-10,-40)(-10,-40)(10,-40)

}
\put(110,-80){
\put(-15,-50){$\tilde{D}_5$}
\put(-15,25){\circle{10}}
\put(15,25){\circle{10}}
\put(0,10){\circle{10}}
\put(0,-10){\circle{10}}
\put(-15,-25){\circle{10}}
\put(15,-25){\circle{10}}
\qbezier(0,5)(0,0)(0,-5.5)
\qbezier(5,12)(8,15)(14,21)
\qbezier(-5,12)(-8,15)(-14,21)
\qbezier(-5,-12)(-8,-15)(-14,-21)
\qbezier(5,-12)(8,-15)(14,-21)
}

\put(50,-30){
\qbezier(0,0)(-10,0)(-20,0)
\put(0,0){\qbezier(0,0)(-2,.5)(-3,3)\qbezier(0,0)(-2,-.5)(-3,-3)}
\qbezier(0,0)(10,-15)(20,-30)\qbezier(0,0)(10,15)(20,30)
\put(20,30){\qbezier(0,0)
(-1.52543 ,- 1.38675)(  -4.16025 ,- 0.83205)
\qbezier(0,0)( -0.693375, - 1.94145)( 0.83205 ,- 4.16025)
}
\put(20,-30)
{
\qbezier(0,0)(-0.693375 , 1.94145)(0.83205 , 4.16025)
\qbezier(0,0)(-1.52543 ,1.38675)( -4.16025, 0.83205)
}
}

\put(130,15){\qbezier(0,0)(35,15)(70,30) 
\put(35,15){
\qbezier(0,0)(-2.03525 ,- 0.328266)( -3.93919 , 1.57568)
\qbezier(0,0)( -1.64133, - 
 1.24741) \qbezier( -1.57568 ,- 3.93919)
}
}
\put(180,-80){
\qbezier(0,0)(-15,0)(-30,0)
\qbezier(0,0)(11,-14)(22,-28)
\qbezier(0,0)(11,14)(22,28)
\qbezier(0,0)(-12,19)(-36,57)
\put(-13,0){\qbezier(0,0)(-2,.5)(-3,3)\qbezier(0,0)(-2,-.5)(-3,-3)}
\put(11,-14){\qbezier(0,0)(-0.842484 , 1.88155 )( 0.50549 ,4.21242)\qbezier(0,0)( -1.6288,1.26373)( -4.21242, 0.50549)}
\put(11,14){\qbezier(0,0)(-1.6288 ,- 1.26373 )( -4.21242 ,- 0.50549)\qbezier(0,0)( -0.842484, - 
 1.88155)( 0.50549 ,- 4.21242) }
 \put(-24,38){\qbezier(0,0)( -0.645242 , 1.95797)( 0.934488 , 4.13845 ) 
 \qbezier(0,0)( -1.49073,  1.42398 )( -4.13845 , 0.934488)  }
}

\put(100,20){
\put(-20,30){$\tilde{A}_5$}
\qbezier(-20,-5)(-20,-10)(-20,-15)
\qbezier(20,-5)(20,-10)(20,-15)
\put(0,-40){\circle{10}}
\put(-20,0){\circle{10}}
\put(20,0){\circle{10}}
\put(-20,-20){\circle{10}}
\put(20,-20){\circle{10}}
\put(0,20){\circle{10}}
\qbezier(-3,17)(-5,15)(-17,3)
\qbezier(3,17)(5,15)(17,3)
\qbezier(-3,-37)(-5,-35)(-17,-23)
\qbezier(3,-37)(5,-35)(17,-23)
}

\put(230,50){
\put(-20,30){$\tilde{A}_6$}
\qbezier(-20,-5)(-20,-10)(-20,-15)
\qbezier(20,-5)(20,-10)(20,-15)

\put(-20,0){\circle{10}}
\put(20,0){\circle{10}}
\put(-20,-20){\circle{10}}
\put(20,-20){\circle{10}}
\put(0,20){\circle{10}}
\qbezier(-3,17)(-5,15)(-17,3)
\qbezier(3,17)(5,15)(17,3)
\put(15,-40){\circle{10}}
\put(-15,-40){\circle{10}}
\qbezier(-19,-26)(-18,-30.5)(-17,-35)
\qbezier(19,-26)(18,-30.5)(17,-35)
\qbezier(-10,-40)(-10,-40)(10,-40)

}

\put(230,-130){
\put(10,15){$\tilde{E}_6$}
\put(0,25){\circle{10}}
\put(0,7){\circle{10}}
\multiput(-30,-10)(15,0){5}{\circle{10}}
\multiput(0,0)(0,18){2}{\qbezier(0,2)(0,-3)(0,-5)}
\multiput(-15,0)(15,0){4}{\qbezier(-10,-10)(-7,-10)(-5,-10)}
}

\put(230,-35){
\put(10,-15){$\tilde{D}_6$}
\put(-15,25){\circle{10}}
\put(15,25){\circle{10}}
\put(0,10){\circle{10}}
\put(0,-10){\circle{10}}
\put(-15,-45){\circle{10}}
\put(15,-45){\circle{10}}
\put(0,-30){\circle{10}}
\qbezier(0,5)(0,0)(0,-5.5)
\qbezier(5,12)(8,15)(14,21)
\qbezier(-5,12)(-8,15)(-14,21)
\qbezier(-5,-32)(-8,-35)(-14,-41)
\qbezier(5,-32)(8,-35)(14,-41)
\qbezier(0,-15)(0,-18)(0,-25.5)
}

\end{picture}
\end{center}
\caption{{\bf Conjectured singular fiber enhancements of a   $\SU(5)$ GUT}. 
 Starting with codimension one in the base, the codimension increases from left to right. 
 Thinking in terms of dual graphs, the rank of the associated Dynkin diagrams increases   as we move in codimension.  
  }
 \label{su5GuTConjIntro}
\end{figure}
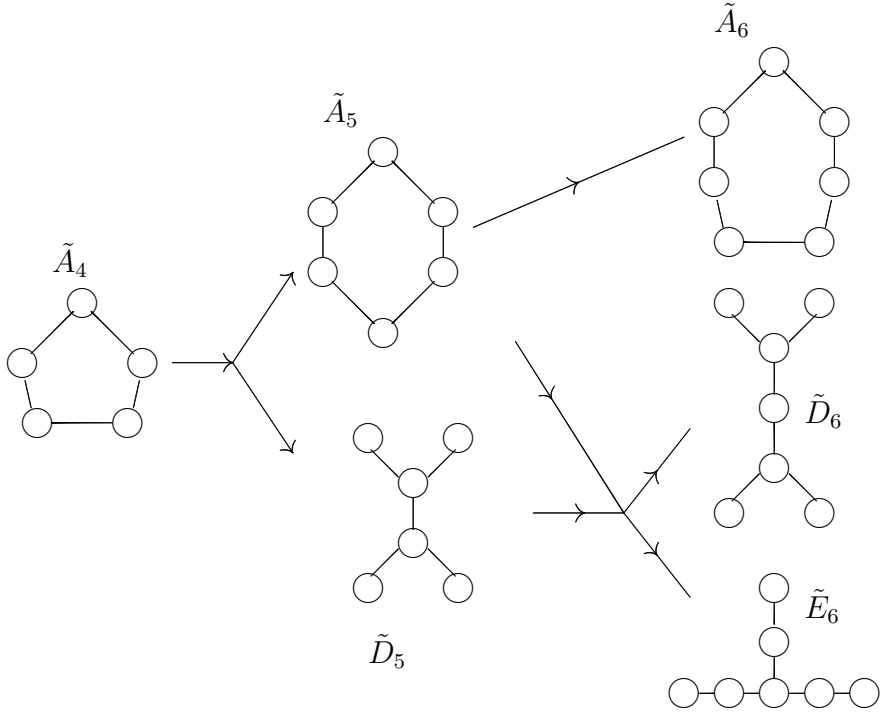

The enhancement to $\tilde{D}_5$,  $\tilde{A}_5$ in codimension-two and $\tilde{D}_6$ in codimension-three are  well understood   in  perturbative  string theory since they can be realized by  simple configurations of D-branes and orientifolds. However, the enhancement to $\tilde{E}_6$ necessary for the description  up-type quarks Yukawa couplings does not have a description in perturbative type IIB string theory. It is a typical example of   strongly coupled  phenomena which requires a F-theory treatment. One of the early attractive aspect of  F-theory phenomenology is the belief that these  successive  enhancements which are crucial for the physics  are realized very naturally by the geometry of a split $\tilde{A}_4$.  This is a highly non-trivial statement:\\

{\em The discriminant should have two components, one with a  singular fiber of type  $\tilde{A}_4$ and another with a nodal curve. The intersection of these two components is composed of two curves over which the singular fiber will degenerate further to $\tilde{A}_5$ and $\tilde{D}_5$. Finally, the two curves should intersect at two different types of points leading respectively to an additional degeneration of the fiber to $\tilde{D}_6$ and $\tilde{E}_6$.} \\

Since we assume the existence of a section, the elliptic fibration admits a birational equivalent  (singular) Weierstrass model and Tate's algorithm provide a systematic way to specialize the coefficients of  the Weierstrass model to impose a  split $\tilde{A}_4$ fiber over a specific  divisor of the base. The  Weierstrass model will be singular, but the split $\tilde{A}_4$ fiber will naturally emerge after resolving its codimension-one singularities. There will be several left-over singularities in higher codimensions in the base. In the context of $\SU(5)$ Unification, these extra-singularities are a blessing in disguise since they open the door to  a sequence of  enhancements of the singular fiber that can  provide the  matter fields and Yukawa couplings required for a realistic physical model.   
 The structure of the discriminant locus of a $\SU(5)$ model  has been carefully analyzed by
 Andreas and Curio \cite{andreas}. They  gave several indications  supporting the existence of the successive  enhancements demanded by $\SU(5)$ phenomenology ($\tilde{A}_5$ and $\tilde{D}_5$ in codimension-one, $\tilde{D}_{6}$ and $\tilde{E}_6$ in codimension-three). Their analysis also indicates  that there might be an additional   $\tilde{A}_6$  enhancement in codimension-three. All together, we have the following list of degenerations of singular fibers as we move in codimension in the base:
$$\tilde{A}_4\rightarrow (\tilde{A}_5, \tilde{D}_5)\rightarrow (\tilde{A}_6, \tilde{D}_6, \tilde{E}_6)$$
Starting from a fiber of type $\tilde{A}_4$, one can  retrieve the same structure  by  assuming that all singular fibers are extended ADE Dynkin diagrams and  as we move in higher codimension, the fiber enhancement is realized by increasing the rank of the dual ADE graph by one unit each time we increase the codimension by one unit.
Starting from $\tilde{A}_4$ in codimension-one, we can then expect fibers with dual graphs $\tilde{A}_5$ and $\tilde{D}_5$ in codimension-two since they are the only extended ADE Dynkin diagrams of rank 5.  In the same spirit, we can expect dual graphs fibers with rank 6 in codimension-3, namely  $\tilde{A}_6$, $\tilde{D}_6$ and $\tilde{E}_6$.

Although the fiber geometry is the backbone of any F-theory model, there has been no systematic analysis of the actual fiber structure of $\SU(5)$ models.
This is mostly due to a common misconception in the physics literature according to which the fiber  structure (even in higher codimension in the base) follows  from Tate's algorithm. Tate's algorithm is useful to identify the singular fiber over a component of the discriminant locus of a Weierstrass model without the need to perform the explicit desingularization by just analyzing the vanishing order of the coefficients of the Weierstrass model.
 Unfortunately, Tate's algorithm does not apply to codimension-two or higher. It is easy to construct examples where a singularity of the discriminant locus does not generate a change in the fiber structure of a smooth elliptic fibration or where the change that it generates does not correspond to what is expected by ``applying''  Tate's algorithm. It follows that the  fiber structure of $\SU(5)$ Grand Unified Theories in F-theory is very much   a conjecture although it is at the core of many phenomenological constructions\footnote{ For particular toric examples see \cite{ExtraCite1, ExtraCite2, ExtraCite2}.}.
 From the point of view of resolution of singularities, there is no reason to assume that the  fiber structure of an $\SU(5)$ model follows the conjectured singularities: the singular elliptic fibers in codimension-two and three in the base don't have to be (extended) ADE Dynkin diagrams and the rank does not have to increase with the codimension.  
Since the $\SU(5)$ model has a clean definition using a (singular) Weierstrass model, studying its fiber geometry  is a well posed mathematical problem that can be attacked systematically by constructing an explicit desingularization. We will provide desingularizations by  constructing explicit resolutions of  singularities. We ask the resolution to be small (and therefore crepant) in order to preserve the Calabi-Yau condition. We also ask the resolution to preserve the flatness of the fibration in order to ensure that all fibers have the same dimension.

\subsection{Mathematical interests of the $\SU(5)$ model}

Mathematically, the conjectured fiber geometry of the  $\SU(5)$ model in F-theory is an interesting structure  to analyze for several reasons. 
Although singular fibers in codimension-one of an elliptic fibration are well understood and classified, singular fibers above higher codimension loci are only understood under conditions that are evaded by many interesting physical models. %The main condition that comes at mind is having simple normal crossing divisor as the discriminant locus.
Resolution of singularities is a branch of  mathematics where examples have always play a crucial role. The $\SU(5)$ model, with its sophisticated cascade of degenerations of singular fibers is a beautiful geometry to explore. As we will see along the paper, studying the details of its resolution provides a surprisingly  rich  geometry. 

In order to put things into perspective, let us quickly review some  history on singular fibers of elliptic fibrations. We will have a more detailed analysis in section \ref{ellipticfibrations}. 
In the 1960s, Kodaira \cite{Kodaira}  and N\'eron \cite{Neron} have studied the singular fibers that can occur for an   elliptic surface. Their analysis can be generalized to case of singular fibers above  codimension-one points of an elliptic fibration over a higher dimensional base (see for example \cite{Fujita}). 
Miranda has studied the fiber structure of elliptic threefolds  by providing a systematic method to obtain flat elliptic fibrations from  (singular) Weierstrass models \cite{Miranda}. His results were later generalized to $n$-folds by Szydlo \cite{Szydlo}. We will call {\emph Miranda models} the elliptic fibrations with a simple normal crossing divisors  obtained by Miranda (and generalized by Szydlo) by an explicit regularization of Weierstrass models. 
In their regularization of Weierstrass models, Miranda and Szydlo  blow-up the base for example to get rid of  collisions of fibers admitting different values of the $j$-invariant or  to avoid collisions that do not admit small resolutions. 
 In F-theory, the elliptic fibrations are Calabi-Yau spaces and blowing-up the base can destroy the Calabi-Yau condition depending on the dimension of the base and the dimension of the centers of the blow-up. Blowing-up the base can also add new divisors in the discriminant locus and therefore change the structure of the gauge groups and  the physics of the model. We will review Miranda models and their collisions of singularities  in sections \ref{Miranda} and \ref{Szydlo}.

The question we address in this paper is a particular case of the following one: 
\begin{problem}
Consider an elliptic fibration over a complex $n$-fold, described by a (singular) Weierstrass model admitting a (split) singularity  of a given Kodaira type at the generic point of a  divisor of the discriminant locus and a fiber of type $I_1$ otherwise.
\begin{enumerate} 
\item Can we obtain a crepant resolution describing a  flat fibration\footnote{ By a crepant resolution we mean that the pull back of the canonical divisor of the singular space is the canonical divisor of the smooth resolution.}? 
\item What  is the tree of fiber enhancements for such a resolution?
\end{enumerate}
\end{problem}
 N\'eron has solved the problem when the base is a curve  by constructing regular models for elliptic surfaces defined by  Weierstrass equations \cite{Neron}. If we remove the condition to have a crepant resolution but ask to have a simple normal crossing divisor as the discriminant locus, the problem has been solved by Miranda \cite{Miranda} in the case of elliptic threefolds and generalized later by  Zlydlo for $n$-folds \cite{Szydlo}. Grassi and Morrison  have attacked the problem for Calabi-Yau elliptic threefolds \cite{GrassiMorrison} under some assumptions on the type of singularities that can occur. They also  discussed the F-theory interpretation of the fibers (gauge group and matter representation) while providing purely mathematical proofs. The analysis of Grassi and Morrison  generalizes the results of \cite{FTheoryTate} where the base is restricted to be an Hirzebruch surface $\mathbb{F}_n$ and the proofs rely heavily on the existence of an  heterotic dual.

 In this paper we treat in details the case of a split $\tilde{A}_4$ singularity  (split  $I_5$) and we assume that the base is a complex  threefold. 
We will construct a crepant resolution and study the resulting singular fibers.
Finding one small resolution will be enough to make some strong statement. In the category of projective algebraic varieties, Batyrev proved in \cite{Batyrev} that any two $n$-folds related by a birational map preserving their canonical class have the same Betti numbers. It follows in particular that distinct small resolutions have the same Betti numbers. Finding a small resolution of the $\SU(5)$ model might tell us important information on the conjectured fiber structure of such spaces.

\subsection{Summary of results}

The $\SU(5)$ model in F-theory is defined by the singular  Weierstrass model:
$$
\mathscr{E}: y^2+ \beta_5  x y z+ \beta_3 w^2 y z^2=x^3+  \beta_4 w x^2 z + \beta_2 w^3 x z^2 + \beta_0 w^5 z^3,
$$
where  $w$ is a section of a line bundle $\mathscr{L}_{su(5)}$ and $\beta_j$ ($j=0,2,3,4,5$)  is a   section of $\mathscr{L}^{\otimes (6-j)}\otimes \mathscr{L}_{su(5)}^{\otimes(j-5)}$. This structure ensures that over a generic point of $D_{\su(5)}: w=0$, we have a singularity leading to a dual graph $\tilde{A}_4$ after a desingularization by blow-ups. The dual graph $\tilde{A}_4$ will be {\em split}: it is composed of curves that are  individually well defined so that they are not subject to monodromies as we circle around the divisor $w=0$.

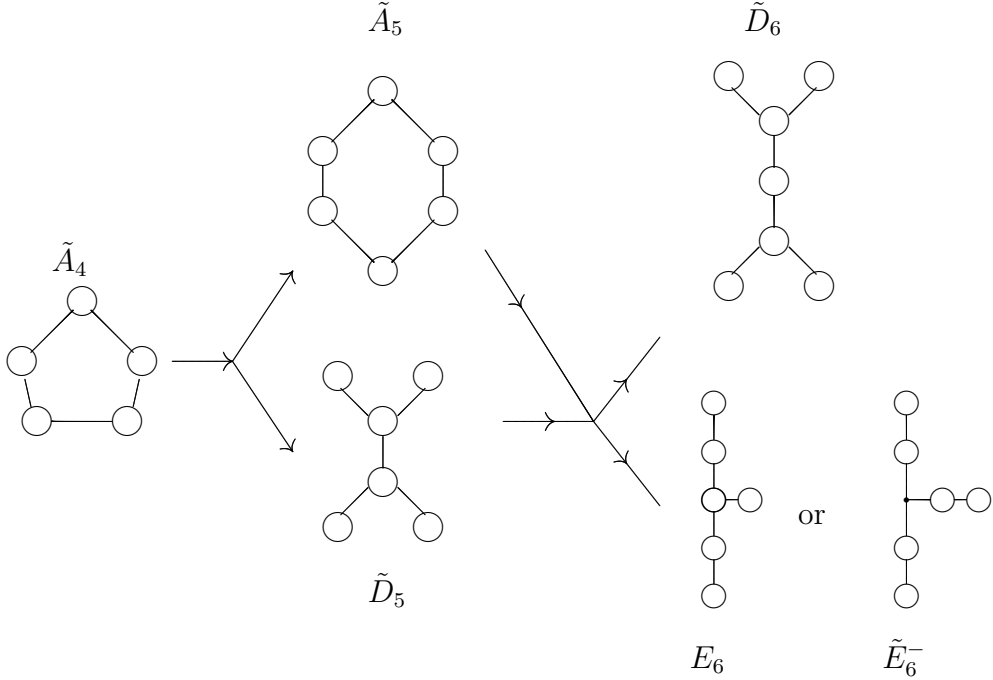
\begin{figure}[tbh]
\begin{center}
\setlength{\unitlength}{.40 mm}
\begin{picture}(190,200)(40,-120)
\put(0,-10){ 
\put(-10,10){$\tilde{A}_4$}
\put(0,0){\circle{10}}
\put(-20,-20){\circle{10}}
\put(20,-20){\circle{10}}
\put(15,-40){\circle{10}}
\put(-15,-40){\circle{10}}
\qbezier(-3,-3)(-5,-5)(-17,-17)
\qbezier(3,-3)(5,-5)(17,-17)
\qbezier(-19,-26)(-18,-30.5)(-17,-35)
\qbezier(19,-26)(18,-30.5)(17,-35)
\qbezier(-10,-40)(-10,-40)(10,-40)

}
\put(100,-60){
\put(-5,-50){$\tilde{D}_5$}
\put(-15,25){\circle{10}}
\put(15,25){\circle{10}}
\put(0,10){\circle{10}}
\put(0,-10){\circle{10}}
\put(-15,-25){\circle{10}}
\put(15,-25){\circle{10}}
\qbezier(0,5)(0,0)(0,-5.5)
\qbezier(5,12)(8,15)(14,21)
\qbezier(-5,12)(-8,15)(-14,21)
\qbezier(-5,-12)(-8,-15)(-14,-21)
\qbezier(5,-12)(8,-15)(14,-21)
}

\put(50,-30){
\qbezier(0,0)(-10,0)(-20,0)
\put(0,0){\qbezier(0,0)(-2,.5)(-3,3)\qbezier(0,0)(-2,-.5)(-3,-3)}
\qbezier(0,0)(10,-15)(20,-30)\qbezier(0,0)(10,15)(20,30)
\put(20,30){\qbezier(0,0)
(-1.52543 ,- 1.38675)(  -4.16025 ,- 0.83205)
\qbezier(0,0)( -0.693375, - 1.94145)( 0.83205 ,- 4.16025)
}
\put(20,-30)
{
\qbezier(0,0)(-0.693375 , 1.94145)(0.83205 , 4.16025)
\qbezier(0,0)(-1.52543 ,1.38675)( -4.16025, 0.83205)
}
}

\put(170,-50){
\qbezier(0,0)(-15,0)(-30,0)
\qbezier(0,0)(11,-14)(22,-28)
\qbezier(0,0)(11,14)(22,28)
\qbezier(0,0)(-12,19)(-36,57)
\put(-13,0){\qbezier(0,0)(-2,.5)(-3,3)\qbezier(0,0)(-2,-.5)(-3,-3)}
\put(11,-14){\qbezier(0,0)(-0.842484 , 1.88155 )( 0.50549 ,4.21242)\qbezier(0,0)( -1.6288,1.26373)( -4.21242, 0.50549)}
\put(11,14){\qbezier(0,0)(-1.6288 ,- 1.26373 )( -4.21242 ,- 0.50549)\qbezier(0,0)( -0.842484, - 
 1.88155)( 0.50549 ,- 4.21242) }
 \put(-24,38){\qbezier(0,0)( -0.645242 , 1.95797)( 0.934488 , 4.13845 ) 
 \qbezier(0,0)( -1.49073,  1.42398 )( -4.13845 , 0.934488)  }
}

\put(100,40){
\put(-5,40){$\tilde{A}_5$}
\qbezier(-20,-5)(-20,-10)(-20,-15)
\qbezier(20,-5)(20,-10)(20,-15)
\put(0,-40){\circle{10}}
\put(-20,0){\circle{10}}
\put(20,0){\circle{10}}
\put(-20,-20){\circle{10}}
\put(20,-20){\circle{10}}
\put(0,20){\circle{10}}
\qbezier(-3,17)(-5,15)(-17,3)
\qbezier(3,17)(5,15)(17,3)
\qbezier(-3,-37)(-5,-35)(-17,-23)
\qbezier(3,-37)(5,-35)(17,-23)
}

\put(210,-60){
\setlength{\unitlength}{.32 mm}
\put(0,20){
\put(-10,-110){$E_6$}
\put(0,0){\circle{10}}
\qbezier(0,-5)(0,-7)(0,-15)
\put(0,-20){\circle{10}}
\put(0,-20){
\put(0,-20){\circle{10}}
{\put(0,-20){\circle{10}}
\put(0,-40){\circle{10}}}
\put(0,-60){\circle{10}}
\multiput(0,0)(0,-20){3}{\qbezier(0,-5)(0,-10)(0,-15)}
\put(15,-20){\circle{10}}
\qbezier(5,-20)(7,-20)(10,-20)
}
}

\put(35,-30){or}

\put(80,0){
\put(-10,-90){$\tilde{E}^-_6$}
\put(0,20){\circle{10}}
\qbezier(0,15)(0,13)(0,5)
\put(0,0){\circle{10}}
{\put(0,-40){\circle{10}}}
\put(0,-60){\circle{10}}
\put(0,0){\qbezier(0,-5)(0,-20)(0,-35)}
\put(0,-40){\qbezier(0,-5)(0,-10)(0,-15)}
\put(15,-20){{\circle{10}} \put(-20,0){\qbezier(-5,0)(2,0)(5,0) }}
\put(30,-20){\circle{10} \put(-20,0){\qbezier(0,0)(2,0)(5,0) }}
\put(0,-20){\circle*{2}}
}

}

\put(230,40){
\put(-10,40){$\tilde{D}_6$}
\put(-15,25){\circle{10}}
\put(15,25){\circle{10}}
\put(0,10){\circle{10}}
\put(0,-10){\circle{10}}
\put(-15,-45){\circle{10}}
\put(15,-45){\circle{10}}
\put(0,-30){\circle{10}}
\qbezier(0,5)(0,0)(0,-5.5)
\qbezier(5,12)(8,15)(14,21)
\qbezier(-5,12)(-8,15)(-14,21)
\qbezier(-5,-32)(-8,-35)(-14,-41)
\qbezier(5,-32)(8,-35)(14,-41)
\qbezier(0,-15)(0,-18)(0,-25.5)
}

\end{picture}
\end{center}
\caption{\footnotesize{\bf Degeneration of singular fibers of a  small resolution of the $\SU(5)$ model}. 
 Starting with codimension-one in the base, the codimension increases from left to right. 
 In comparaison with the conjecture fiber structure, there are no fibers of type $\tilde{A}_6$ and the fibers of type $\tilde{E}_6$ are replaced by (non-Kodaira) fibers of type  $E_6$ or  $\tilde{E}^-_6(:=T^-_{3,3,3})$ according to the choice of the resolution.    We obtain six different resolutions, four of which have fibers of type  $E_6$ in codimension-three while two have  fibers of type $\tilde{E}^-_6$.}
 \label{su5resolved.short}
\end{figure}

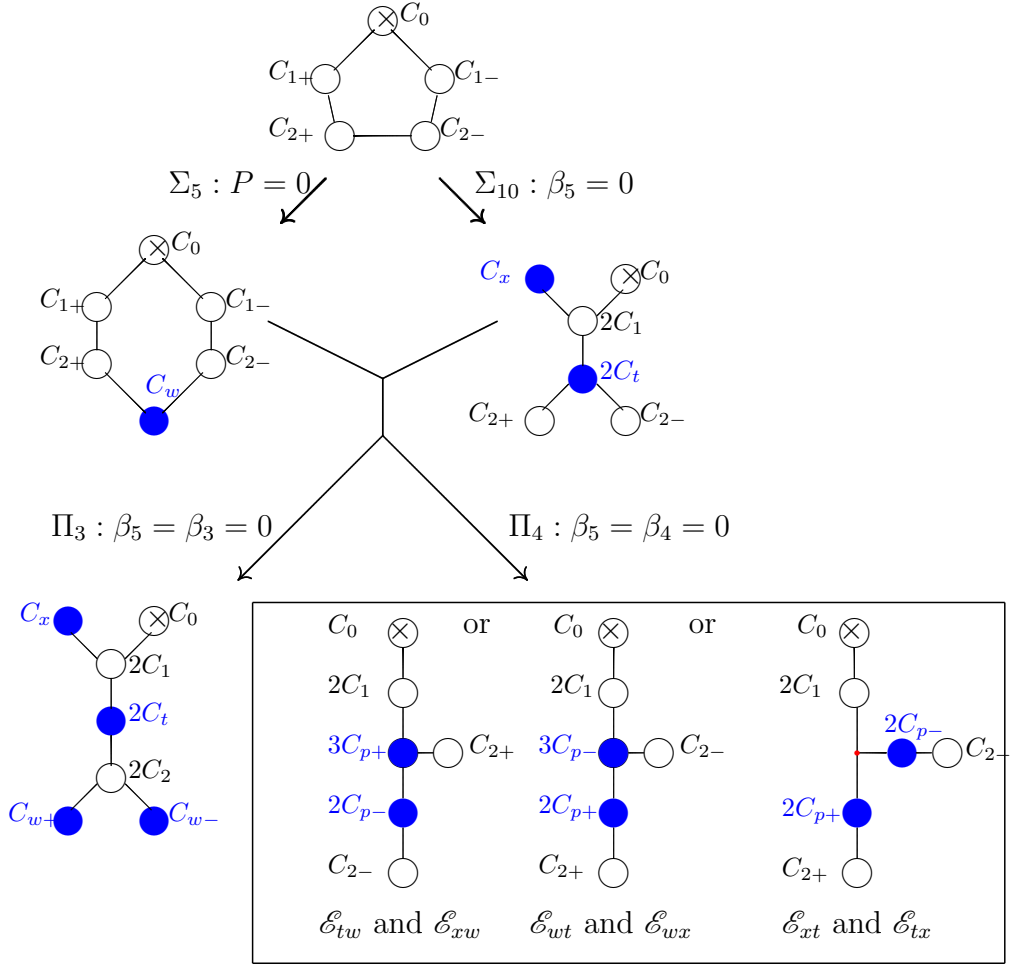
\begin{figure}[ htb ]
\setlength{\unitlength}{.38 mm}
\begin{picture}(25,340)(-65,-300)

\put(100,25){
\put(6,0){\footnotesize $C_0$}
\put(-3,-2){{$\times$}}
\put(25,-20){\footnotesize $C_{1-}$}
\put(-40,-20){\footnotesize $C_{1+}$}
\put(20,-40){\footnotesize $C_{2-}$}
\put(-40,-40){\footnotesize $C_{2+}$}
\put(0,0){\circle{10}}
\put(-20,-20){\circle{10}}
\put(20,-20){\circle{10}}
\put(15,-40){\circle{10}}
\put(-15,-40){\circle{10}}
\qbezier(-3,-3)(-5,-5)(-17,-17)
\qbezier(3,-3)(5,-5)(17,-17)
\qbezier(-19,-26)(-18,-30.5)(-17,-35)
\qbezier(19,-26)(18,-30.5)(17,-35)
\qbezier(-10,-40)(-10,-40)(10,-40)

\put(35,-70){
\linethickness{.3mm}
\qbezier(.5,-.5)(-7,7)(-15,15)
\qbezier(.5,-.5)(-1.1,2)(.2,4)
\qbezier(.5,-.5)(-2,1.1)(-4,0)
\put(-3,10){$\Sigma_{10}:\beta_5=0$}
}

\put(-35,-70){
\linethickness{.3mm}
\qbezier(-.5,-.5)(10,10)(15,15)
\qbezier(-.5,-.5)(1.1,2)(-.2,4)
\qbezier(-.5,-.5)(2,1.1)(4,0)
\put(-40,10){$\Sigma_5: P=0$}
}
}

\put(170,-90){
\put(-15,25){\color{blue}\circle*{10}}
\put(15,25){\circle{10}}
\put(12.5,23){{$\times$}}
\put(0,10){\circle{10}}
\put(0,-10){\color{blue}\circle*{10}}
\put(-15,-25){\circle{10}}
\put(15,-25){\circle{10}}
\put(6,7){\footnotesize $ 2C_{1}$}
\put(6,-10){\footnotesize \color{blue}$2C_t$}
\put(20,-25){\footnotesize $C_{2-}$}
\put(-40,-25){\footnotesize $C_{2+}$}
\put(20,25){\footnotesize $C_{0}$}
\put(-36,25){\footnotesize\color{blue} $C_x$}
\qbezier(0,5)(0,0)(0,-5.5)
\qbezier(5,12)(8,15)(14,21)
\qbezier(-5,12)(-8,15)(-14,21)
\qbezier(-5,-12)(-8,-15)(-14,-21)
\qbezier(5,-12)(8,-15)(14,-21)

}

\put(20,-75){
\put(6,20){\footnotesize $C_0$}
\put(-3,18){{$\times$}}
\put(25,0){\footnotesize $C_{1-}$}
\put(-40,0){\footnotesize $C_{1+}$}
\put(25,-20){\footnotesize $C_{2-}$}
\put(-40,-20){\footnotesize $C_{2+}$}
\qbezier(-20,-5)(-20,-10)(-20,-15)
\qbezier(20,-5)(20,-10)(20,-15)

\put(0,-40){\color{blue}\circle*{10}}
\put(-20,0){\circle{10}}
\put(20,0){\circle{10}}
\put(-20,-20){\circle{10}}
\put(20,-20){\circle{10}}
\put(0,20){\circle{10}}

\put(-3,-30){\footnotesize\color{blue} $C_{w}$}
\qbezier(-3,17)(-5,15)(-17,3)
\qbezier(3,17)(5,15)(17,3)
\qbezier(-3,-37)(-5,-35)(-17,-23)
\qbezier(3,-37)(5,-35)(17,-23)

}

\put(5,-210){
\put(-15,25){\color{blue}\circle*{10}}
\put(15,25){\circle{10}}
\put(12,23){{$\times$}}
\put(0,10){\circle{10}}
\put(0,-10){\color{blue}\circle*{10}}
\put(-15,-45){\color{blue}\circle*{10}}
\put(15,-45){\color{blue}\circle*{10}}
\put(0,-30){\circle{10}}
\put(6,-30){\footnotesize $ 2C_{2}$}
\put(6,7){\footnotesize $ 2C_{1}$}
\put(6,-10){\footnotesize \color{blue}$2C_{t}$}
\put(20,-45){\footnotesize \color{blue}$C_{w-}$}
\put(-36,-45){\footnotesize \color{blue}$C_{w+}$}
\put(20,25){\footnotesize $C_{0}$}
\put(-32,25){\footnotesize\color{blue} $C_x$}
\qbezier(0,5)(0,0)(0,-5.5)
\qbezier(5,12)(8,15)(14,21)
\qbezier(-5,12)(-8,15)(-14,21)
\qbezier(-5,-32)(-8,-35)(-14,-41)
\qbezier(5,-32)(8,-35)(14,-41)
\qbezier(0,-15)(0,-18)(0,-25.5)
}

\put(65,-210){
\setlength{\unitlength}{.40 mm}
\put(40,20){
\put(0,0){\circle{10}}
\put(-5,-2){{$\times$}}
\put(-25,0){\footnotesize $C_0$}
\put(20,0){or}
\qbezier(0,-5)(0,-7)(0,-15)
\put(-25,-20){\footnotesize $2 C_{1}$}
\put(0,-20){\circle{10}}
\put(0,-20){
\put(0,-20){\circle{10}}
{\color{blue}\put(0,-20){\circle*{10}}
\put(0,-40){\circle*{10}}}
\put(0,-60){\circle{10}}
\multiput(0,0)(0,-20){3}{\qbezier(0,-5)(0,-10)(0,-15)}
\put(15,-20){\circle{10}}
\qbezier(5,-20)(7,-20)(10,-20)
\put(22,-20){\footnotesize $C_{2+}$}
{\color{blue} \put(-25,-20){\footnotesize $3C_{p+}$}
\put(-25,-40){\footnotesize $2C_{p-}$}}
\put(-25,-60){\footnotesize $C_{2-}$}
\put(-28,-80){{$\mathscr{E}_{tw} \ \text{and}\ \mathscr{E}_{xw}$}}
}
}

\put(110,20){

\put(0,0){\circle{10}}
\put(-5,-2){{$\times$}
}
\put(-20,0){\footnotesize $C_0$}
\qbezier(0,-5)(0,-7)(0,-15)
\put(25,0){{or}}
\put(-21,-20){\footnotesize $2 C_{1}$}
\put(0,-20){\circle{10}}
\put(0,-20){
\put(0,-20){\circle{10}}
{\color{blue}\put(0,-20){\circle*{10}}
\put(0,-40){\circle*{10}}}
\put(0,-60){\circle{10}}
\multiput(0,0)(0,-20){3}{\qbezier(0,-5)(0,-10)(0,-15)}
\put(15,-20){\circle{10}}
\qbezier(5,-20)(7,-20)(10,-20)
\put(22,-20){\footnotesize $C_{2-}$}
{\color{blue} \put(-25,-20){\footnotesize $3C_{p-}$}
\put(-25,-40){\footnotesize $2C_{p+}$}}
\put(-25,-60){\footnotesize $C_{2+}$}
\put(-28,-80){{$\mathscr{E}_{wt} \  \text{and}\  \mathscr{E}_{wx}$}}
}

}

\put(190,20){
\put(0,0){\circle{10}}
\put(-5,-2){{$\times$}}
\put(-19,0){\footnotesize $C_0$}
\qbezier(0,-5)(0,-7)(0,-15)
\put(-25,-20){\footnotesize $2 C_{1}$}
\put(0,-20){\circle{10}}
\put(-2.5,-20){
{\color{red}
\color{blue}\put(0,-40){\circle*{10}}}
\put(0,-60){\circle{10}}
\put(0,0){\qbezier(0,-5)(0,-20)(0,-35)}
\put(0,-40){\qbezier(0,-5)(0,-10)(0,-15)}
\put(15,-20){{\color{blue}\circle*{10}} \put(-20,0){\qbezier(-5,0)(2,0)(5,0) }}
\put(30,-20){\circle{10} \put(-20,0){\qbezier(0,0)(2,0)(5,0) }}
\put(36,-21){\footnotesize $C_{2-}$}
{\color{blue} \put(9,-13){\footnotesize $2C_{p-}$}
\put(-25,-41){\footnotesize $2C_{p+}$}}
\put(-25,-61){\footnotesize $C_{2+}$}
\color{red}\put(0,-20){\circle*{2}}
\color{black}
\put(-25,-80){{$\mathscr{E}_{xt}\  \text{and}\  \mathscr{E}_{tx}$}}
}
}

\put(-10,0){
\linethickness{.2mm}
\put(0,30){\line(1,0){250}}
\put(250,30){\line(0,-1){120}}
\put(0,30){\line(0,-1){120}}
\put(0,-90){\line(1,0){250}}
}
}

\put(50,-170){
\linethickness{.2mm}
\qbezier(-.5,-.5)(17,17)(50,50)
\qbezier(-.5,-.5)(1.1,2)(-.2,4)
\qbezier(-.5,-.5)(2,1.1)(4,0)
\put(-66,15){$\Pi_3:\beta_5=\beta_3=0$}

\put(50,50)
{\qbezier(0,0)(0,10)(0,20)}

\put(50,70){
\qbezier(0,0)(20,10)(40,20)
\qbezier(0,0)(-20,10)(-40,20)
}

\put(100,0){
\linethickness{.2mm}
\qbezier(.5,-.5)(-17,17)(-50,50)
\qbezier(.5,-.5)(-1.1,2)(.2,4)
\qbezier(.5,-.5)(-2,1.1)(-4,0)
\put(-6,15){$\Pi_4:\beta_5=\beta_4=0$}
}

}

\end{picture}
\caption{  \footnotesize{Fiber degeneration of a small resolution of the   $\SU(5)$ model. 
The node $C_0$ is the one that touches the section.  The nodes $C_{1\pm}$ and $C_{2\pm}$ are coming from the resolution of the singularities over a generic point of the divisor $D_{\su(5)}$. 
The remaining nodes  $C_{x}$, $C_{w}$ and $C_t$ are obtained from the resolution of the higher codimensional singularities. We have 6 possible resolutions $\mathscr{E}_{xw}$, $\mathscr{E}_{wx}$, $\mathscr{E}_{xt}$, $\mathscr{E}_{tw}$ and $\mathscr{E}_{wt}$.\label{res.1}
 }
}
\end{figure}

 In this paper, we construct a  crepant resolution of the $\SU(5)$ model geometry. The resolution we obtain defines a {\em flat elliptic fibration} so that all the fibers have the same dimension.  The singular Weierstrass equation defining the $\SU(5)$ model admits a  quite  complicate discriminant, but we can resolve all the  singularities in two easy steps by working directly with the Weierstrass equation.
In the first step, we resolve the singularity that generates the $\tilde{A}_4$ fibers over a generic point of the divisor $D_{\su(5)}$. This is done  by two successive blow-ups of the singular locus. The generic fiber over $D_{\su(5)}$ is then a Kodaira fiber of type $I_5$ with its dual graph the extended Dynkin diagram $\tilde{A}_4$. After the resolution of the generic singularity, there are  still some left-over singularities in codimension-two and codimension-three in the base. In a second step we define a simultaneous resolution of these higher codimension singularities. One cannot just blow-up the singular locus since in codimension-two and three it will destroy the flatness of the fibration by introducing higher dimensional components on the fibers. We construct instead a small resolution by using the fact that  after resolving the singularities over the codimension-one locus, the remaining singularities are  described by a binomial variety. The binomial variety admits as its singular locus  three lines of conifold singularities  intersecting at a common point where the singularity worsen. There are six small resolutions of this geometry related to each other by a network of flop transitions that form a dihedral group $\mathrm{Dih}_6$ of twelve elements. These binomial transformations are induced from the symmetry of the  binomial variety before the resolution. The small resolution of the binomial geometry leads to two possible types of fiber enhancement $\tilde{A}_4\rightarrow (\tilde{A}_5,\tilde{D}_5)\rightarrow (\tilde{D}_6, E_6)$ or $\tilde{A}_4\rightarrow (\tilde{A}_5,\tilde{D}_5)\rightarrow (\tilde{D}_6, E_6^{-})$. The two possibilities differ in codimensio-three by the fiber over the points where a $\tilde{E}_6$ enhancement is usually expected. Here we see that at these points , we have instead a projective Dynkin diagram $E_6$ or a a  fiber with dual graph  $\tilde{E}_6^{-} :=T^-_{2,2,2}$. The $\tilde{E}^{-}_6$ diagram is not ADE and can be better described as a $\tilde{E}_6$ diagram with the central node contracted to a point. The physical meaning of such fibers is not clear at all. We note that they appear at points in the base where the string coupling is of order 1, more precisely $\tau=\mathrm{exp}(\pi i/3 )$. We also note that the fiber $E_6$ we obtain in codimension-three, is really an exotic version of a $\tilde{D}_5$ fiber since once we remove the node that touches the section, we get a dual graph $D_5$. Interestingly, in the degeneration of singular fibers from codimension-two to codimension-three, the rank of the dual graph increases only for the $\tilde{D}_6$ fiber while  in the case of the the $E_6$ and the $\tilde{E}^{-}_6$ fibers have there is no increase of rank. 

The results of this paper can then be summarized by the following propositions and the fiber degenerations  are 
illustrated in figures \ref{su5resolved.short} and \ref{res.1}.
\begin{prop}[Resolution in codimension-one]\label{prop.Codim1}
The resolution of the singularities in codimension-one is obtained by two successive blow-ups leading to a split $I_5$ fiber over  the divisor $D_{\su(5)}$. 
\end{prop}

\begin{prop}[The binomial geometry]\label{Prop.Binomial}
After resolving the singularities in codimension-one, we  are left with  higher codimensional singularities all  visible in a unique patch where the variety is locally described by the following   affine  binomial variety in $\mathbb{C}[x,w,t,y,s]$: 
$$
y s -  x w t =0, 
$$
where  $s=y+\beta_5+w\beta_3$ and $t=x+\beta_4+w \beta_2 + w^2 \beta_0$. The singular locus $y=s=xw=wt=xs=0$ of the binomial variety is composed of  three lines of conifold singularities meeting at a common point where the singulary enhances. 
\end{prop}
\begin{prop}[A network of small resolutions]\label{Prop.Network}
Using the binomial geometry, we determine six small resolutions $\{{\mathscr{E}}_{xw},{\mathscr{E}}_{xt} ,{\mathscr{E}}_{wt}, {\mathscr{E}}_{wx},{\mathscr{E}}_{tx}, {\mathscr{E}}_{wt}\}$: 
$$
{\mathscr{E}}_{u_1 u_2 }
\begin{cases}
u_1 \alpha_- -\sigma_- y=0, \\
u_2 \alpha_+ -\sigma_+ s=0, \\
\alpha_-\alpha_+ -\sigma_-\sigma_+ u_3=0
\end{cases}
$$
where $(u_1, u_2, u_3)$ is  a permutation of   $(x,w,t)$ and $[\alpha_-:\sigma_-]\times[\alpha_+:\sigma_+]$ are projective coordinates of $\mathbb{F}_0=\mathbb{CP}^1\times\mathbb{CP}^1$.
The singular fibers\footnote{
This describe all the fibers above the divisor $D_{\su(5)}$ with the exception of the node $C_0$ that touches the section. 
} are at   $wx=0$.  
\end{prop}

\begin{prop}[ Fiber enhancements in codimension-two]  \label{Prop.Codim2}  
In codimension-two, we have an enhancement of the fiber $I_5$ to a fiber with dual graph $\tilde{D}_5$   over the curve  $\Sigma_{10}:\beta_5=0$  and an enhancement to a fiber with dual graph $\tilde{A}_6$ along the curve $\Sigma_5:P=0$.  
\end{prop}

\begin{prop}[ Fiber enhancements in codimension-three]\label{Prop.Fiber}
In codimension- three, we have a  fiber enhancement at the intersection of the two curves $\Sigma_{5}$ and $\Sigma+{10}$. This intersection contains two types of points that we call $\Pi_3$ and $\Pi_3$  ( $\Sigma_5\cap\Sigma_{10}=\Pi_3\cup \Pi_4$):  
\begin{enumerate}
\item Over  $\Pi_3:\beta_5=\beta_3=0$, we get a fiber with dual graph $\tilde{D}_6$. 
\item  Over   $\Pi_4:\beta_5=\beta_4=0$, we get   exotic fibers which don't increase the number of nodes of the graph. The precise dual graph of these exotic fibers  depends on the choice of the resolution. For the  resolutions ${\mathscr{E}}_{xt}$ and ${\mathscr{E}}_{xt}$ the fiber is of a new type that we call $\tilde{E}^-_6$ while it is a projective $E_6$ fiber  for the other 4 small resolutions. 
The fiber $\tilde{E}_6^-$ is not an ADE fiber.  It is a bouquet  of three 2-chains intersecting at a common point. 
\item Over  $\Pi_7: P=R=0$, there is  no  enhancement, the fiber is still a $\tilde{A}_5$.
\end{enumerate}
\end{prop}

\begin{prop}[  Flop transitions]\label{Prop.Floptransitions}
All the six different resolutions are connected to each other by a network of  flop transitions forming a discrete dihedral group $Dih_{6}$ of 12 elements.
\end{prop}

Finally, we can rule out the existence of the conjectured resolution in a large class of varieties: 

\begin{prop}[  Birational invariance and a no-go theorem]\label{Prop.Nogo}
In the category of smooth projective complex algebraic varieties, two varieties related by a birational map that does not change the canonical class have the same Betti numbers as proven by Batyrev \cite{Batyrev}. It follows immediately that it is impossible to find a resolution of singularities that will have the same fibers as those obtained in the six small resolutions $\tilde{E}_6$, but will  have  fibers of type $\tilde{E}_6$ over the points $\Pi_4:\beta_5=\beta_4=0$ or fibers of type $\tilde{A}_6$ over $\Pi_7: P=R=0$. 
\end{prop}
If the Weierstrass model we start with is modified by specializing some of the sections $\beta_i$,   one can end up with a very different structure of singular fibers. We are currently exploring such options.

 \subsection{Structure of the paper}

In section  \ref{ellipticfibrations}, we review several mathematical results on the structure of elliptic fibrations; Weierstrass models,  Kodaira classification, Tate algorithm and Miranda models. 
 In section \ref{SU5}, we introduce in details the specific Weierstrass model that describes $\SU(5)$ models. 
We will also review  the  conjecture on its fiber geometry in section \ref{conjecture}. In section \ref{codimone}, we resolve the singularities in codimension-one over the base  and present the  underlying binomial geometry in section \ref{binomial}. We also discuss the toric structure of the binomial variety and study its small resolutions both torically and algebraically. 
We will also analyze the network of flop transitions between the different possible small resolutions we have obtained. 
In section \label{SU5res}, we present the small resolution of the codimension-two and codimension-three singularities of the $\SU(5)$ by exploiting the resolution of the binomial variety. We also analyze the  resulting fiber geometry. 
Finally we conclude in section \ref{conclusions}.

%\clearpage 

\section{Elliptic fibrations and collisions of singularities \label{ellipticfibrations}}
In this section, we review some important results on  elliptic fibrations and their singular fibers in different codimensions. This will put into perspective the conjectured fiber structure of $\SU(5)$ models in F-theory. In subsection \ref{KodairaTate}, we review Kodaira classification  and  Tate algorithm. Section \ref{Miranda} is an introduction to Miranda models. Miranda models describe certain elliptic threefolds which are flat fibrations with a restricted list of possible collisions of singular fibers. They provide (counter)examples to several common believes on the nature of colliding singularities. In particular, over loci of  codimension two or higher in the base, the singular fibers don't have to be Kodaira fibers and the rank of the singular fiber at a collision of singularities does not necessary increase, it can even decrease. We also quickly describe in section \ref{Szydlo} the generalization of Miranda models to $n$-fold by Szydlo.  We will start by a quick review on elliptic curves over $\mathbb{C}$.

\subsection{Elliptic fibrations and Weierstrass models}
An elliptic curve is a smooth complex curve of genus one with a marked point on it.  Topologically it is a two-torus $T^2=S^1\times S^1$ with a  selected point. It is also an Abelian variety  with the mark point as its neutral element. 
Modulo a similitude transformation, an elliptic curve over the complex number is always equivalent to the complex torus   $\mathbb{C}/  (\mathbb{Z}+\tau\mathbb{Z})$ described as the quotient  of the complex plane by the double-lattice generated by   $1$ and the complex number $\tau$ (the {period}) .  Geometrically, the period  $\tau$,  characterizes the shape of the complex-torus and by convention, is restricted to live on the  upper-half plane $\mathscr{H}=\{ \tau\in \mathbb{C} |  \   \mathrm{Im}(\tau)>0\}$. 
Two 2-tori  are equivalent modulo similitude if their  period ratios  are related by a modular transformation:
\begin{equation}
\tau \mapsto \frac{a \tau + b }{c\tau +d}, \quad \begin{pmatrix} a & b \\ c & d \end{pmatrix}\in \Sl(2,\mathbb{Z}).
\end{equation} 

\begin{figure}[thb]
\begin{center}
\setlength{\unitlength}{1.6 mm}
\begin{picture}(15,9)(0,0)
\put(0,0){
\qbezier(0,0)(4,0)(15,0)
\qbezier(0,0)(0,4)(0,10)
\qbezier(0,0)(3,3)(6,6)
\qbezier(6,6)(10,6)(14,6)
\qbezier(8,0)(11,3)(14,6)
\put(15,0){\qbezier(0,0)(-.5,.2)(-.7,.7)
\qbezier(0,0)(-.5,-.2)(-.7,-.7)
}
\multiput(3,3)(8,0){2}{\qbezier(-.7,0)(0,0)(.7,0)}
\multiput(3.7,0)(6,6){2}{\qbezier(-.7,-.7)(0,0)(.7,.7)}
\multiput(4.3,0)(6,6){2}{\qbezier(-.7,-.7)(0,0)(.7,.7)}
\put(5,6.8){ \text{\small $\tau$}}
\put(13,6.8){ \text{\small $\tau+1$}}
\put(8,-2){\text{\small $1$}}
\put(0,-2){\text{\small $0$}}
\put(0,10){\qbezier(0,0)(.2,-.5)(.7,-.7)\qbezier(0,0)(-.2,-.5)(-.7,-.7)}
}
\end{picture}
\caption{A torus seen as the quotient $\mathbb{C}/ (\mathbb{Z}+\tau \mathbb{Z})$.} 
\end{center}
\end{figure}
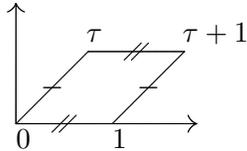
Using the Riemann-Roch theorem, it is easy to show that an elliptic curve can always be expressed by a cubic equation in  $\mathbb{CP}^2$. When the characteristic is different from  2 and 3, the cubic equation  can be reduced to a (reduced) Weierstrass equation:
\begin{equation}
y^2z= x^3 + F x z^2 + G z^3,
\end{equation}
where $[x:y:z]$ are the projective coordinates of $\mathbb{CP}^2$. We take   $z=0$ to be the line at infinity. It cuts the  Weierstrass equation at the rational point   $[0:1:0]$ which is considered as the origin of the elliptic curve. The elliptic curve is regular if and only if  $4 F^3+27 G^2\neq 0$.The moduli space of  complex tori  modulo similitudes is given by the quotient of the upper-half plane by  modular transformations. There is a function, called the {\em Klein $j$-invariant} which does not change under modular transformations and maps the moduli space of complex  tori modulo similitudes to the  complex plane $\mathbb{C}$. Two elliptic curves are isomorphic if and only if they have the same $j$-invariant.  The $j$-invariant can be computed from the Weierstrass equation as follows:
\begin{equation}
j(\tau)=1728 \frac{4 F^3}{4 F^3 + 27 G^2}\quad \text{or equivalently } \quad j(\tau)=1728 -  \frac{27  G^2}{4 F^3 + 27 G^2}. \label{RWF}
\end{equation}
It admits the following expression as a Laurent series:
\begin{equation}
j(\tau)=\frac{1}{q}+744+ \sum_{n>0} c_n q^n, \quad c_n \in \mathbb{N}, \quad q=\exp(2\pi \mathrm{i}  \tau).
\end{equation}
The modular group admits as a {\em fundamental domain}  the closure of the region :
$$R_{\Gamma}=\{ \tau \in \mathscr{H}: |\tau+\overline{\tau}|<1 \    \text{and}\   |\tau|>1\},$$
 with a $\mathbb{Z}_2$  identification on the boundary given by $\tau \sim -\overline{\tau}$.  
 When we have to make a choice between two points on the boundary, we will take the one with negative real part. 
It is common to use the  following normalization: 
\begin{equation}
J(\tau)=\frac{1}{1728}j(\tau)=\frac{4 F^3}{4 F^3 + 27 G^2}.
\end{equation} 
We recall some additional properties of the  $J$-invariant:
\begin{equation}
 J(\mathrm{i})=1, \quad J(\mathrm{e}^{\frac{2\pi}{3} \mathrm{i}})=0, \quad  J(-\overline{\tau})=\overline{J(\tau)},  
 \quad \lim_{Im(\tau)\rightarrow+ \infty} |J(\tau)|=\infty. 
 \end{equation}
If is useful to also include tori admitting an infinite value for the  $j$-invariant. This corresponds to allowing an infinite value for the imaginary part of $\tau$. By the action of the modular group, we should then also  include  all the cusps (rational points of the real line). The $j$-invariant is then an isomorphism between the moduli space of tori (modulo similitude)  and the complex sphere $\mathbb{CP}^1$:
\begin{equation}
J: \overline{\mathscr{H}}\rightarrow \mathbb{CP}^1: J(\tau)\mapsto [U,V]=[4 F^3:4 F^3 + 27 G^2],
\end{equation}
where $[U:V]$ denotes the projective coordinates of $\mathbb{CP}^1$.
 In F-theory, since the imaginary part of $\tau$ is the inverse of the string coupling, infinite value of $\tau$ corresponds to the weak coupling limit of F-theory\cite{Sen:1997gv}.

 An {\em elliptic fibration} is a proper surjective morphism  $\pi: Y\rightarrow B$ between complex varieties with connected fibers such that the general fiber is  an elliptic curve. Here we consider elliptic fibrations with a smooth section.  In F-theory,  the section identifies the  physical space seen in type IIB string theory while the elliptic fiber  determines the value of the  type IIB axio-dilaton field.  An elliptic fibration with a section is birationally equivalent to a  (singular) {\em Weierstrass model}.  

Let $\mathscr{L}$ be a line bundle defined over $B$. We define  $E=\mathscr{O}\oplus\mathscr{L}^{\otimes(-2)}\oplus \mathscr{L}^{\otimes(-3)}$, and we consider the projectization $\mathbb{P}(E)$, with tautological line bundle $\mathscr{O}(1)$. A {
 Weierstrass model} is defined as an hypersurface $Y$ of $\mathbb{P}(E)$, with equation:
\begin{equation}
Y:\quad z y^2 +  a_1  x y z + a_3  y z^2 -(x^3+ a_2 x^2 z + a_4 x z^2+ a_6 z^3)=0.\label{ai1}
\end{equation}
Here $z$, $x$ and $y$ are sections of $\mathscr{O}(1)$, $\mathscr{O}(1)\otimes \mathscr{L}^{\otimes (2)}$ and $\mathscr{O}(1)\otimes \mathscr{L}^{\otimes (3)}$, respectively. 
Each variable  $a_i$ ($i=1,2,3,4,6)$ is a section of  the line bundle $\mathscr{L}^{\otimes (i)}$.  
One can always write the Weierstrass model in its reduced form  by completing the square in $y$ and the cube in $x$ in the expression of the generalized Weierstrass model \eqref{ai1}. The section of the fibration is given by $z=0$ and defines a divisor in $Y$ isomorphic to $B$. The conormal bundle of the section is isomorphic to $\mathscr{L}$.
Following Deligne's formulaire \cite{Formulaire},  it is useful to introduce the following variables: 
\begin{equation}\label{Deligne}
\begin{cases}
b_2 & = a_1^2 + 4 a_2, \quad 
b_4  = a_1 a_3 + 2 a_4, \quad
b_6 =a_3^2+ 4 a_6,\\
c_4 &= b_2^2-24 b_4,\quad 
c_6 =-b_2^3 + 36 b_2 b_4 -216 b_6 , \\
\Delta &= -\frac{1}{4}b_2^2 (b_2 b_6 -b_4^2)-8 b_4^3 -27 b_6^2 + 9 b_2 b_4 b_6,
\end{cases}
\end{equation}
from which we can express the  reduced Weierstrass  form \eqref{RWF} with:  
\begin{equation}
F= -\frac{1}{48} c_4 , \quad G=- \frac{1}{864} c_6.  
\end{equation}
In terms of the reduced Weierstrass model, the discriminant $\Delta$ is simply: 
\begin{equation}
\Delta=-16(4 F^3+27 G^2).\label{discr}
\end{equation}
The discriminant locus is the set of points of the base over which the elliptic fiber is not regular.
An elliptic curve in the Tate's form \eqref{ai1} always admits the following $\mathbb{Z}_2$ symmetry, which corresponds to the inverse law of the torus group\cite{Formulaire}:
\begin{equation}\label{z2}
(x,y)\mapsto (x, -y-a_1 x-a_3).
\end{equation}
This involution plays an central role  in the description of orientifolds in F-theory. It is related to the discrete operator $(-1)^{F_L}\Omega$ defined in perturbative string theory. Here  $F_L$ is the 
left-hand fermion number and $\Omega$ is the worlsheet parity inversion. From the point  of view of the modular group, it is just  minus the identity ($-\mathrm{I}_2$) and acts trivially on the period \cite{Sen:1997gv}.

\subsection{ Kodaira classification and Tate algorithm \label{KodairaTate}} 

 In the early 1960s,  Kodaira has classified singular fibers of (minimal) elliptic surfaces \cite{Kodaira}. 
Kodaira's classification can be  extended to singular fibers over codimension-one points in the base modulo mild assumptions. He identified  8 types of singular fibers, including two infinite series ($I_1, I_n (n>0),  II, III, IV, I_n^*, II^*, III^*$). 
Each fiber is constituted of intersecting rational curves and the dual graph of such singular fibers are closely  related to   simply-laced  (affine) Dynkin diagrams. In Kodaira list, there are only two  irreducible singular fibers : nodal curves (type $I_1$) and  cusp (type $II$). All the remaining fibers are reducible. There are three that are not extended ADE diagrams: type $I_2$ is composed of  two rational curves meeting transversally at two points, type $III$ consists of wo rational curves tangent at a point and type $IV$ is composed of three rational curves meeting transversally at a common point (type $IV$). The remaining types are extended ADE Dynkin diagram: the dual graph  $\tilde{A}_{n}$ gives type $I_n$ ($n\leq 3$), the dual graph $\tilde{D}_{4+n}$ corresponds to type $I^*_{n}$, the dual graphs  $\tilde{E}_6$, $\tilde{E}_7$ and $\tilde{E}_8$ correspond respectively to type $IV^*$,  $III^*$ and  $II^*$.  
In 1964, Andr\'e N\'eron has reproduced Kodaira's list by explicitly constructing regular models for elliptic surfaces defined by   Weierstrass models\cite{Neron}.  
In 1975, Tate,  using N\'eron results, 
 has developed an algorithm which gives the type of singular fibers of an elliptic surface by manipulating the coefficients of the  Weierstrass equation \cite{Tate}. 

In F-theory, Tate's algorithm has been refined to determine the gauge groups (even   non-simply laced ones) associated to seven-branes wrapping  divisors of the discriminant locus \cite{FTheoryTate}. Non-simply laced Dynkin diagrams are included in a Kodaira fibers those nodes that are exchanged by the action of  monodromies. Some specific examples also extended  Tate's algorithm to certain codimensions two loci in order to predict the matter contents  at the intersection of two divisors of the discriminant locus.    
Unfortunately, in the physics literature, Tate's algorithm is often used outside of its domain of validity.  This has created important  misconceptions on the structure of singular fibers over points located in higher codimensions in the base.

\begin{table}[hbt]
\begin{center}
\begin{tabular}{|c|c|c| c | l  |c|c|}
\hline 
{\small Type }&  ${\rm ord}(F)$ & ${\rm ord}(G)$ & ${\rm ord}(\Delta)$   
                                                   
                                                                                          & $J$  &  
                                                                                          
                                                                                          {\footnotesize Monodromy}                & Fiber \\
                               \hline 
$I_0$ & $\geq 0$  & $\geq 0$ &  {\small $0$} & 
{\small
$\mathbb{R}$
}
 & $\mathrm{I}_2$ & Smooth torus \\
\hline
$I_1$ & $0$ & $0$ &  {\small $1$}  & $\infty$ &
$
\begin{pmatrix}
1& 1\\
0 & 1
\end{pmatrix}
$

 &
{  \setlength{\unitlength}{1 mm}
\begin{picture}(10,10)(-16,-4)
\put(-5,0){
\qbezier(-5,0)(4,6)(5,0)
\qbezier(-5,0)(4,-6)(5,0)
}
\put(-10,0){\qbezier(0,0)(-1,-1)(-3,-3) 
\qbezier(0,0)(-1,1)(-3,3) 
\put(-26,0){\footnotesize (Nodal curve)}
}
\end{picture}}
 
\\
\hline
$I_n$ &{\small $0$} &  {\small $0$} & {\small $n>1$ }  & $\infty$ & 
{\small $
\begin{pmatrix}
1& n\\
0 & 1
\end{pmatrix}
$
}
 & 
{  \setlength{\unitlength}{.9 mm}
\begin{picture}(35,13)(-5,-2)
\put(-6,6){\circle{4}}
\put(-1,1){\line(-1,1){4}}
\put(0,0){\circle{4}}
\put(2.2,0){\line(1,0){2}}
\put(6.4,0){\circle{4}}
\put(8,0){\line(1,0){2}}
\put(12,0){\circle{4}}
\put(14,0){\line(1,0){1}}
\put(16,0){\line(1,0){1}}
\put(18,0){\line(1,0){1}}
\put(20,0){\line(1,0){1}}
\put(22,0){\circle{4}}
\put(23,2){\line(1,1){4}}
\put(28,7){\circle{4}}
\put(-6.6,5.3){\tiny{$1$}}
\put(-.5,-.5){\tiny{$1$}}
\put(5.9,-.5){\tiny{$1$}}
\put(12,-.5){\tiny{$1$}}
\put(21.4,-.5){\tiny{$1$}}
\put(-.5,6.6){\tiny{$1$}}
\put(5.9,6.6){\tiny{$1$}}
\put(12,6.6){\tiny{$1$}}
\put(27.4,6.6){\tiny{$1$}}

\put(0,7){\circle{4}}
\put(2.2,7){\line(1,0){2}}
\put(6.4,7){\circle{4}}
\put(8,7){\line(1,0){2}}
\put(12,7){\circle{4}}
\multiput(14,7)(2,0){6}{\line(1,0){1}}
\put(-4,7){\line(1,0){2}}
\end{picture}}
 \\
\hline
$II$ & $\geq 1$ &  $1$ & $2$   &$0$& 
{\small $
\begin{pmatrix}
1& 1\\
-1 & 0
\end{pmatrix}
$

}
& 
{  \setlength{\unitlength}{1 mm}
\begin{picture}(10,10)(-10,-4)
\put(0,0){\qbezier(0,0)(2.7,.7)(5,5)
\qbezier(0,0)(2.7,-.7)(5,-5)
\put(-30,-.7){\footnotesize {Cuspidial curve}   }
}
\end{picture}}
 \\
\hline
$III$ & {\small$1$ } & {\small $\geq 2$ }  & {\small  $3$} &  {\small $1$} &
{\small $
\begin{pmatrix}
0 & 1\\
-1 & 0
\end{pmatrix}
$
}
  & 
{  \setlength{\unitlength}{.7 mm}
\begin{picture}(10,10)(-2,-2.5)
\put(0,0){
\qbezier(0,5)(4.242,4.242)(5,0)
\qbezier(0,-5)(4.242,-4.242)(5,0)}
\put(10,0){\qbezier(0,5)(-4.242,4.242)(-5,0)
\qbezier(0,-5)(-4.242,-4.242)(-5,0)
}
\end{picture}}
 \\
\hline
$IV$ &  {\small $\geq 2$} & {\small $2$} & {\small $4$}  &    {\small $0$} &
{\small $
\begin{pmatrix}
0 & 1\\
-1 & -1
\end{pmatrix}
$
}
 &  
{  \setlength{\unitlength}{1 mm}
\begin{picture}(10,10)(-5,-2)
\put(0,0){\qbezier(-2.5,-4.33)(0,0)(2.5,4.33)
\qbezier(-2.5,4.33)(0,0)(2.5,-4.33)
\qbezier(-5,0)(0,0)(5,0)
}
\end{picture}}
\\
\hline
$I^*_n$ & {\small $2$} & {\small $\geq 3$ } & {\small $n+6$} & {\small  $\infty$} & 
{\small $
\begin{pmatrix}
-1& -n\\
0 &- 1
\end{pmatrix}
$
}
  &  
{  \setlength{\unitlength}{.9 mm}
\begin{picture}(35,13)(-5,-2)
\put(-6,6){\circle{4}}
\put(-6,-6){\circle{4}}
\put(-1,-1){\line(-1,-1){4}}
\put(-1,1){\line(-1,1){4}}
\put(0,0){\circle{4}}
\put(2.2,0){\line(1,0){2}}
\put(6.4,0){\circle{4}}
\put(8,0){\line(1,0){2}}
\put(12,0){\circle{4}}
\put(14,0){\line(1,0){1}}
\put(16,0){\line(1,0){1}}
\put(18,0){\line(1,0){1}}
\put(20,0){\line(1,0){1}}
\put(22,0){\circle{4}}
\put(23,-1){\line(1,-1){4}}
\put(23,2){\line(1,1){4}}
\put(28,7){\circle{4}}
\put(28,-6){\circle{4}}
\put(-6.6,5.3){\tiny{$1$}}
\put(-6.6,-6.4){\tiny{$1$}}
\put(-.5,-.5){\tiny{$2$}}
\put(5.9,-.5){\tiny{$2$}}
\put(12,-.5){\tiny{$2$}}
\put(21.4,-.5){\tiny{$2$}}
\put(27.4,6.6){\tiny{$1$}}
\put(27.4,-6.6){\tiny{$1$}}
\end{picture}}
\\*
 \cline{2-4}  &  {\small $\geq 2$} &{\small  $3$} &{\small  $n+6$} &  %$I_n ^*$ & 
& &  \\
\hline
$IV^*$ & {\small  $\begin{matrix}\\   \geq 3\\  \\  \end{matrix}$} &{\small  $4$ }& {\small $8$ }&    {\small $0$ 
}& 
{\small $
\begin{pmatrix}
-1& -1\\
1 & 0
\end{pmatrix}
$
}

&
\setlength{\unitlength}{1 mm}
\begin{picture}(25,16)(2,-10)
\put(0,0){\circle{4}}
\put(2.2,0){\line(1,0){2}}
\put(6.4,0){\circle{4}}
\put(8.6,0){\line(1,0){2}}
\put(12.6,0){\circle{4}}
\put(14.6,0){\line(1,0){2}}
\put(18.8,0){\circle{4}}
\put(21,0){\line(1,0){2}}
\put(25.2,0){\circle{4}}
\put(12.6,-6.2){\circle{4}}
\put(12.6,-12.4){\circle{4}}
\put(12.6,-8.4){\line(0,-1){2}}
\put(12.6,-2){\line(0,-1){2}}
\put(12.2,-13.4){\tiny{$1$}}
\put(24.7,-.5){\tiny{$1$}}
\put(12.2,-.5){\tiny{$3$}}
\put(-.5,-.5){\tiny{$1$}}
\put(5.9,-.5){\tiny{$2$}}
\put(18.3,-.5){\tiny{$2$}}
\put(12.2,-6.4){\tiny{$2$}}
\end{picture}

 \\
\hline
$III^*$ &
{\small  $\begin{matrix}\\   3\\   \\  \end{matrix}$ }& {\small  $\geq 5$} & {\small $9$}&  {\small $1$} 
& {\small  $
\begin{pmatrix}
0& -1\\
1 & 0
\end{pmatrix}
$}
 &
\setlength{\unitlength}{1 mm}
\begin{picture}(35,10)(4,-4)
\put(0,0){\circle{4}}
\put(2.2,0){\line(1,0){2}}
\put(6.4,0){\circle{4}}
\put(8.6,0){\line(1,0){2}}
\put(12.6,0){\circle{4}}
\put(14.6,0){\line(1,0){2}}
\put(18.8,0){\circle{4}}
\put(21,0){\line(1,0){2}}
\put(25,0){\circle{4}}
\put(31.4,0){\circle{4}}
\put(37.8,0){\circle{4}}
\put(27.2,0){\line(1,0){2}}
\put(33.6,0){\line(1,0){2}}
\put(18.8,-6.2){\circle{4}}
\put(18.8,-2){\line(0,-1){2}}
\put(24.7,-.5){\tiny{$3$}}
\put(12.2,-.5){\tiny{$3$}}
\put(-.5,-.5){\tiny{$1$}}
\put(5.9,-.5){\tiny{$2$}}
\put(18.3,-.5){\tiny{$4$}}
\put(31,-.5){\tiny{$2$}}
\put(37,-.5){\tiny{$1$}}
\put(18.3,-7){\tiny{$2$}}
\end{picture}
\\
\hline
$II^*$ & {\small  $\begin{matrix}\\  \geq 4\\  \\    \end{matrix}$} & {\small  $5$} & {\small  $10$} & {\small  $0$ }& 
{\small
$
\begin{pmatrix}
0& -1\\
1 & 1
\end{pmatrix}
$
}
&
\setlength{\unitlength}{.9 mm}
\begin{picture}(50,12)(-2,-6)
\put(0,0){\circle{4}}
\put(2.2,0){\line(1,0){2}}
\put(6.4,0){\circle{4}}
\put(8.6,0){\line(1,0){2}}
\put(12.6,0){\circle{4}}
\put(14.6,0){\line(1,0){2}}
\put(18.8,0){\circle{4}}
\put(21,0){\line(1,0){2}}
\put(25,0){\circle{4}}
\put(31.4,0){\circle{4}}
\put(37.8,0){\circle{4}}
\put(44,0){\circle{4}}
\put(27.2,0){\line(1,0){2}}
\put(33.6,0){\line(1,0){2}}
\put(40,0){\line(1,0){2}}
\put(31.4,-6.2){\circle{4}}
\put(31.4,-2){\line(0,-1){2}}
\put(24.7,-.5){\tiny{$5$}}
\put(12.2,-.5){\tiny{$3$}}
\put(-.5,-.5){\tiny{$1$}}
\put(5.9,-.5){\tiny{$2$}}
\put(18.3,-.5){\tiny{$4$}}
\put(31,-.5){\tiny{$6$}}
\put(37,-.5){\tiny{$4$}}
\put(43.5,-.5){\tiny{$2$}}
\put(31,-7){\tiny{$3$}}
\end{picture}

\\
\hline
\end{tabular}
\end{center}
\caption{ \footnotesize{\bf Kodaira Classification of  singular fibers of an elliptic fibration.} 
The fiber of type $I^*_0$ is special among  its family $I^*_n$ because its $J$-invariant can take any value  in $\mathbb{C}$. A  fiber of type $I_n$ or $I^*_n$ ($n>0$) has a pole of order $n$ over the divisor on which it is defined. 
}\label{table.KodairaTate}
\end{table}

\begin{table}[bht]
\begin{center}
\begin{tabular}{|c|c|c|c|c|c|c|c|}
\hline
type  &  group  & \quad $ a_1$\quad  &
\quad $a_2$\quad  & \quad $a_3$ \quad & \quad $ a_4 $ \quad& \quad $ a_6$ \quad & $\Delta$ \\
\hline \hline $I_0 $  &  ---  & $ 0 $  & $ 0 $  & $ 0 $  & $ 0 $  &
$ 0$  & $0$ \\
 $I_1 $  &  ---  & $0 $  & $ 0 $  & $ 1 $  & $ 1 $  &
$ 1 $  & $1$ \\ 
$I_2 $  & $SU(2)$  & $ 0 $  & $ 0 $  & $ 1 $  & $ 1
$  & $2$  & $
2 $ \\ $I_{3}^{ns} $  &  unconven.  & $0$  & $0$  & $2$  & $2$  & $3$  & $3$ \\
$I_{3}^{s}$  & unconven.  & $0$  & $1$  & $1$  & $2$  & $3$  & $3$ \\
$I_{2k}^{ns}$  & $ Sp(k)$  & $0$  & $0$  & $k$  & $k$  & $2k$  & $2k$ \\
$I_{2k}^{s}$  & $SU(2k)$  & $0$  & $1$  & $k$  & $k$  & $2k$  & $2k$ \\
$I_{2k+1}^{ns}$  & unconven.  &  $0$  & $0$  & $k+1$  & $k+1$  &
$2k+1$  & $2k+1$
\\ $I_{2k+1}^s$  & $SU(2k+1)$  & $0$  & $1$  & $k$  & $k+1$  & $2k+1$  & $2k+1$
\\ $II$  &  ---  & $1$  & $1$  & $1$  & $1$  & $1$  & $2$ \\ $III$  & $SU(2)$  & $1$
 & $1$  & $1$  & $1$  & $2$  & $3$ \\ $IV^{ns} $  & unconven.  & $1$  & $1$  & $1$
 & $2$  & $2$  & $4$ \\ $IV^{s}$  & $SU(3)$  & $1$  & $1$  & $1$  & $2$  & $3$  & $4$
\\ $I_0^{*\,ns} $  & $G_2$  & $1$  & $1$  & $2$  & $2$  & $3$  & $6$ \\
$I_0^{*\,ss}$  & $SO(7)$  & $1$  & $1$  & $2$  & $2$  & $4$  & $6$
\\ $I_0^{*\,s} $  & $SO(8)^*$  & $1$  & $1$  & $2$  & $2$  & $4$  &
$6$ \\ $I_{1}^{*\,ns}$
 & $SO(9)$  & $1$  & $1$  & $2$  & $3$  & $4$  & $7$ \\ $I_{1}^{*\,s}$  & $SO(10) $
 & $1$  & $1$  & $2$  & $3$  & $5$  & $7$ \\ $I_{2}^{*\,ns}$  & $SO(11)$  & $1$  & $1$
 & $3$  & $3$  & $5$  & $8$ \\ $I_{2}^{*\,s}$  & $SO(12)^*$  & $1$  & $1$  & $3$
 & $3$  & $5$ & $8$\\
$I_{2k-3}^{*\,ns}$  & $SO(4k+1)$  & $1$  & $1$  & $k$  & $k+1$
 & $2k$  & $2k+3$ \\ $I_{2k-3}^{*\,s}$  & $SO(4k+2)$  & $1$  & $1$  & $k$  & $k+1$
 & $2k+1$  & $2k+3$ \\ $I_{2k-2}^{*\,ns}$  & $SO(4k+3)$  & $1$  & $1$  & $k+1$
 & $k+1$  & $2k+1$  & $2k+4$ \\ $I_{2k-2}^{*\,s}$  & $SO(4k+4)^*$  & $1$  & $1$
 & $k+1$  & $k+1$  & $2k+1$
 & $2k+4$ \\ $IV^{*\,ns}$  & $F_4 $  & $1$  & $2$  & $2$  & $3$  & $4$
 & $8$\\ $IV^{*\,s} $  & $E_6$  & $1$  & $2$  & $2$  & $3$  & $5$  &  $8$\\
$III^{*} $  & $E_7$  & $1$  & $2$  & $3$  & $3$  & $5$  &  $9$\\
$II^{*} $
 & $E_8\,$  & $1$  & $2$  & $3$  & $4$  & $5$  &  $10$ \\
 non-min  &  ---  & $ 1$  & $2$  & $3$  & $4$  & $6$  & $12$ \\
\hline
\end{tabular}
\end{center}
 \caption{ {\bf F-theory Tate's algorithm}.   
The superscript (s/ns/ss) stands for
(split/non-split/semi-split), meaning that  (there is/there is not/ there is  a partial)  monodromy action by
an outer automorphism on the vanishing cycles along the singular  locus. }\label{table.NeronTate}
\end{table}

\clearpage

\subsection{  Miranda models and collisions of singular fibers\label{Miranda}} 

Singular fibers of an elliptic fibration  over loci in  codimension-two or higher codimension in base are not well understood,  except when strong conditions like simple normal crossing of the discriminant locus and preservation of the  $j$-invariant at collisions   are imposed. Under such conditions, Miranda has analyzed in 1983 the  singular fibers for elliptic threefolds defined by Weierstrass models \cite{Miranda}. His construction was   generalized to $n$-folds by Szydlo in 1999  \cite{Szydlo}. Miranda showed that it is always possible to obtain a regular model for an elliptic threefold defined by a  Weierstrass equation. 
We will refer to such elliptic fibrations as {\em Miranda models}. 
The strategy of Miranda consists of  blowing-up the base over intersections of divisors that would lead to ``bad collisions'' of singularities,  then pull-back the fibrations to the new base and he then desingularizes  the total space.  After enough blow-ups, the discriminant locus is a simple normal crossing divisor with only a very limited number of collisions admitting  small resolutions. A Miranda model only has the following 8  collisions:
\begin{align}
\begin{tabular}{l c l}
$J=\infty$  & : & $I_n+I_m\quad\    \quad\quad\rightarrow\quad\quad I_{n+m}$  \\
& & $ I_{2n}+I_{m}^*\quad\quad\quad \rightarrow \quad \quad I_{n+m}^*$\\
& & $ I_{2n+1}+I_{m}^*\quad \quad \rightarrow\quad\quad I^{*+}_{n+m +1}$
  \setlength{\unitlength}{1 mm}
\begin{picture}(25,15)(-10,-4)
\put(-6,6){\circle{4}}
\put(-6,-6){\circle{4}}
\put(-1,-1){\line(-1,-1){4}}
\put(-1,1){\line(-1,1){4}}
\put(0,0){\circle{4}}
\put(2.2,0){\line(1,0){2}}
\put(6.4,0){\circle{4}}
\put(8,0){\line(1,0){2}}
\put(12,0){\circle{4}}
\put(14,0){\line(1,0){1}}
\put(16,0){\line(1,0){1}}
\put(18,0){\line(1,0){1}}
\put(22,0){\circle{4}}
\put(-6.6,5.3){\tiny{$1$}}
\put(-6.6,-6.4){\tiny{$1$}}
\put(-.5,-.5){\tiny{$2$}}
\put(5.9,-.5){\tiny{$2$}}
\put(12,-.5){\tiny{$2$}}
\put(21.4,-.5){\tiny{$2$}}
\put(-2,-7){\footnotesize (  a total of $n+m+5$ nodes)}
\end{picture}
\\
  $J=0$  & $:$ &  $II+IV$
   \setlength{\unitlength}{1 mm}
\begin{picture}(25,8)(-24,-1)
\put(-15,0){$\longrightarrow$}
\put(0,0){\circle{4}}
\put(2.2,0){\line(1,0){2}}
\put(6.4,0){\circle{4}}
\put(-.5,-.5){{\tiny$1$}}
\put(5.9,-.5){{\tiny$2$}}
\end{picture}
  \\
  & & $ II+I_0^*$ 
     \setlength{\unitlength}{1 mm}
\begin{picture}(25,8)(-25,-2)
\put(-15,0){$\longrightarrow$}
\multiput(0,0)(6.4,0){3}{\circle{4}}
\multiput(2.2,0)(6.4,0){2}{\line(1,0){2}}
\put(-.5,-.5){{\tiny$1$}}
\put(5.9,-.5){{\tiny$2$}}
\put(12.3,-.5){{\tiny$3$}}
\end{picture}
  \\
  & & $ IV+I_0^*$
       \setlength{\unitlength}{1 mm}
\begin{picture}(25,8)(-24,-2)
\put(-15,0){$\longrightarrow$}
\multiput(0,0)(6.4,0){5}{\circle{4}}
\multiput(2.2,0)(6.4,0){4}{\line(1,0){2}}
\put(-.5,-.5){{\tiny$1$}}
\put(5.9,-.5){{\tiny$2$}}
\put(12.3,-.5){{\tiny$3$}}
\put(18.7,-.5){{\tiny$4$}}
\put(25.1,-.5){{\tiny$2$}}
\end{picture}
  \\
  & & $II+IV^* $ 
      \setlength{\unitlength}{1 mm}
\begin{picture}(25,8)(-22,-2)
\put(-15,0){$\longrightarrow$}
\multiput(0,0)(6.4,0){4}{\circle{4}}
\multiput(2.2,0)(6.4,0){3}{\line(1,0){2}}
\put(-.5,-.5){{\tiny$1$}}
\put(5.9,-.5){{\tiny$2$}}
\put(12.3,-.5){{\tiny$4$}}
\put(18.7,-.5){{\tiny$2$}}
\end{picture}
  \\
     $J=1$   &   $:$ &  $III+I^*_0$
         \setlength{\unitlength}{1 mm}
\begin{picture}(25,8)(-23,-2)
\put(-15,0){$\longrightarrow$}
\multiput(0,0)(6.4,0){5}{\circle{4}}
\multiput(2.2,0)(6.4,0){4}{\line(1,0){2}}
\put(-.5,-.5){{\tiny$1$}}
\put(5.9,-.5){{\tiny$2$}}
\put(12.3,-.5){{\tiny$3$}}
\put(18.7,-.5){{\tiny$2$}}
\put(25.1,-.5){{\tiny$1$}}
\end{picture}
    \end{tabular}
\end{align}
Only the first two collisions (namely $I_n+I_m$ and $I_{2n}+I^*_{m}$) leads to  Kodaira fibers. The others are not Kodaira fibers but admit $D_k$ and $A_k$ {\emph projective} Dynkin diagrams as dual graphs. 
The new fiber of type $I^{*+}_n$ has the shape of a diagram of a projective Dynkin diagram of type  $D_{n+4}$.
We summarize all the allowed collisions of Miranda's models  in  table \ref{MirandaCol2}. The last column of that table shows the fiber that would have been obtained if one had resolved an elliptic  surface with base a general smooth curve passing through the collision. It  illustrates  that abusing Tate's algorithm already in  codimension-two leads to a misinterpretation of the fiber structure and that the  rank of the ADE fiber does not necessary increase at a collision, it can actually decrease.

\begin{table}[bth]
\begin{center}
\begin{tabular}{|c|c|c|c|}
\hline 
& & & \\
{\footnotesize $j$-inv} &{ \small Collision} & Dual graph  & $\begin{matrix}\text{ if the base was a smooth curve }\\ \text{ through the collision point}\end{matrix}$ \\
& & &\\
\hline 
$\infty$& 
$
\begin{matrix}
\\   
\text{\small $I_{M_1}+I_{M_2}$}\\ \\ 
\end{matrix}
$
 &

{  \setlength{\unitlength}{.9 mm}
\begin{picture}(35,13)(-5,-2)
\put(-6,6){\circle{4}}
\put(-1,1){\line(-1,1){4}}
\put(0,0){\circle{4}}
\put(2.2,0){\line(1,0){2}}
\put(6.4,0){\circle{4}}
\put(8,0){\line(1,0){2}}
\put(12,0){\circle{4}}
\put(14,0){\line(1,0){1}}
\put(16,0){\line(1,0){1}}
\put(18,0){\line(1,0){1}}
\put(20,0){\line(1,0){1}}
\put(22,0){\circle{4}}
\put(23,2){\line(1,1){4}}
\put(28,7){\circle{4}}
\put(-6.6,5.3){\tiny{$1$}}
\put(-.5,-.5){\tiny{$1$}}
\put(5.9,-.5){\tiny{$1$}}
\put(12,-.5){\tiny{$1$}}
\put(21.4,-.5){\tiny{$1$}}
\put(-.5,6.6){\tiny{$1$}}
\put(5.9,6.6){\tiny{$1$}}
\put(12,6.6){\tiny{$1$}}
\put(27.4,6.6){\tiny{$1$}}

\put(0,7){\circle{4}}
\put(2.2,7){\line(1,0){2}}
\put(6.4,7){\circle{4}}
\put(8,7){\line(1,0){2}}
\put(12,7){\circle{4}}
\multiput(14,7)(2,0){6}{\line(1,0){1}}
\put(-4,7){\line(1,0){2}}
\put(0,-5){\footnotesize $I_{M_1+M_2}$}
%\put(-2,5){\tiny $\tilde{A}_n$}
\end{picture}} & same \\
\hline 

 $\infty$& 
 $
 \begin{matrix}
 
 I_{2n}+I_{m}^*\\  \\
 %\\\text{$M_2$ even}
 \end{matrix}
 $ &
\setlength{\unitlength}{1 mm}
\begin{picture}(35,20)(-5,-7)
\put(-6,6){\circle{4}}
\put(-6,-6){\circle{4}}
\put(-1,-1){\line(-1,-1){4}}
\put(-1,1){\line(-1,1){4}}
\put(0,0){\circle{4}}
\put(2.2,0){\line(1,0){2}}
\put(6.4,0){\circle{4}}
\put(8,0){\line(1,0){2}}
\put(12,0){\circle{4}}
\put(14,0){\line(1,0){1}}
\put(16,0){\line(1,0){1}}
\put(18,0){\line(1,0){1}}
\put(20,0){\line(1,0){1}}
\put(22,0){\circle{4}}
\put(23,-1){\line(1,-1){4}}
\put(23,2){\line(1,1){4}}
\put(28,7){\circle{4}}
\put(28,-6){\circle{4}}
\put(-6.6,5.3){\tiny{$1$}}
\put(-6.6,-6.4){\tiny{$1$}}
\put(-.5,-.5){\tiny{$2$}}
\put(5.9,-.5){\tiny{$2$}}
\put(12,-.5){\tiny{$2$}}
\put(21.4,-.5){\tiny{$2$}}
\put(27.4,6.6){\tiny{$1$}}
\put(27.4,-6.6){\tiny{$1$}}
\put(5,-7){\footnotesize $I^\star_{n+m}$}
\end{picture}
& \setlength{\unitlength}{1 mm}
\begin{picture}(35,20)(-5,-7)
\put(-6,6){\circle{4}}
\put(-6,-6){\circle{4}}
\put(-1,-1){\line(-1,-1){4}}
\put(-1,1){\line(-1,1){4}}
\put(0,0){\circle{4}}
\put(2.2,0){\line(1,0){2}}
\put(6.4,0){\circle{4}}
\put(8,0){\line(1,0){2}}
\put(12,0){\circle{4}}
\put(14,0){\line(1,0){1}}
\put(16,0){\line(1,0){1}}
\put(18,0){\line(1,0){1}}
\put(20,0){\line(1,0){1}}
\put(22,0){\circle{4}}
\put(23,-1){\line(1,-1){4}}
\put(23,2){\line(1,1){4}}
\put(28,7){\circle{4}}
\put(28,-6){\circle{4}}
\put(-6.6,5.3){\tiny{$1$}}
\put(-6.6,-6.4){\tiny{$1$}}
\put(-.5,-.5){\tiny{$2$}}
\put(5.9,-.5){\tiny{$2$}}
\put(12,-.5){\tiny{$2$}}
\put(21.4,-.5){\tiny{$2$}}
\put(27.4,6.6){\tiny{$1$}}
\put(27.4,-6.6){\tiny{$1$}}

\put(5,-7){ $I^\star_{2n+m}$}
\end{picture}
\\
\hline 
$\infty$& 
  $\begin{matrix} 
  I_{2n+1}+I_{m}^* \\  \\ 
  %\\\text{$M_2$ odd}
  \end{matrix}$ &  
  \setlength{\unitlength}{1 mm}
\begin{picture}(25,20)(-4,-10)
\put(-6,6){\circle{4}}
\put(-6,-6){\circle{4}}
\put(-1,-1){\line(-1,-1){4}}
\put(-1,1){\line(-1,1){4}}
\put(0,0){\circle{4}}
\put(2.2,0){\line(1,0){2}}
\put(6.4,0){\circle{4}}
\put(8,0){\line(1,0){2}}
\put(12,0){\circle{4}}
\put(14,0){\line(1,0){1}}
\put(16,0){\line(1,0){1}}
\put(18,0){\line(1,0){1}}
\put(22,0){\circle{4}}
\put(-6.6,5.3){\tiny{$1$}}
\put(-6.6,-6.4){\tiny{$1$}}
\put(-.5,-.5){\tiny{$2$}}
\put(5.9,-.5){\tiny{$2$}}
\put(12,-.5){\tiny{$2$}}
\put(21.4,-.5){\tiny{$2$}}
\put(0,-7){\footnotesize $I^{*+}_{n+m}$}
\put(-10,-11){ \footnotesize $(n+m+4)$ components}
\end{picture}
& 

\setlength{\unitlength}{1 mm}
\begin{picture}(35,20)(-5,-7)
\put(-6,6){\circle{4}}
\put(-6,-6){\circle{4}}
\put(-1,-1){\line(-1,-1){4}}
\put(-1,1){\line(-1,1){4}}
\put(0,0){\circle{4}}
\put(2.2,0){\line(1,0){2}}
\put(6.4,0){\circle{4}}
\put(8,0){\line(1,0){2}}
\put(12,0){\circle{4}}
\put(14,0){\line(1,0){1}}
\put(16,0){\line(1,0){1}}
\put(18,0){\line(1,0){1}}
\put(20,0){\line(1,0){1}}
\put(22,0){\circle{4}}
\put(23,-1){\line(1,-1){4}}
\put(23,2){\line(1,1){4}}
\put(28,7){\circle{4}}
\put(28,-6){\circle{4}}
\put(-6.6,5.3){\tiny{$1$}}
\put(-6.6,-6.4){\tiny{$1$}}
\put(-.5,-.5){\tiny{$2$}}
\put(5.9,-.5){\tiny{$2$}}
\put(12,-.5){\tiny{$2$}}
\put(21.4,-.5){\tiny{$2$}}
\put(27.4,6.6){\tiny{$1$}}
\put(27.4,-6.6){\tiny{$1$}}
\put(5,-7){\footnotesize $I^\star_{2n+m+1}$}
\end{picture}
\\
\hline
$0$& 
$\begin{matrix} 
II+IV\\ \\ 
\end{matrix}
$ & 
\setlength{\unitlength}{1 mm}
\begin{picture}(25,20)(-5,-7)
\put(0,0){\circle{4}}
\put(2.2,0){\line(1,0){2}}
\put(6.4,0){\circle{4}}
\put(-.5,-.5){{\tiny$1$}}
\put(5.9,-.5){{\tiny$2$}}
\end{picture}
     &

\setlength{\unitlength}{1 mm}
\begin{picture}(25,20)(-5,-10)
\put(0,0){\circle{4}}
\put(2.2,0){\line(1,0){2}}
\put(6.4,0){\circle{4}}
\put(6.4,6.2){\circle{4}}
\put(6.4,-6.2){\circle{4}}
\put(8.6,0){\line(1,0){2}}
\put(12.6,0){\circle{4}}
\put(6.4,2){\line(0,1){2}}
\put(6.4,-2){\line(0,-1){2}}
\put(12.2,-.5){\tiny{$1$}}
\put(-.5,-.5){\tiny{$1$}}
\put(5.9,-.5){\tiny{$2$}}
\put(5.8,5.6){\tiny{1 }}
\put(5.8,-6.6){\tiny{1 }}
\put(-8,-7){$I_0^\star$}
\end{picture}
\\

\hline 
$0$& 
$\begin{matrix}
II+I^*_0\\ \\
\end{matrix}$ &      \setlength{\unitlength}{1 mm}
\begin{picture}(25,20)(-5,-7)
\put(0,0){\circle{4}}
\put(2.2,0){\line(1,0){2}}
\put(6.4,0){\circle{4}}
\put(8.6,0){\line(1,0){2}}
\put(12.6,0){\circle{4}}
\put(-.5,-.5){\tiny{$1$}}
\put(5.8,-.5){\tiny{$2$}}
\put(12,-.5){\tiny{$3$}}
\end{picture}
 &

\setlength{\unitlength}{1 mm}
\begin{picture}(25,20)(0,-14)
\put(0,0){\circle{4}}
\put(2.2,0){\line(1,0){2}}
\put(6.4,0){\circle{4}}
\put(8.6,0){\line(1,0){2}}
\put(12.6,0){\circle{4}}
\put(14.6,0){\line(1,0){2}}
\put(18.8,0){\circle{4}}
\put(21,0){\line(1,0){2}}
\put(25.2,0){\circle{4}}
\put(12.6,-6.2){\circle{4}}
\put(12.6,-12.4){\circle{4}}
\put(12.6,-8.4){\line(0,-1){2}}
\put(12.6,-2){\line(0,-1){2}}
\put(12.2,-13.4){\tiny{$1$}}
\put(24.7,-.5){\tiny{$1$}}
\put(12.2,-.5){\tiny{$3$}}
\put(-.5,-.5){\tiny{$1$}}
\put(5.9,-.5){\tiny{$2$}}
\put(18.3,-.5){\tiny{$2$}}
\put(12.2,-6.4){\tiny{$2$}}
\put(0,-10){IV$^\star$}
\end{picture}
\\
\hline 
$0$& 
$\begin{matrix}
II+IV^*\\
\\
\end{matrix}
$ &

\setlength{\unitlength}{1 mm}
\begin{picture}(25,10)(0,-7)
\put(0,0){\circle{4}}
\put(2.2,0){\line(1,0){2}}
\put(6.4,0){\circle{4}}
\put(8.6,0){\line(1,0){2}}
\put(12.6,0){\circle{4}}
\put(14.6,0){\line(1,0){2}}
\put(18.8,0){\circle{4}}
\put(21,0){\line(1,0){2}}
\put(25.2,0){\circle{4}}
\put(-.5,-.5){\tiny{$1$}}
\put(5.9,-.5){\tiny{$2$}}
\put(12.1,-.5){\tiny{$3$}}
\put(18.2,-.5){\tiny{$4$}}
\put(24.6,-.5){\tiny{$2$}}
\end{picture}

&

\setlength{\unitlength}{1 mm}
\begin{picture}(45,14)(0,-10)
\put(0,0){\circle{4}}
\put(2.2,0){\line(1,0){2}}
\put(6.4,0){\circle{4}}
\put(8.6,0){\line(1,0){2}}
\put(12.6,0){\circle{4}}
\put(14.6,0){\line(1,0){2}}
\put(18.8,0){\circle{4}}
\put(21,0){\line(1,0){2}}
\put(25,0){\circle{4}}
\put(31.4,0){\circle{4}}
\put(37.8,0){\circle{4}}
\put(44,0){\circle{4}}
\put(27.2,0){\line(1,0){2}}
\put(33.6,0){\line(1,0){2}}
\put(40,0){\line(1,0){2}}
\put(31.4,-6.2){\circle{4}}
\put(31.4,-2){\line(0,-1){2}}
\put(24.7,-.5){\tiny{$5$}}
\put(12.2,-.5){\tiny{$3$}}
\put(-.5,-.5){\tiny{$1$}}
\put(5.9,-.5){\tiny{$2$}}
\put(18.3,-.5){\tiny{$4$}}
\put(31,-.5){\tiny{$6$}}
\put(37,-.5){\tiny{$4$}}
\put(43.5,-.5){\tiny{$2$}}
\put(31,-7){\tiny{$3$}}
\put(5,-7){ II$^\star$}
\end{picture}
 \\
\hline
$0$&  
$\begin{matrix}
IV+I^*_0\\
\\
\end{matrix}
$ &

\setlength{\unitlength}{1 mm}
\begin{picture}(25,10)(-5,-7)
\put(0,0){\circle{4}}
\put(2.2,0){\line(1,0){2}}
\put(6.4,0){\circle{4}}
\put(8.6,0){\line(1,0){2}}
\put(12.6,0){\circle{4}}
\put(14.6,0){\line(1,0){2}}
\put(18.8,0){\circle{4}}
\put(-.5,-.5){\tiny{$1$}}
\put(5.9,-.5){\tiny{$2$}}
\put(12.1,-.5){\tiny{$4$}}
\put(18.2,-.5){\tiny{$2$}}
\end{picture}

&

\setlength{\unitlength}{1 mm}
\begin{picture}(45,14)(0,-8)
\put(0,0){\circle{4}}
\put(2.2,0){\line(1,0){2}}
\put(6.4,0){\circle{4}}
\put(8.6,0){\line(1,0){2}}
\put(12.6,0){\circle{4}}
\put(14.6,0){\line(1,0){2}}
\put(18.8,0){\circle{4}}
\put(21,0){\line(1,0){2}}
\put(25,0){\circle{4}}
\put(31.4,0){\circle{4}}
\put(37.8,0){\circle{4}}
\put(44,0){\circle{4}}
\put(27.2,0){\line(1,0){2}}
\put(33.6,0){\line(1,0){2}}
\put(40,0){\line(1,0){2}}
\put(31.4,-6.2){\circle{4}}
\put(31.4,-2){\line(0,-1){2}}
\put(24.7,-.5){\tiny{$5$}}
\put(12.2,-.5){\tiny{$3$}}
\put(-.5,-.5){\tiny{$1$}}
\put(5.9,-.5){\tiny{$2$}}
\put(18.3,-.5){\tiny{$4$}}
\put(31,-.5){\tiny{$6$}}
\put(37,-.5){\tiny{$4$}}
\put(43.5,-.5){\tiny{$2$}}
\put(31,-7){\tiny{$3$}}
\put(5,-7){ II$^\star$}
\end{picture}

\\
\hline 
$1728$& 
$\begin{matrix} 
III+I^*_0\\
\\
\end{matrix}
$ &

\setlength{\unitlength}{1 mm}
\begin{picture}(25,10)(0,-6)
\put(0,0){\circle{4}}
\put(2.2,0){\line(1,0){2}}
\put(6.4,0){\circle{4}}
\put(8.6,0){\line(1,0){2}}
\put(12.6,0){\circle{4}}
\put(14.6,0){\line(1,0){2}}
\put(18.8,0){\circle{4}}
\put(21,0){\line(1,0){2}}
\put(25.2,0){\circle{4}}
\put(-.5,-.5){\tiny{$1$}}
\put(5.9,-.5){\tiny{$2$}}
\put(12.1,-.5){\tiny{$3$}}
\put(18.2,-.5){\tiny{$2$}}
\put(24.6,-.5){\tiny{$1$}}
%\put(5,-7){ II$^\star$}
\end{picture}

&

\setlength{\unitlength}{1 mm}
\begin{picture}(35,14)(0,-8)
\put(0,0){\circle{4}}
\put(2.2,0){\line(1,0){2}}
\put(6.4,0){\circle{4}}
\put(8.6,0){\line(1,0){2}}
\put(12.6,0){\circle{4}}
\put(14.6,0){\line(1,0){2}}
\put(18.8,0){\circle{4}}
\put(21,0){\line(1,0){2}}
\put(25,0){\circle{4}}
\put(31.4,0){\circle{4}}
\put(37.8,0){\circle{4}}
\put(27.2,0){\line(1,0){2}}
\put(33.6,0){\line(1,0){2}}
\put(18.8,-6.2){\circle{4}}
\put(18.8,-2){\line(0,-1){2}}
\put(24.7,-.5){\tiny{$3$}}
\put(12.2,-.5){\tiny{$3$}}
\put(-.5,-.5){\tiny{$1$}}
\put(5.9,-.5){\tiny{$2$}}
\put(18.3,-.5){\tiny{$4$}}
\put(31,-.5){\tiny{$2$}}
\put(37,-.5){\tiny{$1$}}
\put(18.3,-7){\tiny{$2$}}
\put(5,-7){ III$^\star$}
\end{picture}
\\
\hline 
\end{tabular}
\end{center}
{ 
\caption{\label{MirandaCol2}
{\bf Colliding singularities} in an elliptic threefold as constructed by Miranda.  The non-Kodaira fiber  $I^{*+}_n$ has  the shape of a diagram of type $D_{n+4}$.
The last column shows the fiber that would be obtained for an elliptic with base a smooth curve passing through the point of collision. %The last column is what would be predicted by ``applying'' Tate algorithm in higher codimension. 
}
}
\end{table}
\clearpage 
 
\clearpage

\subsection{Szydlo generalization of  Miranda models\label{Szydlo}}

Assuming  the same conditions as Miranda,  Szydlo has analyzed the  general case of collisions in higher codimensions. Interestingly, starting from codimension-three,  the only collisions possible are those with  $J=\infty$ ( type $I_n$ and $I_n^*$) with the following restrictions:   there are at most one fiber of type $I^*_n$ and at most one fiber of type $I_{2m+1}$, the number of fiber of type $I_{2n}$ is  bounded by the codimension of the collision. Taking this into account we have the following 4 types of collisions:
\begin{align}
\begin{tabular}{l  l l l}
$J=\infty$  & : & $I_{2n_1}+\cdots I_{2n_k}$ &  $\longrightarrow I_{2n},\quad n=n_1+\cdots+ n_k$\\
& & $ I_{2 n_1}+\cdots I_{2 n_k}+I_{2r+1}$ & $ \longrightarrow I_{2n+2r+1},$\\
& & $ I_{2 n_1}+\cdots I_{2 n_k}+I_{m}^*$ & $ \longrightarrow I_{n+m+1}^*,$\\
& & $I_{2 n_1}+\cdots I_{2 n_k}+ I_{2 r+1}+I_{m}^*$ & $\longrightarrow  I^{*+}_{n+r+m+1}$.
    \end{tabular}
\end{align}

 The resolution of  the singularities at the collisions  depends on some discrete choices. In particular,  the order in which the blow-ups are performed is crucial for the final result. For example Miranda and Szydlo don't have the same results for the collision 
 $IV+I^*_0$ 
  and the justification can be traced back to different conventions on how to order the blow-ups:
\begin{align}
\begin{tabular}{c|c}
(Miranda) & (Szydlo)\\
\hline
$IV+I^*_0:\quad
{\setlength{\unitlength}{1 mm}
\begin{picture}(20,10)(0,-2)
\IIVo 
\end{picture}}
\quad$&   $\quad I^*_0+IV:\quad
{\setlength{\unitlength}{1 mm}
\begin{picture}(20,10)(0,-2)
\IIVob 
\end{picture}}
$
\end{tabular}
\end{align}

\subsection{F-theory vs Miranda models\label{vs}}

The results of Miranda and Szydlo  provide an algorithm to obtain regular models from singular Weierstrass models satisfying some conditions like  simple normal crossing for the discriminant locus.
Miranda models have been used in F-theory in cases where blowing-up the base did not destroy the Calabi-Yau condition. 
  However, for applications of F-theory to Grand Unification, the assumptions of Miranda and Szydlo are very naturally  violated\footnote{Even in a purely geometric context one can argue that asking for normal crossing is too strong since the discriminant locus of a general Weierstrass model has always a cusp singularity and this is not really an invitation for normal crossing. But the condition of normal crossing often occurs in mathematics since it makes life easier in proving theorems. However, in physics, one does not always the freedom to choose his assumptions.  }.  For example, the enhancement $\tilde{D}_5\rightarrow \tilde{E}_6$  which is very natural in the context of Grand Unified Theories is not allowed  within Miranda or Szydlo's framework since it involves fiber of different $J$-invariant  ($J=\infty\rightarrow J=0$). Even the transition $\tilde{A}_4\rightarrow \tilde{D}_5$, which is  allowed in perturbative type IIB,  is not allowed in a Miranda's collision since the only fiber of type  $I^*_n$ that can be obtained in codimension-two have $n\geq 2$. These examples illustrate  that in F-theory, one has to  include  wilder singularities in  the discriminant locus and relax the requirement of a well behaved  $J$-invariant. But at the same time, one has to be careful with blowing-up the base since it can lead to violation of  the Calabi-Yau condition. 
Finally, we would like to point out that even the flatness condition is not required in F-theory. Flatness of the fibration is equivalent here to the requirement that all the fibers are of pure dimension one. But there are interesting physical phenomena when a fiber contains a higher dimensional component  \cite{Codim3Sing}. 
We have seen that the assumption of  flatness, normal crossing and well defined $J$-invariant would eliminate interesting physics. However, once we remove these very friendly  conditions,   it is not clear how one will achieve a classification at all.

\section{Geometric engineering of $\SU(5)$ models in F-theory }\label{SU5}
In section \ref{ellipticfibrations} we have reviewed the basic notions of elliptic fibrations . In this short section, we will present the construction of the $\SU(5)$ model in F-theory and introduce formally the conjecture on its singular fibers.

An  $\SU(5)$ Grand Unified Theory is geometrically engineered in  F-theory by a Weierstrass model admitting  a split $I_5$ fiber  over a general point of a divisor of the discriminant locus: 
   \[
\text{split }\  I_{5}\Longrightarrow \mathrm{SU}(5)\  \text{GUT}.
\]
We recall that  a  Weierstrass model  $Y\rightarrow B$ over the base $B$ is given by  the following cubic equation written in a  $\mathbb{P}^2$-bundle over the base $B$:
\begin{equation}
z y^2 +  a_1  x y z + a_3  y z^2 =x^3+ a_2 x^2 z + a_4 x z^2+ a_6 z^3,\label{ai}
\end{equation}
where  the variables  $x$, $y$ and $z$ are the projective coordinates of a $\mathbb{P}^2$-bundle. 
 Each variable  $a_i$ ($i=1,2,3,4,6)$ is a section of  the line bundle $\mathscr{L}^{\otimes( -i)}$. 
  Supposed that $D_{\su(5)}$ is the divisor over which we want to have a split $I_5$ singular fiber.  
We will assume that $D_{\su(5)}$ is the zero locus of a section $w$ of a line bundle  $\mathscr{L}_{\su(5)}$:
\begin{equation}
D_{\su(5)}:\quad w=0.
\end{equation}
We can now consider the coefficients  of the Weierstrass equation as polynomials in $w$. 
It follows from table \ref{table.NeronTate} that after  resolving  the singularity, we will have a  split $I_{5}$ fiber  over $w=0$ if we impose the following specialization  of the  coefficients $a_i$:
\begin{equation}
 a_1=\beta_5  ,\quad a_2=\beta_4 w , \quad a_3= \beta_3 w^2,\quad  a_4= \beta_2 w^3, \quad a_6= \beta_0 w^5.\label{su5cond}
\end{equation}
Here, each  $\beta_j$ ($j=0,2,3,4,5$) is a section of the line bundle $\mathscr{L}^{\otimes(6-j)}\otimes \mathscr{L}_{\su(5)}^{\otimes(j-5)}$. We assume each $\beta_j$ does not vanish identically over  $w=0$. 
The resulting (singular) Weierstrass model  is given by the following equation: 
\begin{equation}
\mathscr{E}: z y^2 +  \beta_5  x y z + \beta_3 w^2 y z^2 -(x^3+ \beta_4 w x^2 z + \beta_2 w^3 x z^2+ \beta_0 w^5 z^3)=0,\label{Wsp}
\end{equation}
where  $w$ and $\beta_j$ ($j=0,2,3,4,5$) are respectively sections of $\mathscr{L}_{\su(5)}$ and $\mathscr{L}^{\otimes(6-j)}\otimes \mathscr{L}_{\su(5)}^{\otimes(j-5)}$.
The  Weierstrass model admits the following factorization of its discriminant: 
\begin{equation}
\Delta= -w^5 \Delta'.
\end{equation}
The discriminant locus   is the union of two divisors, namely  $D_{\su(5)}:w=0$ and   $D:\Delta'=0$ where  
\begin{equation}
\Delta'=\beta_5^4 P + w \beta_5^2 (8 \beta_4 P + \beta_5 R)+ w^2 (16 \beta^2_3 \beta_4^3 + \beta_5 Q)+w^3 S + w^4 T + w^5 U. 
\end{equation}
Here   $P,R,Q,S,T$ and $U$ are polynomials in $\beta_j$ ($j=0,2,3,4,5$): 
\begin{align}
P &=\beta^2_3 \beta_4 -\beta_2 \beta_3 \beta_5 + \beta_0 \beta_5^2, \quad 
R=4 \beta_0 \beta_4\beta_5-\beta^3_3 -\beta^2_2 \beta_5
,     \\
Q &=-2(18 \beta _3^3 \beta _4+8 \beta _2 \beta _3 \beta _4^2-15 \beta _2 \beta _3^2 \beta _5+4 \beta _2^2 \beta _4 \beta _5-24 \beta _0 \beta _4^2 \beta _5+18 \beta _0 \beta _3 \beta _5^2),  \nonumber \\
S &=27 \beta _3^4-72 \beta _2 \beta _3^2 \beta _4-16 \beta _2^2 \beta _4^2+64 \beta _0 \beta _4^3
  +96 \beta _2^2 \beta _3 \beta _5-144 \beta _0 \beta _3 \beta _4 \beta _5-72 \beta _0 \beta _2 \beta _5^2,
  \nonumber   \\
T &=8 \left(8 \beta _2^3+27 \beta _0 \beta _3^2-36 \beta _0 \beta _2 \beta _4\right),\quad 
U =432 \beta _0^2.   \nonumber 
\end{align}
The reduced Weierstrass model corresponding to equation \eqref{Wsp} is  given by:
 \begin{align}
c_4 &=-48 w^3 \beta _2+16 w^2 \beta _4^2-24 w^2 \beta _3 \beta _5+8 w \beta _4 \beta _5^2+\beta _5^4,   \nonumber   \\
c_6&=-864 w^5 \beta _0-216 w^4 \beta _3^2+288 w^4 \beta _2 \beta _4-64 w^3 \beta _4^3 \\
& \quad +\beta_5(144 w^3 \beta_3 \beta _4 +72 w^3 \beta _2 \beta _5-48 w^2 \beta _4^2 +36 w^2 \beta _3 \beta _5-12 w \beta _4 \beta _5^2-\beta _5^5).  \nonumber   
\end{align}

 \subsection{Conjectured fiber geometry \label{conjecture}}

\begin{figure}[bht]
\begin{center}
\setlength{\unitlength}{.25 mm}
\begin{picture}(190,225)(20,-140)
\put(0,-10){ 
\put(-10,10){$\tilde{A}_4$}
\put(0,0){\circle{10}}
\put(-20,-20){\circle{10}}
\put(20,-20){\circle{10}}
\put(15,-40){\circle{10}}
\put(-15,-40){\circle{10}}
\qbezier(-3,-3)(-5,-5)(-17,-17)
\qbezier(3,-3)(5,-5)(17,-17)
\qbezier(-19,-26)(-18,-30.5)(-17,-35)
\qbezier(19,-26)(18,-30.5)(17,-35)
\qbezier(-10,-40)(-10,-40)(10,-40)

}
\put(110,-80){
\put(-15,-55){$\tilde{D}_5$}
\put(-15,25){\circle{10}}
\put(15,25){\circle{10}}
\put(0,10){\circle{10}}
\put(0,-10){\circle{10}}
\put(-15,-25){\circle{10}}
\put(15,-25){\circle{10}}
\qbezier(0,5)(0,0)(0,-5.5)
\qbezier(5,12)(8,15)(14,21)
\qbezier(-5,12)(-8,15)(-14,21)
\qbezier(-5,-12)(-8,-15)(-14,-21)
\qbezier(5,-12)(8,-15)(14,-21)
}

\put(50,-30){
\qbezier(0,0)(-10,0)(-20,0)
\put(0,0){\qbezier(0,0)(-2,.5)(-3,3)\qbezier(0,0)(-2,-.5)(-3,-3)}
\qbezier(0,0)(10,-15)(20,-30)\qbezier(0,0)(10,15)(20,30)
\put(20,30){\qbezier(0,0)
(-1.52543 ,- 1.38675)(  -4.16025 ,- 0.83205)
\qbezier(0,0)( -0.693375, - 1.94145)( 0.83205 ,- 4.16025)
}
\put(20,-30)
{
\qbezier(0,0)(-0.693375 , 1.94145)(0.83205 , 4.16025)
\qbezier(0,0)(-1.52543 ,1.38675)( -4.16025, 0.83205)
}
}

\put(130,15){\qbezier(0,0)(35,15)(70,30) 
\put(35,15){
\qbezier(0,0)(-2.03525 ,- 0.328266)( -3.93919 , 1.57568)
\qbezier(0,0)( -1.64133, - 
 1.24741) \qbezier( -1.57568 ,- 3.93919)
}
}
\put(180,-80){
\qbezier(0,0)(-15,0)(-30,0)
\qbezier(0,0)(11,-14)(22,-28)
\qbezier(0,0)(11,14)(22,28)
\qbezier(0,0)(-12,19)(-36,57)
\put(-13,0){\qbezier(0,0)(-2,.5)(-3,3)\qbezier(0,0)(-2,-.5)(-3,-3)}
\put(11,-14){\qbezier(0,0)(-0.842484 , 1.88155 )( 0.50549 ,4.21242)\qbezier(0,0)( -1.6288,1.26373)( -4.21242, 0.50549)}
\put(11,14){\qbezier(0,0)(-1.6288 ,- 1.26373 )( -4.21242 ,- 0.50549)\qbezier(0,0)( -0.842484, - 
 1.88155)( 0.50549 ,- 4.21242) }
 \put(-24,38){\qbezier(0,0)( -0.645242 , 1.95797)( 0.934488 , 4.13845 ) 
 \qbezier(0,0)( -1.49073,  1.42398 )( -4.13845 , 0.934488)  }
}

\put(100,20){
\put(-20,30){$\tilde{A}_5$}
\qbezier(-20,-5)(-20,-10)(-20,-15)
\qbezier(20,-5)(20,-10)(20,-15)
\put(0,-40){\circle{10}}
\put(-20,0){\circle{10}}
\put(20,0){\circle{10}}
\put(-20,-20){\circle{10}}
\put(20,-20){\circle{10}}
\put(0,20){\circle{10}}
\qbezier(-3,17)(-5,15)(-17,3)
\qbezier(3,17)(5,15)(17,3)
\qbezier(-3,-37)(-5,-35)(-17,-23)
\qbezier(3,-37)(5,-35)(17,-23)
}

\put(230,50){
\put(-20,30){$\tilde{A}_6$}
\qbezier(-20,-5)(-20,-10)(-20,-15)
\qbezier(20,-5)(20,-10)(20,-15)

\put(-20,0){\circle{10}}
\put(20,0){\circle{10}}
\put(-20,-20){\circle{10}}
\put(20,-20){\circle{10}}
\put(0,20){\circle{10}}
\qbezier(-3,17)(-5,15)(-17,3)
\qbezier(3,17)(5,15)(17,3)
\put(15,-40){\circle{10}}
\put(-15,-40){\circle{10}}
\qbezier(-19,-26)(-18,-30.5)(-17,-35)
\qbezier(19,-26)(18,-30.5)(17,-35)
\qbezier(-10,-40)(-10,-40)(10,-40)

}

\put(230,-130){
\put(10,15){$\tilde{E}_6$}
\put(0,25){\circle{10}}
\put(0,7){\circle{10}}
\multiput(-30,-10)(15,0){5}{\circle{10}}
\multiput(0,0)(0,18){2}{\qbezier(0,2)(0,-3)(0,-5)}
\multiput(-15,0)(15,0){4}{\qbezier(-10,-10)(-7,-10)(-5,-10)}
}

\put(230,-35){
\put(10,-15){$\tilde{D}_6$}
\put(-15,25){\circle{10}}
\put(15,25){\circle{10}}
\put(0,10){\circle{10}}
\put(0,-10){\circle{10}}
\put(-15,-45){\circle{10}}
\put(15,-45){\circle{10}}
\put(0,-30){\circle{10}}
\qbezier(0,5)(0,0)(0,-5.5)
\qbezier(5,12)(8,15)(14,21)
\qbezier(-5,12)(-8,15)(-14,21)
\qbezier(-5,-32)(-8,-35)(-14,-41)
\qbezier(5,-32)(8,-35)(14,-41)
\qbezier(0,-15)(0,-18)(0,-25.5)
}

\end{picture}
\end{center}
\caption{{\bf Conjectured singular fiber enhancements of a   $\SU(5)$ GUT}. 
 Starting with codimension one in the base, the codimension increases from left to right. 
 Thinking in terms of the dual graph, the rank of the associated Dynkin diagram increases by 1  as we move in codimension.  
  }
 \label{su5GuTConj}
\end{figure}
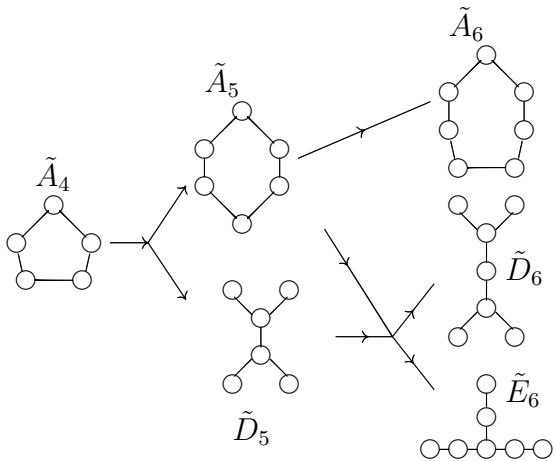

The support of the discriminant locus $\Delta$  of the elliptic fibration that we have introduced to describe an $\SU(5)$ GUT is  the union of two divisors, namely  $D_{\su(5)}: w=0$ (of multiplicity 5) and $D': \Delta'=0$ (of multiplicity one). 
Over a general point of $D_{\su(5)}$, we have a split $I_5$ fiber corresponding to a $\mathrm{SU}(5)$ gauge group. 
 Over  a general point of $D'$ we have a  $I_1$ fiber (a nodal curve): 
 \begin{equation}
 D_{\su(5)}: w=0 \quad (I_5), \qquad D': \Delta'=0 \quad (I_1).
 \end{equation}
 Since in general $\Delta'$ does not factorize   further,  we have 5 D7-branes wrapping  the divisor $D_{\su(5)}:w=0$ and one D7-brane wrapping the  divisor $D':\Delta'=0$.  
The divisor $D'$ is actually singular  and all its singularities are supported on the codimension-2 locus  $w= P \beta_5=0$ which is  the union of two curves of $D_{\su(5)}$, namely $\Sigma_{5}:w=P=0$ and $\Sigma_{10}:w=\beta_5=0$.  Interestingly the singular locus of $D'$ is exactly its intersection with $D_{\su(5)}$:
We have 
 \begin{equation}
D_{\su(5)}\cap D'= \Sigma_5 \cup \Sigma_{10}.
 \end{equation}
 It is conjectured that the $I_5$ fiber  enhances to a $I_6$ fiber  over $\Sigma_5$ and to a $I^*_1$ fiber  over $\Sigma_{10}$ (see for example \cite{andreas,Hayashi:2009ge}): 
  \begin{equation}
\Sigma_{5}: w=P=0 \quad (I_6 ?), \qquad \Sigma_{10}: w=\beta_5=0 \quad (I^*_1 ?).
 \end{equation}
In codimension-three, at the intersection of $\Sigma_{5}$ and $\Sigma_{10}$ the singularity is expected to enhance further.   The intersection of $\Sigma_{5}$ and $\Sigma_{10}$ is the union of two codimension-three loci  in the base, $\Pi_3$ and $\Pi_4$:
 \begin{equation}
\Sigma_{5}\cap \Sigma_{10}=
 \Pi_3  \cup \Pi_4.
 \end{equation}
 Over $ \Pi_3 $ and $ \Pi_4 $, it is conjectured that the fiber enhances to a fiber  $I^*_2$  and  $IV^*$ (or more generally to a fiber with dual graph $\tilde{D}_6$ and $\tilde{E}_6$):
 \begin{equation}
 \Pi_3 : w=\beta_3=\beta_5=0\quad (I^*_2 ?)\qquad \cup \quad\quad \Pi_4:w=\beta_4=\beta_5=0 \quad (IV^*?).
 \end{equation}
The curve  $\Sigma_{5}$ contains additional points  $\Pi_7:w=P=R=0$ over which the singular fiber $I_6$ is conjectured  to enhance further to a fiber of type  $I_7$  \cite{andreas,Hayashi:2009ge}:
 \begin{equation}
 \Pi_7 : w=P=R=0\quad (I_7?).
 \end{equation}
These points are also codimension-three points in the base. there are not coming from the collisions of two curves,  but they contain as a proper subset the points $\Pi_3$. The  conjectured tree of singular fibers enhancement is summarized in  figure  \ref{su5GuTConj} and figure \ref{su5GUT.fib.enhancement}.

\begin{center} 
 \begin{table}[thb]
 \begin{tabular}{cc} \phantom{ddddd}&  
 \xymatrix{
  &*+{ 
    \begin{tabular}{|c|}
  \hline
  $\Delta=D_{\su(5)}\cup D'$\\
  \hline {\footnotesize $w^5 \Delta' =0$}    \\ \hline   \end{tabular}
   }\ar[dl] \ar[dr] 
  &  \\
    *+{
  \begin{tabular}{|c|}
  \hline
  {\footnotesize $D_{\su(5)}:   I_5 \  \tilde{A}_4$}\\
  \hline {\footnotesize $w=0$}    \\ \hline   \end{tabular}
  }\ar[dr] &  &  *+{
  \begin{tabular}{|c|}
  \hline
  $  D':\  I_1\  \mathrm{U}(1)$\\
  \hline {\footnotesize $\Delta'=0$}\\
  \hline
\end{tabular}
    } \ar[dl] & 
  \\
&   *+[F] +{D_{\su(5)} \cap D'=\Sigma_{5}\cup \Sigma_{10}}\ar[dl]\ar[dr] &   \\ 
   *+{ 
     \begin{tabular}{|c|}
  \hline
  $\Sigma_{5}:   I_6\quad  \tilde{A}_5$\\
  \hline {\footnotesize $w=P=0$}    \\ \hline   \end{tabular}  
  } \ar[dr] \ar[ddd]&  &  *+{
    \begin{tabular}{|c|}
  \hline
  $\Sigma_{10}:   I^*_1\quad \tilde{D}_5$\\
  \hline {\footnotesize $w=\beta_5=0$}    \\ \hline   \end{tabular}
    }\ar[dl]  \\
     &     *+[F]+{\Sigma_{5} \cap \Sigma_{10} =\Pi_3\cup \Pi_3}\ar[dd]\ar[ddr]& \\
 &  & &  & \\
 *+{
   \begin{tabular}{|c|}
  \hline
  $\Pi_7:   I_7\quad \tilde{A}_6$\\
  \hline {\footnotesize $w=P=R=0$}    \\ \hline   \end{tabular}
 } \ar[r] &  *+{
   \begin{tabular}{|c|}
  \hline
  $\Pi_3:   I^*_2\quad \tilde{D}_6$\\
  \hline {\footnotesize $w=\beta_3=\beta_5=0$}    \\ \hline   \end{tabular}
 } & *+{ 
    \begin{tabular}{|c|}
  \hline
  $\Pi_4:   IV^*\quad \tilde{E}_6$\\
  \hline {\footnotesize $w=\beta_5=\beta_4=0$}    \\ \hline   \end{tabular}
 }
 }
 \end{tabular}
\caption{ Conjectured singular fiber enhancements  of  a general $\SU(5)$  GUT  model.
\label{su5GUT.fib.enhancement}}
\end{table}
\end{center}

\section{Resolution over codimension-one loci in the base \label{codimone}}

In this section we resolve the singularities that project to codimension-one loci in the base. 
The variety $\mathscr{E}$ is singular at $x=y=w=0$. Fiberwise, we have a singular point on the elliptic fiber above the divisor $D_{\su(5)}:w=0$ in the base. The divisor $D_{\su(5)}$ is a component of the discriminant locus of multiplicity 5. The resolution is obtained by two successive blow-ups after which the  singular fiber is replaced by a cycle of 5 rational curves defining the Kodaira type $I_5$ over $D_{\su(5)}$. 

\subsection{Blowing-up the codimension-1 singularity}
Our starting point is the defining Weierstrass equation of the $\SU(5)$ GUT model:
\begin{equation}
\mathscr{E}: z y^2 +  \beta_5  x y z + \beta_3 w^2 y z^2 -(x^3+ \beta_4 w x^2 z + \beta_2 w^3 x z^2+ \beta_0 w^5 z^3)=0.
\end{equation}
The support of all the singularities of the variety $\mathscr{E}$ is: 
\begin{equation}
Sing(\mathscr{E}): \quad x=y=w=0.
\end{equation}
 Over the divisor $D_{\su(5)}: w=0$,  the elliptic fiber degenerate into the nodal curve: 
\begin{equation}
C_0: z y^2+  \beta_5  x y z -x^3=0,
\end{equation}
This nodal curve degenerate further to a cuspidial curve over the codimension-two locus $\beta_5=w=0$ of the base. 

We will blow-up this singular locus $x=y=w=0$. For that matter we introduce  the projective coordinates $[U_1,U_2,U_3]$ of a $\mathbb{CP}^2$ together with the following relations: 
\begin{equation}
x U_2 = y U_1, \quad x U_3 =w U_1, \quad y U_3  = w U_2.
\end{equation}
Working first in a  patch defined $\mathscr{U}_1$ defined by $U_1\neq 0$, we have 
 $y= \frac{U_2}{U_1} x, \quad w=\frac{U_3}{U_1} x.$ 
Defining $x_1=x$, $y_1=U_2/ U_1$ and $w_1=U_3/U_1$, the blow-up can be expressed in $\mathscr{U}_1$ by the morphism $\varphi_1: \quad (x,y,w)\mapsto [x_1, x_1  y_1 , x_1  w_1].$
We can proceed in a similar way in the patches  $\mathscr{U}_2$ and $\mathscr{U}_3$ defined respectively by   $U_2\neq 0$ and  $U_3\neq 0$.  Alltogether, the blow-up is computed using the  three morphisms: 
\begin{align}
\varphi_1 &: \quad (x,y,w)\mapsto [x_1, x_1  y_1 , x_1  w_1], \\
\varphi_2 &: \quad (x,y,w)\mapsto [y_2 x_2,   y_2 , y_2 w_2], \\ 
\varphi_3 &: \quad (x,y,w)\mapsto [w_3 x_3,  w_3 y_3 ,  w_3].
\end{align}
The different patches are glued together along their intersection by requiring the morphism to match. 
 For example,  along $\mathscr{U}_1\cap \mathscr{U}_2$, the gluing is based on matching $\varphi_1=\varphi_2$ and gives $(x_1, y_1, w_1)=(x_2 y_2, {1}/{x_2}, {w_2}/{x_2})$. 
 Before the blow-up, the special fiber is obtained by cutting $\mathscr{E}$ along  $w=0$. After the blow-up, we have to keep track of the fiber function in every patch.   
 In the patch $\mathscr{U}_1$, $\mathscr{U}_2$ and $\mathscr{U}_3$,  we get respectively the fiber function $w_1 x_1$,  $w_2y_2$ and  $w_3$. The fact that the irreducible $w$ is replaced by reducible fiber function is responsible for the new cycles that will constitute the  singular  fiber $I_5$.

\begin{table}[h]
 \begin{center}
  \begin{tabular}{c| c l |r}
 Patch & & Proper transform of the defining equation $\mathscr{E}$ & Fibers  \\
 \hline
 $\mathscr{U}_1$ & $\mathscr{E}:$ &$y^2+ w^2 \beta_3 x y+ \beta_5 y=x+w^5 x^3 \beta_0+ w^3 \beta_2 x^2 +
w\beta_4 x $ & $x w=0$\\
  $\mathscr{U}_3$ & $\mathscr{E}:$ &
 $ y^2+\beta_5  x y +w \beta_3 y=w x^3+w^3 \beta_0 +w^2 \beta_2 x +w \beta_4 x^2$ & $w=0$
  \\
 $\mathscr{U}_{3,1}$ & $\mathscr{E}:$ &$ y^2+\beta_5 y+w\beta_3 y= wx^2 +w^3 \beta_0 x + w^2 \beta_2 x +w \beta_4 x$ & $x w=0$\\
  $\mathscr{U}_{3,3}$ & $\mathscr{E}:$ &$  y^2+\beta_5 x   y  + \beta_3 y =
w ^2  x^3  +w  \beta_0 + w  \beta_2 x  +w  \beta_4 x^2  $ & $w=0$
 \end{tabular}
 \end{center}
 \caption{Defining equations  
   for the  blow-ups generating the fiber $I_5$.  \label{I5res.eq} }
 \end{table}

Let us analyze  the fiber structure after the first blow-up. 
 In the patch $\mathscr{U}_1$, the fiber is obtained by cutting along  $w_1 x_1=0$. The component  $w_1=0$  gives the proper transform of the original fiber $C_0$ while 
 $x_1=0$ gives a  new reducible component  
 which is the union of  two  rational curves $C_{1+}$ and $C_{1-}$; each intersecting  $C_0$ at one point.  These two components are disjoint in the patches $\mathscr{U}_1$ and $\mathscr{U}_2$.  The proper transform of the Weierstrass equation $\mathscr{E}$ is smooth in the patches $\mathscr{U}_1$ and $\mathscr{U}_2$.
In the patch  $\mathscr{U}_3$, we don't see $C_0$, but we can see  both  $C_{1+}$
and $C_{1-}$ and their intersection point  $w_3=x_3=y_3=0$, which is the   singular locus of the proper transform of $\mathscr{E}$.  
\begin{equation}\label{comp.U1}
\mathscr{U}_1:
\begin{cases}
 C_0 &:  \quad    w= y^2+\beta_5 y -x=0 \\
C_{1+}&: \quad   x= y=0 \\
 C_{1-} &:\quad   x= y+\beta_5=0.
\end{cases} 
, \quad \
\mathscr{U}_3:
\begin{cases}
C_{1+}&: \quad   w= y=0 \\
 C_{1-} &:\quad   w= y+\beta_5 x =0.
\end{cases} 
\end{equation}
 To  completely smooth the variety up to codimension-two ( codimension-one in the base),  we have to perform an additional blow-up in the patch $\mathscr{U}_3$ at the  $x_3=y_3=w_3=0$. 
 We introduce an additional  $\mathbb{P}^2$ with projective coordinates $[U_{3,1},U_{3,2}, U_{3,3}]$ and the relations 
$$U_{3,2} x_3-U_{3,1} y_3=0, \quad U_{3,2} w_3-U_{3,3} y_3=0,   \quad U_{3,1} w_3 - U_{3,3} x_3=0.$$
 Defining $x_{3,1}=x_3$, $y_{3,1}=U_{3,2}/U_{3,1}$, $w_{3,1}=U_{3,3}/ U_{3,1}$,etc,  the blow-up of $x_3=y_3=w_3=0$ is defined by the following three morphisms:
 \begin{align}
\varphi_{3,1} &: \quad (x_3,y_3,w_3)\mapsto [x_{3,1}, x_{3,1}  y_{3,1} , x_{3,1}  w_{3,1}], \\
\varphi_{3,2} &: \quad (x_3,y_3,w_3)\mapsto [y_{3,2} x_{3,2},   y_{3,2} , y_{3,2} w_{3,2}], \\ 
\varphi_{3,3} &: \quad (x_3,y_3,w_3)\mapsto [w_{3,3} x_{3,3},  w_{3,3} y_{3,3} ,  w_{3,3}].
\end{align}
where $\varphi_{3,k}$ is defined in the patch $\mathscr{U}_{3,k}: U_{3,k}\neq 0$ ($k=1,2,3$). 
The special  fiber in the patch $\mathscr{U}_{3,1}$ is given by the fiber function $w_{31} x_{31}=0$  which gives two reducible  components $ { C}_1$ and $ { C}_2$ that are composed of the rational curves $C_{1\pm}$ and $C_{2\pm}$.
 Alltogether $C_0, C_{1\pm}, C_{2\pm}$ define the fiber $I_5$. Alltogether all the components of the fiber $I_5$ can be seen in the patch $\mathscr{U}_1$ (where the can see $C_0$, the only component touching the section) and in the  $\mathscr{U}_{3,1}$, where we have all the other components:
\begin{align}\label{comp.U31}
\mathscr{U}_{3,1}:
\begin{cases}
C_{1+}&: \quad   w= y=0 \\
 C_{1-} &:\quad   w=y+\beta_5=0\\
C_{2+}&: \quad   x= y=0\\
 C_{2-} &:\quad   x= y+\beta_5  +w \beta_3=0\\
\end{cases} 
\end{align} 
After this blow-up, the variety is smooth in codimension-one. As we will see next, there are singularities above loci  in higher codimensions in the base. But they are all visible in the patch $\mathscr{U}_{3,1}: U_{3,1}\neq 0$ where the components $C_{1\pm}$ and $C_{2\pm}$ are visible. 
The patch $\mathscr{U}_{3,1}$ will play a central role in the rest of this paper since this is the ground where all the additional enhancement will take place. 
The two blow-ups yielding   the fiber $I_5$ are summarized\footnote{To simplify notations, we don't put the indexes and we work just with $(x,y,w)$. This is usually enough when we specify explicitly in which patch we are working.}  in equations \eqref{comp.U1} and \eqref{comp.U31} and in  tables \ref{I5res.eq} and \ref{I5res}.

\begin{table}[!h]
\begin{tabular}{c}
 \xymatrix{
& & {\begin{tabular}{|c|}
\hline
$U:\   $ 
{\small  $C_0$}\\
\hline 
{\small Sing:}\\
{\small  $x=y=w=0$}\\
\hline
\end{tabular}}\ar[dl]\ar[d]\ar[dr]
 \\
& {\begin{tabular}{|c|}
\hline
$\mathscr{U}_1:\  $ \\
\hline 
 {\small $C_{1\pm },\  C_0$}\\
\hline 
\end{tabular}}
& 
{\begin{tabular}{|c|}
\hline
$\mathscr{U}_2:\ $\\   
\hline 
{\small $C_{1\pm}+C_0$}\\
\hline 
\end{tabular}}
&
{\begin{tabular}{|c|}
\hline
$\mathscr{U}_3:\   $
{\small  $C_{1\pm }$}\\
\hline 
{\small Sing:}\\
{\small $x=y=w=0$}\\
\hline
\end{tabular}}
\ar[dl]\ar[d] \ar[dr]\\
&  &
{\begin{tabular}{|c|}
\hline
$\mathscr{U}_{3,1}:\  $\\
\hline 
{\small $C_{2\pm},\ C_{1\pm} $}\\
\hline 
\end{tabular}}
  &
 {\begin{tabular}{|c|}
\hline
$\mathscr{U}_{3,2}:\  $\\
\hline 
{\small $C_{2\pm},\  C_{1\pm} $}\\
\hline 
\end{tabular}}
  & {\begin{tabular}{|c|}
\hline
$\mathscr{U}_{3,3}:  $
\\
\hline
{\small $C_{2\pm}$}\\
\hline 
\end{tabular}}
}
\end{tabular}
\setlength{\unitlength}{.4 mm}
\begin{picture}(30,10)(-330,-180)
\put(0,0){
\put(6,0){\footnotesize $C_0$}
\put(0,0){\qbezier(-2,-2)(0,0)(2,2)\qbezier(-2,2)(0,0)(2,-2)}
\put(25,-20){\footnotesize $C_{1-}$}
\put(-42,-20){\footnotesize $C_{1+}$}
\put(17,-53){\footnotesize $C_{2-}$}
\put(-20,-53){\footnotesize $C_{2+}$}
\put(0,0){\circle{10}}
\put(-20,-20){\circle{10}}
\put(20,-20){\circle{10}}
\put(15,-40){\circle{10}}
\put(-15,-40){\circle{10}}
\qbezier(-3,-3)(-5,-5)(-17,-17)
\qbezier(3,-3)(5,-5)(17,-17)
\qbezier(-19,-26)(-18,-30.5)(-17,-35)
\qbezier(19,-26)(18,-30.5)(17,-35)
\qbezier(-10,-40)(-10,-40)(10,-40)
}
\end{picture}

\caption{{\bf Atlas of the blow up of the $I_5$ singularity}. Each patch is indicated by its name and we name the components visible from that patch. In case the variety is still singular and need an additional blow-up in a given patch we indicate explicitly the singular locus. 
\label{I5res} 
}
 \end{table}

\subsection{Fiber mutation over higher codimension  singularities}

After resolving the singularity above $D_{\su(5)}$,  there are still extra singularities left in higher codimension in the base. 
With the exception of $C_0$, all the components of the $I_5$ fiber are visible in the patch $\mathscr{U}_{31}$.  Since there are no singularities on $C_0$ and we will see that all the singularities are visible in the patch $\mathscr{U}_{31}$:  
\begin{equation}
\mathscr{E}\   in\    \mathscr{U}_{31}:\quad y^2+\beta_5 y+w\beta_3 y= wx^2 +w^3 \beta_0 x+ w^2 \beta_2 x +w \beta_4 x.
\end{equation}
 In order to understand the higher codimension enhancement of the fiber, it is necessary to first analyze the extra singularities present in higher codimension.       We have summarized the singular structure in table \ref{codimsing}.
We have recovered all the singular loci expected from the analysis of the discriminant with the exception of the points $\Pi_7$. We will see later that there is actually no enhancement of the fiber above these points.

\begin{table}[h]
\begin{center}
\begin{tabular}{|l | l |c|}
\hline 
\multicolumn{1}{|c|}{Sing. Locus of $\mathscr{E}$  in $\mathscr{U}_{3,1}$} & \multicolumn{1}{|c|}{ 
 \small Codim in the base of $\mathscr{E}$} &  {\small 
 $\begin{matrix}
\text{ Located on  the} \\
\text{ components}
\end{matrix}
$ 
 }
 \\
\hline 
$L_t:\    x=y=w=\beta_5=0$ & 2 $\quad(\Sigma_{10}: \beta_5=0$ in  $D_{\su(5)}$) &  $C_{1}\cap  C_{2+}\cap C_{2-}$ \\
$L_x:\   x+\beta_4=y=w=\beta_5=0$ & 2  $\quad(\Sigma_{10}:\beta_5=0$ in  $D_{\su(5)}$) &  $C_{1}$ \\
\hline
$
L_w \  
\begin{cases}
y=x=w\beta_3+\beta_5=0\\
w^2 \beta_0+w \beta _2+ \beta_4=0
\end{cases}$  &2  $\quad(\Sigma_{5}:=P=0$ in   $D_{\su(5)}$)   & $C_{2+}\cap C_{2-}$\\
\hline
$
p_3\  
\begin{cases}
x=y=\beta_5=\beta_3=0\\
w^2 \beta_0+w \beta _2+ \beta_4=0
\end{cases}
$ & 3 $\quad(\Pi_3: \beta_5=\beta_3=0$  in $D_{\su(5)}$)& $C_{2}$\\
\hline
$p_4:\   x=y=w=\beta_5=\beta_4=0$ & 3 $\quad(\Pi_4:\beta_5=\beta_4=0$ in $D_{\su(5)}$)& $C_{1}\cap C_{2}$\\
\hline
\end{tabular}
\caption{{\bf Singular loci  in higher codimensions} 
\footnotesize{The first column indicates the singular loci, the second column expressed the location of the singularity in the base and mention the codimension. Finally 
the third column indicate which components of the $I_5$ contain  the singularity. We denote by $C_1$ (resp. $C_2$) the node obtain when $C_{1\pm}$ (resp. $C_{2\pm}$) coincide.} 
\label{codimsing}
}
\end{center}
\end{table}

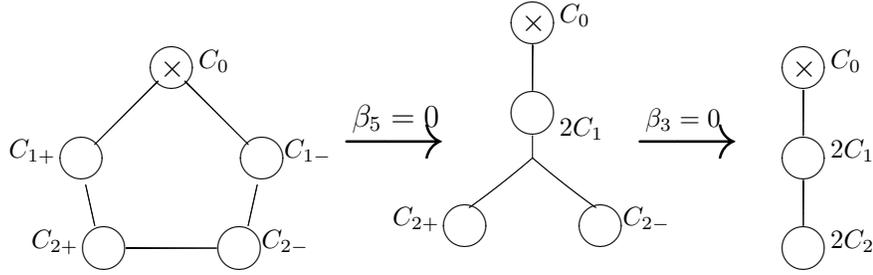
\begin{figure}[!h]
\begin{center}
\setlength{\unitlength}{.6 mm}
\begin{picture}(160,50)(-30,-40)
\put(-10,0){
\put(6,0){\footnotesize $C_0$}
\put(-2.5,-2){{$\times$}}
\put(25,-20){\footnotesize $C_{1-}$}
\put(-36,-20){\footnotesize $C_{1+}$}
\put(20,-40){\footnotesize $C_{2-}$}
\put(-31,-40){\footnotesize $C_{2+}$}
\put(0,0){\circle{10}}
\put(-20,-20){\circle{10}}
\put(20,-20){\circle{10}}
\put(15,-40){\circle{10}}
\put(-15,-40){\circle{10}}
\qbezier(-3,-3)(-5,-5)(-17,-17)
\qbezier(3,-3)(5,-5)(17,-17)
\qbezier(-19,-26)(-18,-30.5)(-17,-35)
\qbezier(19,-26)(18,-30.5)(17,-35)
\qbezier(-10,-40)(-10,-40)(10,-40)
\put(40,-13){{$\beta_5=0$}}
\put(35,-20){  \text{\huge$\longrightarrow$}  }
}
\put(70,10){
\put(0,0){\circle{10}}
\put(-2.5,-2){{$\times$}}
\put(0,-20){\circle{10}}
\put(15,-45){\circle{10}}
\put(-15,-45){\circle{10}}
\put(6,0){\footnotesize $C_0$}
\put(6,-25){\footnotesize $2 C_{1}$}
\put(20,-45){\footnotesize $C_{2-}$}
\put(-31,-45){\footnotesize $C_{2+}$}
\put(0,-25){\line(0,-1){5}}
%\put(0,-25){\circle*{2}}
\qbezier(0,-5)(0,-7)(0,-15)
\qbezier(0,-30)(-7,-36)(-14, -41)
\qbezier(0,-30)(7,-36)(14, -41)
}

\put(95,0){
\put(0,-13){\footnotesize{$\beta_3=0$}}
\put(-5,-20){  \text{\Huge$\longrightarrow$}  }
}
\put(130,0){
\put(0,0){\circle{10}}
\put(-2.5,-2){{$\times$}}
\put(0,-20){\circle{10}}
\put(0,-40){\circle{10}}
\put(6,0){\footnotesize $C_0$}
\put(6,-20){\footnotesize $2 C_{1}$}
\put(6,-40){\footnotesize $2C_{2}$}
\qbezier(0,-5)(0,-7)(0,-15)
\qbezier(0,-25)(0,-27)(0,-35)
}

\end{picture}
\end{center}
\caption{Higher co-dimension specialization of the $I_5$ singularity. 
As we go to higher codimension, there are  singularities over which the $I_5$ fiber changes its shape due to component superposition.
\label{table.I5trans} }
\end{figure}

The $I_5$ fiber changes  dramatically  over the loci in the base corresponding to higher dimensional singularities. For example, over $\Sigma_{10}:\beta_5=0$ and  the components  $C_{1+}$ and $C_{1-}$ coincide and form a unique component $C_1$  which intersect  $C_{2+}$  and $C_{2-}$ at a common point $L_t$.  There is another singularity $L_x$ sitting on $C_1$ away from $C_{2\pm}$ (as long as $\beta_4\neq 0$). 
Over $\Pi_3:\beta_5=\beta_3=0$, $C_{2+}$ and $C_{2-}$ also coincide with each other. The mutation associated with the specialization $\beta_5=0$ and $\beta_5=\beta_3=0$ is  illustrated in figure \ref{table.I5trans}.  Over $\Sigma_5: P=0$, the fiber keeps its $I_5$ structure but it develops a singularity  at $L_w$ corresponding to the intersection of $C_{2+}\cap C_{2-}$ over the locus $\Sigma_5$.

\section{The  binomial geometry of $\SU(5)$ GUTs \label{binomial}}

In this section, we introduce the binomial geometry that naturally appears after resolving the singularities of the $\SU(5)$ in codimension one. We will study carefully the  singularities of the binomial variety and their desingularization by explicit resolutions. We will do it both using the natural toric description of the binomial variety and a direct algebraic description.The binomial variety we are dealing can be seen as  a higher dimension generalization of the  conifold. We will review therefore the small resolution of a conifold and Atiyah's flop in some details.  For  a conifold singularity, there are two small resolutions connected by a flop.  For the binomial variety we consider,  we will have six small resolutions related by a networks of flop transitions forming a dihedral group of order 12. We will use this analysis to fully resolve the  $\SU(5)$ model and describe its fiber in every codimensions.

In order to resolve the higher codimension singularity of the elliptic fibration, we will work in the patch $\mathscr{U}_{3,1}$ where all these singularities are visible. 
After introducing the variables $s$ and $t$ defined by:
\begin{equation}
s= y+\beta_5+w\beta_3,\quad 
t=x+\beta_4+ w^2\beta_0+w \beta_2, 
\end{equation}
we see that the fibration $\mathscr{E}$ has the structure of an {\em affine binomial variety}: 
\begin{equation}
\mathscr{E}\    in\  \mathscr{U}_{3,1}:\quad y s -w x t=0, \quad  x, w, t, y, s \in \mathbb{C}.
\end{equation}
An affine binomial variety is the vanishing locus of a binomial polynomial( the  sum of two  monomials). In this section we will spend some time understanding this geometry. Binomial ideals are reviewed for example in the lectures of  Bernard Teissier \cite{Teissier}. Binomial varieties have the nice property of being affine toric varieties. Their singularities are considered to be the  simplest class of non-degenerate singularities. 
There are several binomial varieties that naturally occur string theory. 
A cusp ($y^2=x^3$) is an example of a dimension-one  binomial variety, 
 a   surface with a $A_k$ singularity ($xy=z^{k+1}$) is also a binomial variety. 
The conifold ($x y-w z=0$) is probqbly the most famous binomial variety in string theory.
A pinch point singularity is described by  
The Whitney's  umbrella ($x^2=z  y^2$)  and is a  binomial variety that appears naturally when we take the  orientifold limit of F-theory  \cite{CDE,AE1}.  
The binomial variety that we are interested in ($xw t-ys=0$)  is a beautiful example of higher dimensional singular variety. Its singular locus is composed of a bouquet of  three double lines and the singularity enhances at the origin of the bouquet. In other words, we have three conifold lines all intersecting at one point where the singularity worsens. Its  intersection with different linear spaces reproduce the Whitney umbrella, the cusp  or the  $A_2$ singularity. 
We will review the  small resolution of the conifold in the next subsection as a warmup before attacking the resolution of the higher codimension singularities of $\mathscr{E}$. 

\begin{table}[htb]
\begin{center}
\begin{tabular}{|c | c|c|}
\hline
Name &  Defining equation & Dimension \\
\hline 
A curve with a cusp    & $x^2-y^3=0$ & 1 \\
A surface with a $A_k$-point & $xy-z^{k+1}=0$ & 2 \\
The Whitney's umbrella &  $y^2- z w^2=0$ & 2\\
The conifold &   $ xy=zw$ & 3 \\
\hline 
\end{tabular}
\end{center}
\caption{Examples of common  binomial varieties.}
\end{table}
 
\subsection{ Atiyah's flop and the small resolution of the conifold}

 A conifold is the double point singularity of the tip of an affine  quadric cone:
\begin{equation}
\mathscr{C}: u_1 u_2- v_1 v_2=0, \quad u_i, v_i \in \mathbb{C}.
\end{equation}
The quadric cone is smooth in codimension-one and codimension-two but  admits a double point  singularity  at the origin. It can be smoothed by a small resolution or by a blow-up. 
The blow-up $\pi: \hat{\mathscr{C}}\rightarrow \mathscr{C}$  of the origin resolves the singularity by replacing  the origin  by a  ruled surface $\mathbb{F}_0=\mathbb{CP}^1\times\mathbb{CP}^1$ defined by a quadric equation in $\mathbb{CP}^3$. This ruled surface $\mathbb{F}_0$ can be contracted   to a rational curve $\mathbb{CP}^1$ in two different ways corresponding to the two  different rulings of $\mathbb{F}_0$.  Each of these two contractions   $\bar{\pi}_i :\hat{\mathscr{C}}\rightarrow \hat{\mathscr{C}}_i$ ($i=1,2$)  defines  a small resolution of the original quadric cone. These two small resolutions  are  related to each other by a {\em flop transition} $\hat{\mathscr{C}}_1\dashleftarrow\dashrightarrow \hat{\mathscr{C}}_2$.  Alternatively $\hat{\mathscr{C}}_i$  can be  obtained by blowing up the (non-Cartier) Weil divisor $u_1=v_1=0$ or $u_1=v_2=0$:
\begin{equation}\label{sr}
\hat{\mathscr{C}}_1:
\begin{cases}
u_1 \alpha-\sigma v_1=0, \\
 v_2 \alpha - \sigma  u_2=0,
 \end{cases}, \quad 
{\hat{\mathscr{C}}}_2:
\begin{cases}
u_1 \alpha-\sigma v_2=0, \\
 v_1 \alpha - \sigma  u_2=0,
  \end{cases}
\end{equation}
where $u_i, v_j\in \mathbb{C}$ and $[\alpha: \sigma]$ are the projective coordinates of a $\mathbb{CP}^1$. The flop transition of the small resolution of the conifold was  described by Atiyah in 1958 \cite{Atiyah}. 

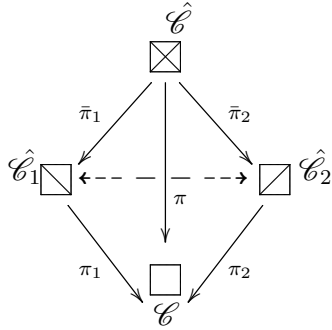
\begin{figure}[htb]
\begin{center}
\setlength{\unitlength}{.4 mm}
\begin{picture}(30,85)(10,0)
\put(0,65){$
\xymatrix{
 & 
 { \setlength{\unitlength}{.4 mm}
\begin{picture}(10,17)(-5,-6)
\put(0,0){\qbezier(-5,-5)(0,-5)(5,-5)
\qbezier(-5,5)(0,5)(5,5)
\qbezier(-5,-5)(-5,0)(-5,5)
\qbezier(5,-5)(5,0)(5,5)
\qbezier(-5,5)(0,0)(5,-5)
\qbezier(-5,-5)(0,0)(5,5)
\put(0,8){$ \hat{\mathscr{C}}$}}
\end{picture}
}
 \ar[dr]^{\bar{\pi}_2}\ar[dl]_{\bar{\pi}_1}\ar[dd]^{\pi} & \\
 { \setlength{\unitlength}{.4 mm}
\begin{picture}(10,12)(-5,-6)
\put(0,0){\qbezier(-5,-5)(0,-5)(5,-5)
\qbezier(-5,5)(0,5)(5,5)
\qbezier(-5,-5)(-5,0)(-5,5)
\qbezier(5,-5)(5,0)(5,5)
\qbezier(-5,5)(0,0)(5,-5)
\put(-16,0){$ \hat{\mathscr{C}}_1$}  
\put(7,-2){$\dashleftarrow - \     - \dashrightarrow$}
}
\end{picture}
}
\ar[dr]_{\pi_1} 
 & & 
 { \setlength{\unitlength}{.4 mm}
\begin{picture}(10,12)(-5,-6)
\put(0,0){\qbezier(-5,-5)(0,-5)(5,-5)
\qbezier(-5,5)(0,5)(5,5)
\qbezier(-5,-5)(-5,0)(-5,5)
\qbezier(5,-5)(5,0)(5,5)
\qbezier(-5,-5)(0,0)(5,5)
\put(8,0){$ \hat{\mathscr{C}}_2$}
}
 \end{picture}}
 \ar[dl]^{\pi_2} 
  \\
 & 
 { \setlength{\unitlength}{.4 mm}
\begin{picture}(10,30)(-5,-20)
\put(0,0){\qbezier(-5,-5)(0,-5)(5,-5)
\qbezier(-5,5)(0,5)(5,5)
\qbezier(-5,-5)(-5,0)(-5,5)
\qbezier(5,-5)(5,0)(5,5)
\put(-4,-15){$\mathscr{C}$}
}
 \end{picture}}
  & }
$}
\end{picture}
%\vspace{1.1cm}
\end{center}
\caption{Flop transition $\hat{\mathscr{C}}_1\dashleftarrow\dashrightarrow \hat{\mathscr{C}}_2$ betwen the two small resolutions of the conifold. }
\end{figure}

\subsection{ Toric description of the  binomial variety}

A binomial variety is always toric. $\mathscr{E}_{bin}: u_1 u_2 u_3 -v_1 v_2 =0$ is an affine  toric variety as it can be seen by describing it as the variety defined by the   semi-group $M$ generated by the four-vectors $m_{u_1}=(1,0,0,0)$, $m_{u_2}=(0,1,0,0)$, $m_{u_3}=(0,0,1,0)$, $m_{v_1}=(0,0,0,1)$ and $m_{v_2}=(1,1,1,-1)$ since they satisfy the relation  
$m_{u_1}+m_{u_2}+m_{u_3}=m_{v_1}+ m_{v_2}$. 
We can also describe its algebraic torus explicitly:  
$\mathscr{E}_*=\{(u_1, u_2, u_3, v_1, v_2)\in \mathscr{E}_{bin}: u_1 u_2 u_3 v_1 v_2 \neq 0).$  It is a 
 four-torus as can be seen by the  isomorphism:
\begin{equation}
\mathbb{T}^4:=(\mathbb{C}^*)^4\rightarrow \mathscr{E}_{*}:\quad (u_1 , u_2 , u_3 , v_1 )\mapsto (u_1, u_2, u_3, v_1 , u_1 u_2 u_3 v_1^{-1} ).
\end{equation}
  Every  lattice point $m=(a, b, c  ,d)\in \mathbb{Z}^4$ defines a  1-parameter subgroup as follows:  
\begin{equation}
\mathbb{C}^*\rightarrow  \mathscr{E}_{*}: \lambda \mapsto \lambda^m=(\lambda^a  , \lambda^b , \lambda^c, \lambda^d , \lambda^{a+b+c-d}
 ).
\end{equation}
These 1-parameter subgroups form a lattice equivalent to  the semi-group $M$ defined above. The dual lattice $N$ of $M$  defines the fan of the toric variety. It can alternatively be defined  as the set of conditions for which the  1-parameter subgroups are well defined  on $\mathscr{E}_{bin}$.  That is,  when the following inequalities hold: 
\begin{equation}
a\geq 0, \quad b\geq 0 ,\quad c\geq 0, \quad d\geq 0 ,\quad  a+b+c -d\geq 0.
\end{equation}
These  inequalities define a cone  $\sigma^\vee$ in $\mathbb{R}^4$ with apex at the origin. The cone $\sigma^\vee$ is generated  by   the rows of the following matrix:

\begin{equation}
\begin{matrix} \bar{a}\\ \bar{b} \\ \bar{c} \\ c\\ b\\  a  \end{matrix}
\begin{pmatrix}
1 & 0 & 0 & 0  \\
0 & 1 & 0 & 0  \\
 0 & 0 & 1 & 0  \\
 1 & 0 & 0 & 1\\
 0 & 1 & 0 & 1 \\
 0 & 0 & 1 & 1 
 \end{pmatrix}
{\setlength{\unitlength}{1.3 mm}
\begin{picture}(20,9)(-10,-9)
\put(0,0){
\qbezier(10,5)(15,2.5)(20,0)
\qbezier(0,0)(5,2.5)(10,5)
\qbezier(0,0)(10,0)(20,0)
\qbezier(0,0)(0,-10)(0,-20)
\put(10,5){
\linethickness{.05mm}
\qbezier(0,0)(0,-10)(0,-20)}
\put(20,0){
\qbezier(0,0)(0,-10)(0,-20)}

\put(0,-20){
\qbezier(0,0)(10,0)(20,0)
\linethickness{.05mm}
\qbezier(10,5)(15,2.5)(20,0)
\qbezier(0,0)(5,2.5)(10,5)
}
\put(-2,0){$a$}\put(-2,-20){$\bar{a}$}
\put(21,0){$b$}\put(21,-20){$\bar{b}$}
\put(10,6){$c$}\put(9.6,-18){$\bar{c}$}
}
\end{picture}
}
\end{equation}
This matrix  above has a very simple geometric interpretation: the rows correspond to the vertices of  a  triangular prism inside the three dimensional hyper-plane: 
$$(x_0,x_1,x_2,x_3)\in \mathbb{R}^4: x_0+x_1+x_2=1 .$$
The cone is completely specified by the prism.
 We can visualize the cone  by  drawing the prism  in a three dimensional space simply  by discarding the first column.  By doing so,  we have to keep in mind that a  point $(x_1,x_2,x_3)\in \mathbb{Z}^3$ in the three dimensional drawing space corresponds to an actual point of the prism  if we can find an $x_0\in  \mathbb{Z}$ such that 
$x_0+x_1+x_2=  1$.The different faces of the cone are also in one-to-one relation with the invariant loci of the torus action. A face of dimension $d$ in the prism corresponds to a face of dimension $(d+1)$ of the cone $\sigma^\vee$ and therefore to a  torus of dimension $4-(d+1)=3-d$. In particular, the prism itself corresponds to an invariant point of the torus action, a facet of the prism  corresponds to an invariant curve,  an edge of the prism corresponds to an invariant  surface of the torus action and a vertex corresponds to a toric divisor.

\subsection{Small resolutions and network of flop transitions}

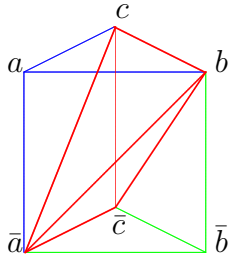
\begin{figure}[thb]
\begin{center}
\setlength{\unitlength}{1.2 mm}
\begin{picture}(15,30)(0,-20)

\put(0,0){

\color{blue}
\linethickness{.1mm}
\put(1,0){
\qbezier(0,0)(5,2.5)(10,5) 
\qbezier(0,0)(10,0)(20,0) 
\qbezier(0,0)(0,-10)(0,-20) 
\color{red}
\linethickness{.2mm}
\qbezier(10,5)(15,2.5)(20,0)
}

\put(10,5){
\color{red}
\linethickness{.05mm}
\qbezier(0,0)(0,-10)(0,-20)} 
\linethickness{.1mm}
\color{green}
\put(20,0){
\qbezier(0,0)(0,-10)(0,-20)
}  

\color{green}
\put(0,-20){
\linethickness{.1mm}
\qbezier(0,0)(10,0)(20,0)  
\linethickness{.1mm}
\qbezier(10,5)(15,2.5)(20,0) 
\color{red}\linethickness{.2mm}
\qbezier(10,5)(15,12.5)(20,20)  
\qbezier(0,0)(5,2.5)(10,5)
}
\color{black}
\put(-2,0){$a$}
\put(-2,-20){$\bar{a}$}
\put(21,0){$b$}
\put(21,-20){$\bar{b}$}
\put(10,6){$c$}\put(9.6,-18){$\bar{c}$}
{\color{red}
\put(10,5){

}

\put(10,5){
\linethickness{.2mm}
\qbezier(0,0)(-5,-12.5)(-10,-25)  

}
{\linethickness{.2mm}\qbezier(20,0)(10,-10)(0,-20)  
}}}

\end{picture}
\caption{The small resolution $T(\bar{a},b)$ of the  binomial variety $y s-w x t =0$ is obtained by a refinement  of the triangular prism $(a,b,c,\bar{a},\bar{b},\bar{c})$  into a sum of three  3 tetrahedra, namely  ${\color{blue}(\bar{a}, a , b, c)}$, {\color{red}$(\bar{a},\bar{c},b,c)$} and  {\color{green}$(\bar{a}, \bar{b},  \bar{c}, b)$}.
 } 
\end{center}
\end{figure}

The singular locus of the variety $\mathscr{E}_{bin}$ is  a bouquet of 3 lines of conifold singularities enhancing at the center of the bouquet where the three lines meet at one point.  In the toric description,  the dual cone of the variety  $\mathscr{E}_{bin}$ is determined by a triangular prism as explained above.    The three  conifold lines of $\mathscr{E}_{bin}$ correspond to the  three rectangular facets of the prism while the center of the bouquet is the interior of the prism. 
A  resolution of all the singularities of $\mathscr{E}_{bin}$ is obtained by  a simplicial refinement of the prism with cells of unit lattice volume. In particular, a {\em small resolution} occurs when  the  refinement  does not add any new vertex. This is because   a vertex corresponds to an invariant   divisor and  by definition a  small resolution does not have any divisor in its exceptional locus. It is possible to get a simplicial subdivision of any rational polytope by performing a succession of {\em star-subdivisions} of its fan.  A star-subdivision  is defined by a choice of a point $v$ and the following algorithm: any cone that contains $v$ is replaced by the joins of its faces with the ray through $v$; each cone not containing $v$ is left unchanged.
A star subdivision defines a proper birational map since after the subdivision, the fan still has the same 
support. If the center of the star subdivision is already a point of the polytope, the birational map is small since there are no new vertices created.

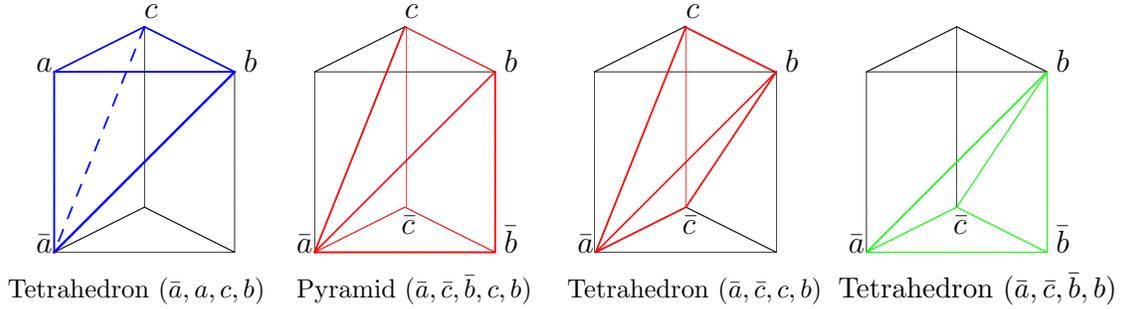
\begin{figure}[bt]
\begin{center}
\setlength{\unitlength}{1.2 mm}
\begin{picture}(100,35)(5,-23)
\put(0,0){
\put(-4,-25){\footnotesize Tetrahedron $(\bar a, a, c, b)$}
\linethickness{.2mm}
\color{blue}
\qbezier(10,5)(15,2.5)(20,0)
\qbezier(0,0)(5,2.5)(10,5)
\qbezier(0,0)(10,0)(20,0)
\qbezier(0,0)(0,-10)(0,-20)
\color{black}
\put(10,5){
\linethickness{.05mm}
\qbezier(0,0)(0,-10)(0,-20)}
\put(20,0){
\linethickness{.05mm}
\qbezier(0,0)(0,-10)(0,-20)}

\put(0,-20){
\linethickness{.05mm}
\qbezier(0,0)(10,0)(20,0)
\linethickness{.05mm}
\qbezier(10,5)(15,2.5)(20,0)
\qbezier(0,0)(5,2.5)(10,5)
}
\put(-2,0){$a$}\put(-2,-20){$\bar{a}$}
\put(21,0){$b$}
\put(10,6){$c$}
{\color{blue}
\put(10,5){\multiput(0,0)(-1,-2.5){10}{\linethickness{.2mm}\qbezier(0,0)(-.25,-0.625)(-.5,-1.25)}}
{\linethickness{.3mm}\qbezier(20,0)(10,-10)(0,-20)}}}

{

\put(30,0){
\put(-2,-25){\footnotesize Pyramid $(\bar a, \bar c, \bar b, c, b)$}
\color{red}\qbezier(10,5)(15,2.5)(20,0) 
\color{black} \linethickness{0.05mm}
\qbezier(0,0)(5,2.5)(10,5) 
\qbezier(0,0)(10,0)(20,0)  
\qbezier(0,0)(0,-10)(0,-20) 
\linethickness{0.2mm}
\put(10,5){
\color{red}
\linethickness{.05mm}
\qbezier(-1,0)(-1,-10)(-1,-20)} 
\color{red}
\put(20,0){
\qbezier(0,0)(0,-10)(0,-20)}

\put(0,-20){
\qbezier(0,0)(10,0)(20,0)
\linethickness{.05mm}
\qbezier(10,5)(15,2.5)(20,0)
\qbezier(0,0)(5,2.5)(10,5)
}
\color{black}
\put(-2,-20){$\bar{a}$}
\put(21,0){$b$}\put(21,-20){$\bar{b}$}
\put(10,6){$c$}\put(9.6,-18){$\bar{c}$}
{\color{red}
\put(10,5){
}

\put(10,5){
\qbezier(0,0)(-5,-12.5)(-10,-25)

}

{\linethickness{.2mm}\qbezier(20,0)(10,-10)(0,-20)}}}

}

{

\put(60,0){
\put(-2,-25){\footnotesize Tetrahedron $(\bar a, \bar c, c, b)$}
\linethickness{.2mm}

\put(1,0){\color{red}\qbezier(10,5)(15,2.5)(20,0) 

\color{black}
\linethickness{.05mm}
\qbezier(0,0)(5,2.5)(10,5) 
\qbezier(0,0)(10,0)(20,0)  
\qbezier(0,0)(0,-10)(0,-20)
}
\put(10,5){
\color{red}
\linethickness{.05mm}
\qbezier(0,0)(0,-10)(0,-20)}
\linethickness{.05mm}
\color{black}
\put(20,0){
\qbezier(0,0)(0,-10)(0,-20)
}  

\color{black}
\put(0,-20){
\linethickness{.05mm}
\qbezier(0,0)(10,0)(20,0)  
\linethickness{.05mm}
\qbezier(10,5)(15,2.5)(20,0)  
\color{red}\linethickness{.2mm}
\qbezier(10,5)(15,12.5)(20,20) 
\qbezier(0,0)(5,2.5)(10,5)
}
\color{black}
\put(-2,-20){$\bar{a}$}
\put(21,0){$b$}
\put(10,6){$c$}\put(9.6,-18){$\bar{c}$}
{\color{red}
\put(10,5){
}

\put(10,5){
\linethickness{.2mm}
\qbezier(0,0)(-5,-12.5)(-10,-25) 

}
{\linethickness{.2mm}\qbezier(20,0)(10,-10)(0,-20) 
}}}

}

{

\put(90,0){
\put(-2,-25){\small Tetrahedron $(\bar a, \bar c, \bar b, b)$}
\linethickness{.05mm}
\color{black}\qbezier(10,5)(15,2.5)(20,0) 
\qbezier(0,0)(5,2.5)(10,5) 
\qbezier(0,0)(10,0)(20,0)  
\qbezier(0,0)(0,-10)(0,-20) 
\put(10,5){

\qbezier(0,0)(0,-10)(0,-20)}
\color{green}

\linethickness{.1mm}
\put(20,0){
\qbezier(0,0)(0,-10)(0,-20)
}

\color{green}
\put(0,-20){
\qbezier(0,0)(10,0)(20,0)   

\qbezier(10,5)(15,2.5)(20,0) 
\qbezier(10,5)(15,12.5)(20,20)  
\qbezier(0,0)(5,2.5)(10,5) 
}
\color{black}
\put(-2,-20){$\bar{a}$}
\put(21,0){$b$}
\put(21,-20){$\bar{b}$}
\put(9.6,-18){$\bar{c}$}
{\color{green}
\put(10,5){
}

\put(10,5){

}
{\linethickness{.2mm}\qbezier(20,0)(10,-10)(0,-20)}}}

}
\color{black}
\end{picture}
\end{center}
\caption{The resolution $T(\bar a, b)$. We split the prism into a tetrahedron $(\bar a, a,b,c)$ and a rectangular pyramid $(\bar a, b,c,\bar b,\bar c)$ using the plane $(\bar a, b,c)$. We then split the rectangular pyramid 
into the two tetrahedra $(\bar a ,\bar c, c,b)$  and $(\bar a, \bar b, \bar c, b)$ using the plane 
$(\bar a,\bar c, b)$. \label{Res.Prism}}
\end{figure}
 
A small resolution of $\mathscr{E}_{bin}$ is given by the classical  subdivision of the prism into three tetrahedra. Such a subdivison can always be expressed as two star-subdivisions with centers $v_1$ and $v_2$ where the two centers are the end points of the  diagonal of one of the three rectangular facets of the prism. We will denote such a resolution $T(v_1,v_2)$.
Let us consider as an  example the resolution $T(\bar{a},b)$ defined by the vertex $\bar{a}$ and $b$. 
This is illustrated in table \ref{Res.Prism}. 
The star-subdivision determines by the vertex $\bar{a}$ generates a subdivision of the prism into a tetrahedron $(\bar{a},a,b,c)$ and the rectangular pyramid $(\bar{a},\bar{c},\bar{b}, b, c)$ as illustrated in the first two picture on the left of table  \ref{Res.Prism}.  
The second star-subdivision is determined by the vertex $b$. It  does not modify the tetrahedron $(\bar{a},a,b,c)$ but it subdivides the rectangular pyramid into two tetrahedra $(\bar{a},\bar{c},b,c)$ and $(\bar{a},\bar{b}, b)$. Alltogether, the small resolution $T(\bar{a},b)$ defines  the following subdivision of the prism into three tetrahedra  as illustrated  in table \ref{Res.Prism}:
\begin{align}
T(\bar{a},b)=\{(\bar{a},a,b, c), \quad (\bar{a},\bar{c},b, c),\quad
(\bar{a}, \bar{b},  \bar{c},  b)\}.
\end{align}

 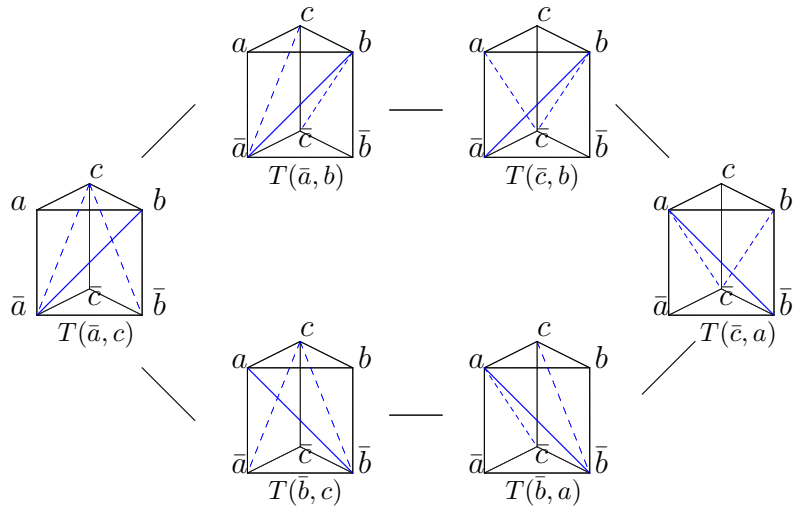
\begin{figure}[tb]
\begin{center}
\setlength{\unitlength}{.7 mm}
\begin{picture}(15,80)(40,-42)
\put(-20,10){
\put(4,-25){\footnotesize $T(\bar a,c)$}
\qbezier(10,5)(15,2.5)(20,0)
\qbezier(0,0)(5,2.5)(10,5)
\qbezier(0,0)(10,0)(20,0)
\qbezier(0,0)(0,-10)(0,-20)
\put(10,5){
\linethickness{.05mm}
\qbezier(0,0)(0,-10)(0,-20)}
\put(20,0){
\qbezier(0,0)(0,-10)(0,-20)}

\put(0,-20){
\qbezier(0,0)(10,0)(20,0)
\linethickness{.05mm}
\qbezier(10,5)(15,2.5)(20,0)
\qbezier(0,0)(5,2.5)(10,5)
}
\put(-5,0){$a$}\put(-5,-20){$\bar{a}$}
\put(22,0){$b$}\put(22,-20){$\bar{b}$}
\put(10,6){$c$}\put(9.6,-18){$\bar{c}$}
\color{blue}
\put(10,5){\multiput(0,0)(1,-2.5){10}{\linethickness{.07mm}\qbezier(0,0)(.25,-0.625)(.5,-1.25)}}
\put(10,5){\multiput(0,0)(-1,-2.5){10}{\linethickness{.07mm}\qbezier(0,0)(-.25,-0.625)(-.5,-1.25)}}
{\linethickness{.15mm}\qbezier(20,0)(10,-10)(0,-20)}

}

\put(20,-20){
\put(4,-25){\footnotesize $T(\bar b,c)$}
\put(0,0){
\qbezier(10,5)(15,2.5)(20,0)
\qbezier(0,0)(5,2.5)(10,5)
\qbezier(0,0)(10,0)(20,0)
\qbezier(0,0)(0,-10)(0,-20)
\put(10,5){
\linethickness{.05mm}
\qbezier(0,0)(0,-10)(0,-20)}
\put(20,0){
\qbezier(0,0)(0,-10)(0,-20)}

\put(0,-20){
\qbezier(0,0)(10,0)(20,0)
\linethickness{.05mm}
\qbezier(10,5)(15,2.5)(20,0)
\qbezier(0,0)(5,2.5)(10,5)
}
\put(-3,0){$a$}\put(-3,-20){$\bar{a}$}
\put(21,0){$b$}\put(21,-20){$\bar{b}$}
\put(10,6){$c$}\put(9.6,-18){$\bar{c}$}
\color{blue}
\put(10,5){\multiput(0,0)(1,-2.5){10}{\linethickness{.07mm}\qbezier(0,0)(.25,-0.625)(.5,-1.25)}}
\put(10,5){\multiput(0,0)(-1,-2.5){10}{\linethickness{.07mm}\qbezier(0,0)(-.25,-0.625)(-.5,-1.25)}}
{\linethickness{.15mm}\qbezier(0,0)(10,-10)(20,-20)}
}

}

\put(20,40){
\put(4,-25){\footnotesize $T(\bar a, b)$}
\put(0,0){
\qbezier(10,5)(15,2.5)(20,0)
\qbezier(0,0)(5,2.5)(10,5)
\qbezier(0,0)(10,0)(20,0)
\qbezier(0,0)(0,-10)(0,-20)
\put(10,5){
\linethickness{.05mm}
\qbezier(0,0)(0,-10)(0,-20)}
\put(20,0){
\qbezier(0,0)(0,-10)(0,-20)}

\put(0,-20){
\qbezier(0,0)(10,0)(20,0)
\linethickness{.05mm}
\qbezier(10,5)(15,2.5)(20,0)
\qbezier(0,0)(5,2.5)(10,5)
}
\put(-3,0){$a$}\put(-3,-20){$\bar{a}$}
\put(21,0){$b$}\put(21,-20){$\bar{b}$}
\put(10,6){$c$}\put(9.6,-18){$\bar{c}$}
\color{blue}
\put(20,0){\multiput(0,0)(-1,-1.5){10}{\linethickness{.07mm}\qbezier(0,0)(-.25,-0.375)(-.5,-.75)}}
\put(10,5){\multiput(0,0)(-1,-2.5){10}{\linethickness{.07mm}\qbezier(0,0)(-.25,-0.625)(-.5,-1.25)}}
{\linethickness{.15mm}\qbezier(20,0)(10,-10)(0,-20)}
}

}

\put(65,40){
\put(4,-25){\footnotesize $T(\bar c,b)$}
\put(0,0){
\qbezier(10,5)(15,2.5)(20,0)
\qbezier(0,0)(5,2.5)(10,5)
\qbezier(0,0)(10,0)(20,0)
\qbezier(0,0)(0,-10)(0,-20)
\put(10,5){
\linethickness{.05mm}
\qbezier(0,0)(0,-10)(0,-20)}
\put(20,0){
\qbezier(0,0)(0,-10)(0,-20)}

\put(0,-20){
\qbezier(0,0)(10,0)(20,0)
\linethickness{.05mm}
\qbezier(10,5)(15,2.5)(20,0)
\qbezier(0,0)(5,2.5)(10,5)
}
\put(-3,0){$a$}\put(-3,-20){$\bar{a}$}
\put(21,0){$b$}\put(21,-20){$\bar{b}$}
\put(10,6){$c$}\put(9.6,-18){$\bar{c}$}
\color{blue}
\put(20,0){\multiput(0,0)(-1,-1.5){10}{\linethickness{.07mm}\qbezier(0,0)(-.25,-0.375)(-.5,-.75)}}
\put(0,0){\multiput(0,0)(1,-1.5){10}{\linethickness{.07mm}\qbezier(0,0)(.25,-0.375)(.5,-.75)}}
{\linethickness{.15mm}\qbezier(20,0)(10,-10)(0,-20)}
}

}

\put(65,-20){
\put(4,-25){\footnotesize $T(\bar b,a)$}
\put(0,0){
\qbezier(10,5)(15,2.5)(20,0)
\qbezier(0,0)(5,2.5)(10,5)
\qbezier(0,0)(10,0)(20,0)
\qbezier(0,0)(0,-10)(0,-20)
\put(10,5){
\linethickness{.05mm}
\qbezier(0,0)(0,-10)(0,-20)}
\put(20,0){
\qbezier(0,0)(0,-10)(0,-20)}

\put(0,-20){
\qbezier(0,0)(10,0)(20,0)
\linethickness{.05mm}
\qbezier(10,5)(15,2.5)(20,0)
\qbezier(0,0)(5,2.5)(10,5)
}
\put(-3,0){$a$}\put(-3,-20){$\bar{a}$}
\put(21,0){$b$}\put(21,-20){$\bar{b}$}
\put(10,6){$c$}\put(9.6,-18){$\bar{c}$}
\color{blue}
\put(10,5){\multiput(0,0)(1,-2.5){10}{\linethickness{.07mm}\qbezier(0,0)(.25,-0.625)(.5,-1.25)}}
\put(0,0){\multiput(0,0)(1,-1.5){10}{\linethickness{.07mm}\qbezier(0,0)(.25,-0.375)(.5,-.75)}}
{\linethickness{.15mm}\qbezier(0,0)(10,-10)(20,-20)}
}

}

\put(100,10){
\put(6,-25){\footnotesize $T(\bar c,a)$}
\put(0,0){
\qbezier(10,5)(15,2.5)(20,0)
\qbezier(0,0)(5,2.5)(10,5)
\qbezier(0,0)(10,0)(20,0)
\qbezier(0,0)(0,-10)(0,-20)
\put(10,5){
\linethickness{.05mm}
\qbezier(0,0)(0,-10)(0,-20)}
\put(20,0){
\qbezier(0,0)(0,-10)(0,-20)}

\put(0,-20){
\qbezier(0,0)(10,0)(20,0)
\linethickness{.05mm}
\qbezier(10,5)(15,2.5)(20,0)
\qbezier(0,0)(5,2.5)(10,5)
}
\put(-3,0){$a$}\put(-3,-20){$\bar{a}$}
\put(21,0){$b$}\put(21,-20){$\bar{b}$}
\put(10,6){$c$}\put(9.6,-18){$\bar{c}$}
\color{blue}
\put(20,0){\multiput(0,0)(-1,-1.5){10}{\linethickness{.07mm}\qbezier(0,0)(-.25,-0.375)(-.5,-.75)}}
\put(0,0){\multiput(0,0)(1,-1.5){10}{\linethickness{.07mm}\qbezier(0,0)(.25,-0.375)(.5,-.75)}}
{\linethickness{.15mm}\qbezier(0,0)(10,-10)(20,-20)}
}

}

\put(10,30){\qbezier(0,0)(-5,-5)(-10,-10)}
\put(10,-30){\qbezier(0,0)(-5,5)(-10,10)}
\put(100,20){\qbezier(0,0)(-5,5)(-10,10)}
\put(105,-15){\qbezier(0,0)(-5,-5)(-10,-10)}
\put(47,29){\qbezier(0,0)(5,0)(10,0)}
\put(47,-29){\qbezier(0,0)(5,0)(10,0)}

\end{picture}
\caption{ The six small resolutions of the binomial variety $y s-w x t=0$. 
Each of the small resolution is obtained by a subdivision of the prism into  3 tetrahedra.    All the possible choices  are connected by a sequence of such  flop transitions on which the symmetric group $\mathscr{S}_3$ acts transitively. 
\label{Res.Prism.Flop} } 
\end{center}
\end{figure}
 
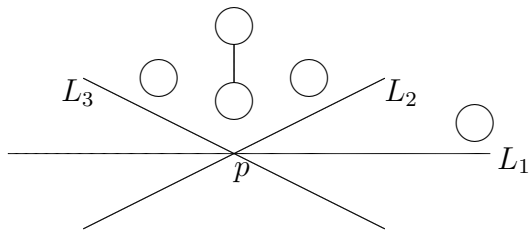
\begin{figure}[thb]
\begin{center}
\setlength{\unitlength}{1 mm}
\begin{picture}(20,30)(0,-5)
\qbezier(-30,0)(0,0)(34,0)
\qbezier(-20,10)(0,0)(20,-10)
\qbezier(-20,-10)(0,0)(20,10)
\put(10,10){\circle{5}}
\put(32,4){\circle{5}}
\put(0,-5){ \put(0,12){\circle{5}}\put(0,22){\circle{5}}\qbezier(0,19.5)(0,17)(0,14.5)}
\put(-10,10){\circle{5}}
\put(35,-2){$L_1$}
\put(20,7){$L_2$}
\put(-23,7){$L_3$}
\put(0,-3){$p$}
\end{picture}
\end{center}
\caption{The exceptional locus of the small resolution of the binomial variety $u_1 u_2 u_3=v_1 v_2$ is a $\mathbb{CP}^1$-bundle with a fiber which  enhances  at the origin to two $\mathbb{CP}^1$  intersecting  at a point.  
\label{str}
}
\end{figure}

 We could have also considered the small resolution 
$$T(\bar{a},c)=\{ (\bar{a},a,b, c), \ (\bar{a},b, c,  \bar{b}),\ (\bar{a}, \bar{b}, \bar{c}, c)\}.$$
The first star-subdivision with center $\bar{a}$ gives the same splitting into a tetrahedon and a rectangular pyramid as in $T(\bar{a},b)$. But the second star-subdivision  uses the other possible splitting of the pyramid into two tehtrahedrons. 
The difference between the resolutions $T(\bar{a},b)$ and $T(\bar{a},c)$ is then a choice on how to split the rectangular pyramid into two tetrahedra. This is exactly the same difference between the two small resolutions of the conifold.  More generally,two  resolutions $T(\bar{u},{v})$ and $T(\bar{p},{q})$  are related by a (conifold) flop transition defined by switching the choice of a diagonal plane  if and only if   $u=p$ or $v=q$.  Altogether,   we have a total of  6 small resolutions 
as  represented in figure \ref{Res.Prism.Flop}
$$
T(\bar{a},{b}), \quad T(\bar{a},{c}), \quad T( \bar{b},{c}),\quad  T(\bar{b},{a}), \quad T(\bar{c},{a}),\quad  T(\bar{c},{b}).
$$
The group $\mathscr{S}_3$ of permutations of 3 letters $\{a,b,c\}$ acts transitively on this set of resolutions.  They are connected to each other by a network of  flop transitions as represented in table \ref{Res.Prism.Flop}. The exceptional locus over a given cone  is  the  union of all the minimal cones intersecting its  relative interior.  For the six resolutions  we obtained, each rectangular  facet admits  as its exceptional locus  a $\mathbb{CP}^1$-bundle  over the singular lines and the  $\mathbb{CP}^1$ fiber enhances to two intersecting  $\mathbb{CP}^1$ at the common intersection of the three lines.
The structure of the exceptional locus is illustrated in figure \ref{str}.

\subsection{Algebraic description of the network of small resolutions }

The toric description shows that a small resolution can be obtained by blowing-up two Weil divisors corresponding to two vertices that are opposite ends of the same diagonal of a given rectangular facet of the prism. This can be easily implemented algebraically. We will derive it slightly differently using an analogy with the small resolution of the conifold.    Consider the defining equation of the binomial variety:  
 \begin{equation}
 u_1 u_2  u_3 = v_1  v_2, \quad  (u_1,u_2,u_3,v_1,v_2)\in \mathbb{C}^5.
 \end{equation}
Assuming for a moment that $v_1v_2\neq 0$, we can rewrite it  as the following  fractional relation
 \begin{equation}
 \frac{u_1}{v_1} \   \frac {u_2}{v_2} \   u_3 = 1. 
 \end{equation}
We then introduce two projective lines $\mathbb{P}^1$ with projective coordinates  $[\alpha_\pm, \sigma_\pm]$ and we identify their affine coordinates with   $u_1/v_1$ and $u_2/ v_2$: 
\begin{equation}
\frac{\sigma_+}{\alpha_+}=\frac{u_1}{v_1}, \quad\quad  \frac{\sigma_-}{\alpha_-}= \frac {u_2}{v_2}. 
\end{equation}
Using the defining equation of the binomial variety, we get the small resolution:
 \begin{align}
 {\mathscr{B}}_{12}:
\begin{cases}
u_1 \alpha_+ -  \sigma_+ v_1=0 ,\\
 u_2 \alpha_-- \sigma_-v_2 =0, \\
\alpha_+\alpha_- -  \sigma_+\sigma_- u_3  =0.
\end{cases} \ \text{where}  \ [\alpha_+:\sigma_+]\times[\alpha_-:\sigma_-]\in  \mathbb{F}_0=\mathbb{CP}^1\times \mathbb{CP}^1.
 \end{align}
Since the Jacobian has maximal rank, this is a smooth variety.
 The exceptional locus over a general point  of $L_1\cup L_2\cup L_3$  is just a $\mathbb{CP}^1$ inside $\mathbb{F}_0=\mathbb{CP}^1\times \mathbb{CP}^1$.  More precisely, $L_1$ gives a $\mathbb{CP}^1$ that represents a ruling of $\mathbb{F}_0$, while $L_2$ gives a $\mathbb{CP}^1$ that represents the other ruling  and $L_3$ is given by a {\em diagonal} $\mathbb{CP}^1$ which is not a ruling of $\mathbb{F}_0$ but  the conic  $\alpha_+\alpha_--\sigma_+\sigma_- u_3=0$ in  $\mathbb{F}_0$.  
 At the  point  $p=L_1\cap L_2=L_2\cap L_3=L_1\cap L_3=L_1\cap L_2\cap L_3$, where the three lines meet, the exceptional locus enhances to two  $\mathbb{CP}^1$ intersecting transversely.
Using this algebraic description, we can also see that there are six different small resolutions. Indeed, the  small resolution ${\mathscr{B}}_{12}$ can be replaced by a different  ``flop  dual"  resolution defined by permuting $(u_1,u_2 ,u_3)$.  In total there are six possibilities.  If we denote by $(i,j,k)$ a permutation of $(1,2,3)$,  we define the resolution $\pi_{ij}: {\mathscr{B}}_{ij}\rightarrow \mathscr{B}$ as follows: 
\begin{equation}
{\mathscr{B}}_{ij}:
\begin{cases}
 u_i \alpha_+  - \sigma_+ v_1 =0, \\
 u_j \alpha_-  -\sigma_- v_2=0, \\
\alpha_+\alpha_-- \sigma_+\sigma_- u_k =0.
\end{cases}
\ \text{where}  \ [\alpha_+:\sigma_+]\times[\alpha_-:\sigma_-]\in \mathbb{F}_0= \mathbb{CP}^1\times \mathbb{CP}^1.
\end{equation}
The  resolution $\pi_{ij}:{\mathscr{B}}_{ij}\rightarrow \mathscr{B}$ is obtained by blowing-up the two  Weil divisors $u_i=v_1=0$ and $u_j=v_2=0$. 
Since these Weil divisors are not Cartier, they actually don't lead to an exceptional divisor, the exceptional locus being in codimension-two. 
 These two divisors correspond to some  vertex $\bar{a}$ and $b$ of the prism and therefore the resolution $\pi_{ij}$ is the same as the toric resolution $T(\bar{a},b)$. The flop transition between the  resolutions $ {\mathscr{B}}_{ij}$ and $ {\mathscr{B}}_{ik}$ can be understood as a geometric procedure under which the $\mathbb{CP}^1$ fiber of the exceptional locus coming from the blow-up of $u_j=v_2=0$ is blown-down and  replaced by the $\mathbb{CP}^1$ coming from the blowup of $u_k=v_2=0$. We can give a similar description for the flop transition between ${\mathscr{B}}_{ij}$  and $ {\mathscr{B}}_{kj}$. Any resolution  ${\mathscr{B}}_{ij}$ can be mapped to to any other ${\mathscr{B}}_{i'j'}$ by a succession of such flop transitions as described by the  loop(compare with figure \ref{Res.Prism.Flop}): 
 \begin{equation}(12)-(13)-(23)-(21)-(31)-(32)-(12).\end{equation} 
 
\section{Small resolutions for $\SU(5)$ GUTs \label{SU5res}}
The small resolutions $\pi_{ij}:\mathscr{B}_{ij}\rightarrow \mathscr{B}$ of the binomial variety  $\mathscr{B}: u_1 u_2 u_3-v_1 v_2=0$ presented in the previous section can be pulled-back to $\mathscr{E}$ to provide small resolutions for all the  higher codimension singularities of the $\SU(5)$ GUT geometry $\mathscr{E}$. This procedure can be summarized by the following diagram:
\begin{equation}
\xymatrix{
{\mathscr{E}}_{ij} \ar[d]^{\hat{\pi}_{ij}}\ar[r] & \mathscr{B}_{ij} \ar[d]^{\pi_{ij}} \\
\mathscr{E} \ar[r]  & \mathscr{B} 
}
\end{equation}
where the map $\mathscr{E}\rightarrow \mathscr{B}$ associated to each  point $p$ of $\mathscr{E}$, the  point $(u_1, u_2, u_3, v_1, v_2)$ of    $\mathscr{B}$ such that:
\begin{equation}
u_1=x, \quad, u_2=w, \ u_3=t, \ v_1=y,\quad  v_2=s.
\end{equation}
We recall that  $t=x+\beta_4+\beta_2 w+ \beta_0 w^2$ and $s=y+\beta_5+\beta_3 w$.  Denoting by  $[\alpha_\pm:\sigma_\pm]$ the  projective coordinates of  two $\mathbb{CP}^1$, the  
6 small  resolutions  are:
 \begin{align}
 \begin{tabular}{lll}
 ${\mathscr{E}}_{wx}$ & $ {\mathscr{E}}_{tw}$ & ${\mathscr{E}}_{xt}$\\
$\begin{cases}
 w \alpha_+  -\sigma_+ y =0 \\
 x \alpha_-  - \sigma_- s =0 \\
\alpha_- \alpha_+  -  \sigma_-\sigma_+  t=0 
\end{cases}$
& 
$\begin{cases}
 t \alpha_+  -\sigma_+ y =0 \\
 w \alpha_-  - \sigma_- s =0 \\
\alpha_- \alpha_+  -  \sigma_-\sigma_+  x=0 
\end{cases} $
& 
$\begin{cases}
 x \alpha_+  -\sigma_+ y =0 \\
 t \alpha_-  - \sigma_- s =0 \\
\alpha_- \alpha_+  -  \sigma_-\sigma_+  w=0 
\end{cases} $
\\   \\
${\mathscr{E}}_{wt}$ & $ {\mathscr{E}}_{xw}$ & $ {\mathscr{E}}_{tx}$\\
$\begin{cases}
 w \alpha_+  -\sigma_+ y =0 \\
 t \alpha_-  - \sigma_- s =0 \\
\alpha_- \alpha_+  -  \sigma_-\sigma_+  x=0 
\end{cases} $
&
$\begin{cases}
 x \alpha_+  -\sigma_+ y =0 \\
 w \alpha_-  - \sigma_- s =0 \\
\alpha_- \alpha_+  -  \sigma_-\sigma_+  t=0 
\end{cases} $
&
$\begin{cases}
 t \alpha_+  -\sigma_+ y =0 \\
 x \alpha_-  - \sigma_- s =0 \\
\alpha_- \alpha_+  -  \sigma_-\sigma_+  w=0
\end{cases} $
\end{tabular}
 \end{align}
The exceptional locus  is a $\mathbb{CP}^1$-bundle over the bouquet $L_x\cup L_w\cup L_t$, but the fiber enhances to two intersecting $\mathbb{CP}^1$ over the center of the bouquet:
$$
\hat{\pi}_{ij}^{-1}(x\in L_x\cup L_w\cup L_t- \{ p_4 \})=
\quad
{\setlength{\unitlength}{1 mm}
\begin{picture}(10,0)(0,0)
{ \put(0,0){\circle{5}} \put(-2,4){$\mathbb{CP}^1$}}
 \end{picture}},\quad 
\hat{\pi}_{ij}^{-1}(p_4)=\quad 
{\setlength{\unitlength}{1 mm}
\begin{picture}(10,10)(0,0)
\put(0,-12){ \put(0,12){\circle{5}}\put(0,22){\circle{5}}\qbezier(0,19.5)(0,17)(0,14.5)
\put(4,11){$\mathbb{CP}^1$}
\put(4,21){$\mathbb{CP}^1$}
}
 \end{picture}}
$$
It is easy to see (for example, by computing the Jacobian) that all the six resolutions  are smooth.   
Although the 6 small resolutions $\mathscr{B}_{ij}$  of the binomial variety $\mathscr{Bin}$ had a similar fiber structure, their uplifts ${\mathscr{E}}_{ij}$  to the $\SU(5)$ geometry would be significantly different from each.
The dependence of $s$ and $t$ on the base through the sections $\beta_0,\beta_2,\beta_3, \beta_4$ and $\beta_5$ will be responsible for the enhancement of fibers of the $\SU(5)$ model. 
In the process of adding the extra nodes coming from the exceptional loci, we  will get new  types of   fibers.

\subsection{Codimension-2  singular   fibers: $\tilde{A}_5$  and   $\tilde{D}_5$ enhancements.}

\begin{figure}[hb t ]
\setlength{\unitlength}{.5 mm}
\begin{picture}(25,60)(-45,-50)
\put(35,-20){\large$\overset{P=0}{\text{\huge $\longrightarrow$}}$}
\put(140,-20){\large$\overset{{ Resolution}}{\text{\huge $\longrightarrow$}}$}

\put(0,0){
\put(6,0){\footnotesize $C_0$}
\put(-3,-2){{$\times$}}
\put(25,-20){\footnotesize $C_{1-}$}
\put(-37,-20){\footnotesize $C_{1+}$}
\put(20,-40){\footnotesize $C_{2-}$}
\put(-32,-40){\footnotesize $C_{2+}$}
\put(0,0){\circle{10}}
\put(-20,-20){\circle{10}}
\put(20,-20){\circle{10}}
\put(15,-40){\circle{10}}
\put(-15,-40){\circle{10}}
\qbezier(-3,-3)(-5,-5)(-17,-17)
\qbezier(3,-3)(5,-5)(17,-17)
\qbezier(-19,-26)(-18,-30.5)(-17,-35)
\qbezier(19,-26)(18,-30.5)(17,-35)
\qbezier(-10,-40)(-10,-40)(10,-40)
}

\put(105,0){
\put(6,0){\footnotesize $C_0$}
\put(-3,-2){{$\times$}}
\put(25,-20){\footnotesize $C_{1-}$}
\put(-37,-20){\footnotesize $C_{1+}$}
\put(20,-40){\footnotesize $C_{2-}$}
\put(-32,-40){\footnotesize $C_{2+}$}
\put(0,0){\circle{10}}
\put(-20,-20){\circle{10}}
\put(20,-20){\circle{10}}
\put(15,-40){\circle{10}}
\put(-15,-40){\circle{10}}
\qbezier(-3,-3)(-5,-5)(-17,-17)
\qbezier(3,-3)(5,-5)(17,-17)
\qbezier(-19,-26)(-18,-30.5)(-17,-35)
\qbezier(19,-26)(18,-30.5)(17,-35)
{\color{red} \qbezier(-10,-40)(-10,-40)(10,-40)}

}

\put(215,-10){
\put(6,20){\footnotesize $C_0$}
\put(-3,18){{$\times$}}
\put(25,0){\footnotesize $C_{1-}$}
\put(-36,0){\footnotesize $C_{1+}$}
\put(25,-20){\footnotesize $C_{2-}$}
\put(-36,-20){\footnotesize $C_{2+}$}
\qbezier(-20,-5)(-20,-10)(-20,-15)
\qbezier(20,-5)(20,-10)(20,-15)

\put(0,-40){\color{blue}\circle*{10}}
\put(-20,0){\circle{10}}
\put(20,0){\circle{10}}
\put(-20,-20){\circle{10}}
\put(20,-20){\circle{10}}
\put(0,20){\circle{10}}

\put(-3,-30){\footnotesize\color{blue} $C_{w}$}
\qbezier(-3,17)(-5,15)(-17,3)
\qbezier(3,17)(5,15)(17,3)
\qbezier(-3,-37)(-5,-35)(-17,-23)
\qbezier(3,-37)(5,-35)(17,-23)

}
\end{picture}
\caption{
The codimension-2 fiber transition $\tilde{A}_4\rightarrow \tilde{A}_5$ above the curve $P=0$ (assuming that $\beta_3\neq 0$). 
This turns the fiber $I_5$ into a Kodaira fiber $I_6$ which corresponds to the extended Dynkin diagram $\tilde{A}_5$. \label{A5}
}
\end{figure}
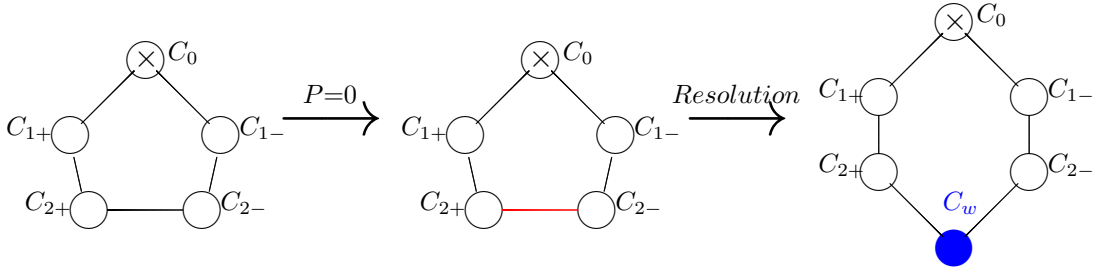

A general fiber of $\mathscr{E}$ over the divisor $D_{\su(5)}$ is a closed chain of 5 $\mathbb{CP}^1$-nodes, namely $C_0$, $C_{1\pm}$ and $C_{2\pm}$.  
In order to analyze the structure of the fiber of the resolutions $\mathscr{E}_{ab}$, we just have to understand what is happening at the singular locus $L_t\cup L_x\cup L_w$. 
We notice that $L_t$ and $L_x$ are both located above the curve $\Sigma_{10}:\beta_5=0$ of the base, whereas $L_w$ is located above $\Sigma_5:P=0$. 
The $I_5$ fiber undergoes a topological change over these two curves. 
Over the locus $\beta_5=0$,  the curves  $C_{1+}$ and $C_{1-}$ coincide to form  a  rational curve  $C_1$ of multiplicity 2. This rational curve $C_1$  intersects   $C_{2+}$ and $C_{2-}$ at a unique point which corresponds to $L_t$. When $\beta_4\neq 0$, there is another singular point  on $C_{1}$ away from $L_t$, this is the singularity $L_x$. The small resolution of $\mathscr{E}_0$  replaces on each fiber over  $\beta_5=0$, the singular points $L_t$ and $L_x$ by the rational curves $C_t$ and $C_x$.  The resulting fiber has the structure of an affine diagram  $\tilde{D}_{5}$. This enhancement is reviewed in table \ref{D5}. 
Above the curve $\Sigma_5$, the singular point $L_w$ corresponds fiber wise to a singular point sitting at the intersection of $C_{2+}$ and $C_{2-}$. Any of the six small resolutions replaces that point by a $\mathbb{CP}^1$ that we call $C_w$. This leads to an  enhancement of the fiber $I_5$ to a fiber $I_6$ with dual graph $\tilde{A}_5$.  This is presented in table \ref{A5}.
The curve $C_x$, $C_t$ and $C_w$ are realized differently for each resolution ${\mathscr{E}}_{ij}$ \footnote{ In general given a curve $C_{\ell}$ and the resolution ${\mathscr{E}}_{ij}$, if $\ell=i$,  
$C_{\ell}$ is defined by $L_{\ell}$ together with the condition $u_i \alpha_+=\alpha_+\alpha_-=0$,  therefore it is defined by the $\mathbb{CP}^1:\alpha_+=0$, that is the line  $\mathbb{CP}^1\times[0:1]$ in $\mathbb{F}_0$. In the same way if $\ell=j$, $C_{\ell}$ we be given by $\alpha_-=0$ corresponding to the line    $[0:1]\times\mathbb{CP}^1$ in $\mathbb{F}_0$. Finally if ($\ell\neq i,j$), then $C_\ell$ is defined by the quadratic equation $\alpha_-\alpha_+-\sigma_-\sigma_+ u_k=0$ in $\mathbb{F}_0$. We call it a diagonal $\mathbb{CP}^1$ since it does not correspond to one of the ruling of $\mathbb{F}_0$. }
.

\begin{figure}[!b h t ]
\setlength{\unitlength}{.5 mm}
\begin{picture}(25,60)(-60,-50)
\put(45,-20){\large$\overset{\beta_5=0}{\text{\huge $\longrightarrow$}}$}
\put(120,-20){\large$\overset{{ Resolution}}{\text{\huge $\longrightarrow$}}$}

\put(0,0){
\put(6,0){\footnotesize $C_0$}
\put(-3,-2){{$\times$}}
\put(25,-20){\footnotesize $C_{1-}$}
\put(-37,-20){\footnotesize $C_{1+}$}
\put(20,-40){\footnotesize $C_{2-}$}
\put(-32,-40){\footnotesize $C_{2+}$}
\put(0,0){\circle{10}}
\put(-20,-20){\circle{10}}
\put(20,-20){\circle{10}}
\put(15,-40){\circle{10}}
\put(-15,-40){\circle{10}}
\qbezier(-3,-3)(-5,-5)(-17,-17)
\qbezier(3,-3)(5,-5)(17,-17)
\qbezier(-19,-26)(-18,-30.5)(-17,-35)
\qbezier(19,-26)(18,-30.5)(17,-35)
\qbezier(-10,-40)(-10,-40)(10,-40)
}
\put(90,0){
\put(0,0){\circle{10}}
\put(-3,-2){{$\times$}}
\put(0,-20){\circle{10}}
\put(15,-40){\circle{10}}
\put(-15,-40){\circle{10}}
\put(6,0){\footnotesize $C_0$}
\put(6,-20){\footnotesize $2 C_{1}$}
\put(20,-40){\footnotesize $C_{2-}$}
\put(-33,-40){\footnotesize $C_{2+}$}
\put(0,-25){\color{red}\circle*{2}}
\put(-5,-20){\color{red}\circle*{2}}
\qbezier(0,-5)(0,-7)(0,-15)
\qbezier(0,-25)(-7,-31)(-14, -36)
\qbezier(0,-25)(7,-31)(14, -36)
}

\put(180,-20){
\put(-15,25){\color{blue}\circle*{10}}
\put(15,25){\circle{10}}
\put(12.5,23){{$\times$}}
\put(0,10){\circle{10}}
\put(0,-10){\color{blue}\circle*{10}}
\put(-15,-25){\circle{10}}
\put(15,-25){\circle{10}}
\put(6,7){\footnotesize $ 2C_{1}$}
\put(6,-10){\footnotesize \color{blue}$ 2C_t$}
\put(21,-25){\footnotesize $C_{2-}$}
\put(-33,-25){\footnotesize $C_{2+}$}
\put(21,25){\footnotesize $C_{0}$}
\put(-29,25){\footnotesize\color{blue} $C_x$}
\qbezier(0,5)(0,0)(0,-5.5)
\qbezier(5,12)(8,15)(14,21)
\qbezier(-5,12)(-8,15)(-14,21)
\qbezier(-5,-12)(-8,-15)(-14,-21)
\qbezier(5,-12)(8,-15)(14,-21)
}

\end{picture}
\caption{ \footnotesize{
The codimension-2 fiber enhancement  $\tilde{A}_4\rightarrow \tilde{D}_5$. It corresponds to the resolution of codimension-2 singularities above the curve $w=\beta_5=0$. 
}
\label{D5}
}
\end{figure}
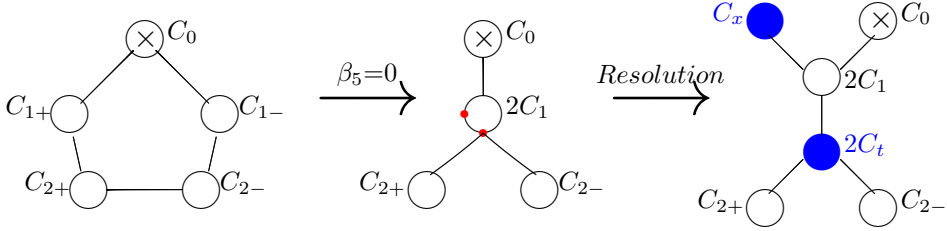

\subsection{An affine $\tilde{D}_6$ enhancement  in codimension-3}

When we were analyzing the structure of the fibers in codimension-2 we have assumed that $\beta_3 \beta_4 \neq 0$ in order to obtain the fiber $I_6$ and that  $\beta_4\neq 0$ to obtain the fiber with dual graph $\tilde{D}_5$. 
If we allow $\beta_3=0$ or $\beta_4=0$ we are in codimension-3 and we can expect a enhancement of the singular fibers. 
The  specialization $\beta_3=\beta_5=0$ and $\beta_4=\beta_5=0$ can be geometrically described as  collisions of singular fibers $I_6$ and $\tilde{D}_5$ located above $P=0$ and $\beta_5=0$. 
 We will first consider the case
 $$\Pi_3:\beta_5=\beta_3=0$$
 and we  will analyze the case $\beta_5=\beta_4=0$ later. As we specialize to $\beta_5=\beta_3=0$ in the $\SU(5)$ divisor,  the curves $C_{1+}$ and $C_{1-}$  (resp. $C_{2+}$ and $C_{2-}$) coincide and determine  a unique curve of multiplicity 2 that we call $C_1$ (resp. $C_2$). 
Each of our 6 small resolutions replaces on each fiber of $\beta_5=\beta_3=0$ the points corresponding to $L_x$ and $L_t$ by   a $\mathbb{CP}^1$ as  it was done already in codimension-2. But something new happens for the locus $L_w$. 
When we specialize to $\beta_3=\beta_5=0$,   the locus $L_w:x=y=\beta_5+\beta_3 w=\beta_4+\beta_2 w + \beta_0 w^2=0$ reduces to  
$ x=y=\beta_5=\beta_3=\beta_4+\beta_2 w + \beta_0 w^2=0$.  This defines two points  $L_{w\pm}$ on the component $C_2$ of the fiber above $\beta_5=\beta_3=0$. Indeed, 
the  equation $\beta_4+\beta_2 w + \beta_0 w^2=0$ gives two different values $w_\pm$ for $w$ in terms of $\beta_4,\beta_2,\beta_0$.  We can think of this as the splitting of the locus $L_{w}$ into two different loci $L_{w\pm}$: 
$$
L_w \xrightarrow{\beta_3=\beta_5=0} L_{w\pm}, \quad w_\pm=\frac{-\beta_2 \pm \sqrt{\beta_2^2-4 \beta_4 \beta_0}}{2\beta_0^2}.
$$
Since we are in codimension-3, we can safely assume that in general $\beta_0\neq 0$ and $\beta_2^2-4 \beta_4 \beta_0\neq 0$ so that we indeed have two values for $w$ which correspond to  two disjoint points on the node $C_2$. 
In the resolution, each of these points $L_{w\pm}$ is replaced by  a $\mathbb{CP}^1$ that we call respectively $C_{w+}$ and $C_{w-}$ \footnote{When $\beta_3\neq 0$, there were a unique value for $w$ since the equation $\beta_5+ \beta_3 w=0$ had a unique solution for $w$. The locus $P=0$ was obtained as a compatibility condition with the other relation $\beta_4+\beta_2 w +\beta_0 w^2=0$ by replacing $w$ by the solution $-\beta_5/ \beta_3$. }.  
Alltogether, we see that for any of our 6 small resolutions, over  $\beta_5=\beta_3=0$ in the $\SU(5)$-divisior,  we get a  fiber which has the structure of an affine Dynkin diagram $\tilde{D}_6$. 
The singular points $L_{w\pm}$  are replaced by $\mathbb{CP}^1$s that are realized as hypersurfaces in  the projective cone $\mathbb{F}_0$ with the same equation as $C_{w}$.

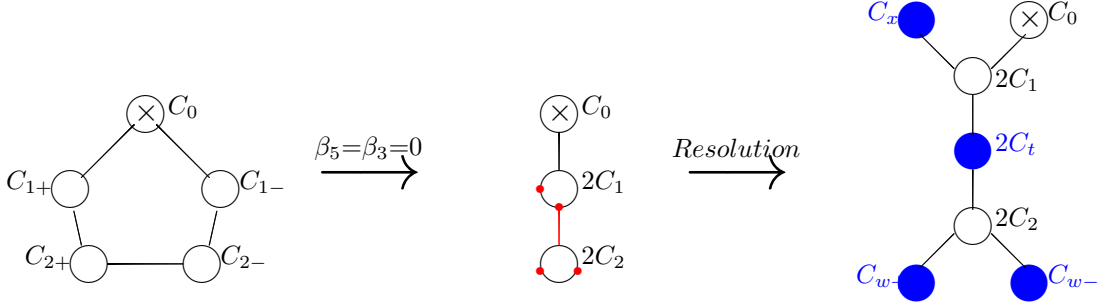
\begin{figure}[!b h t ]
\setlength{\unitlength}{.5 mm}
\begin{picture}(25,100)(-45,-50)
\put(45,-20){\large$\overset{\beta_5=\beta_3=0}{\text{\huge $\longrightarrow$}}$}
\put(140,-20){\large$\overset{{ Resolution}}{\text{\huge $\longrightarrow$}}$}

\put(0,0){
\put(6,0){\footnotesize $C_0$}
\put(-3,-2){{$\times$}}
\put(25,-20){\footnotesize $C_{1-}$}
\put(-37,-20){\footnotesize $C_{1+}$}
\put(20,-40){\footnotesize $C_{2-}$}
\put(-32,-40){\footnotesize $C_{2+}$}
\put(0,0){\circle{10}}
\put(-20,-20){\circle{10}}
\put(20,-20){\circle{10}}
\put(15,-40){\circle{10}}
\put(-15,-40){\circle{10}}
\qbezier(-3,-3)(-5,-5)(-17,-17)
\qbezier(3,-3)(5,-5)(17,-17)
\qbezier(-19,-26)(-18,-30.5)(-17,-35)
\qbezier(19,-26)(18,-30.5)(17,-35)
\qbezier(-10,-40)(-10,-40)(10,-40)
}

\put(110,0){
\put(0,0){\circle{10}}
\put(-3,-2){{$\times$}}
\put(0,-20){\circle{10}}
\put(0,-40){\circle{10}}
\put(6,0){\footnotesize $C_0$}
\put(6,-20){\footnotesize $2 C_{1}$}
\put(6,-40){\footnotesize $2C_{2}$}
\put(0,-25){\color{red}\circle*{2}}
\put(-5,-20){\color{red}\circle*{2}}
\qbezier(0,-5)(0,-7)(0,-15)
\put(0,-20){\color{red}\qbezier(0,-5)(0,-7)(0,-15)}
\put(-5,-42){\color{red}\circle*{2}}
\put(5,-42){\color{red}\circle*{2}}
}

\put(220,0){
\put(-15,25){\color{blue}\circle*{10}}
\put(15,25){\circle{10}}
\put(12,23){{$\times$}}
\put(0,10){\circle{10}}
\put(0,-10){\color{blue}\circle*{10}}
\put(-15,-45){\color{blue}\circle*{10}}
\put(15,-45){\color{blue}\circle*{10}}
\put(0,-30){\circle{10}}
\put(6,-30){\footnotesize $ 2C_{2}$}
\put(6,7){\footnotesize $ 2C_{1}$}
\put(6,-10){\footnotesize \color{blue}$2C_{t}$}
\put(20,-45){\footnotesize \color{blue}$C_{w-}$}
\put(-31,-45){\footnotesize \color{blue}$C_{w+}$}
\put(20,25){\footnotesize $C_{0}$}
\put(-28,25){\footnotesize\color{blue} $C_x$}
\qbezier(0,5)(0,0)(0,-5.5)
\qbezier(5,12)(8,15)(14,21)
\qbezier(-5,12)(-8,15)(-14,21)
\qbezier(-5,-32)(-8,-35)(-14,-41)
\qbezier(5,-32)(8,-35)(14,-41)
\qbezier(0,-15)(0,-18)(0,-25.5)

}

\end{picture}
\caption{ { Fiber  above  $\Pi_3: \beta_5=\beta_3=0$ in the $\SU(5)$ divisor}. 
After the resolution we get a fiber with dual graph an affine Dynkin diagram   $\tilde{D}_{6}$.
}
\end{figure}

%\clearpage

\subsection{Exotic fibers in codimension-three}

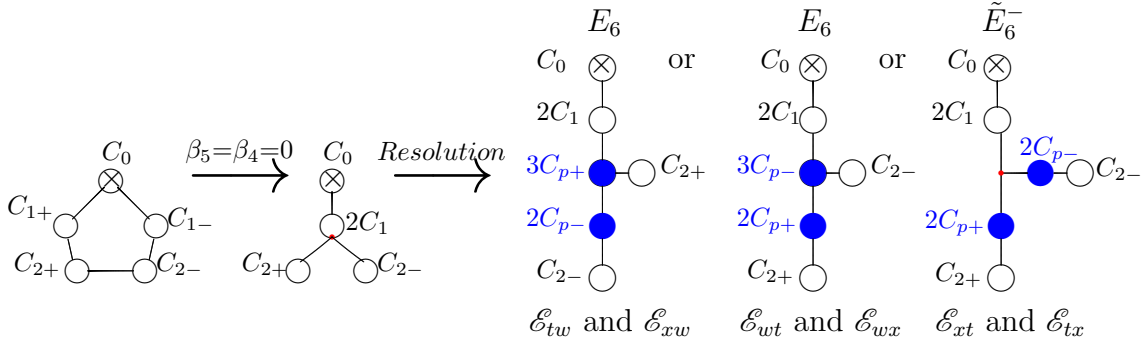
\begin{figure}[   bht ]
\setlength{\unitlength}{.35 mm}
\begin{picture}(25,140)(-45,-80)

\put(-17,-10){
\setlength{\unitlength}{.3 mm}
\put(45,-20){\large$\overset{\beta_5=\beta_4=0}{\text{\huge $\longrightarrow$}}$}
\put(130,-20){\large$\overset{{ Resolution}}{\text{\huge $\longrightarrow$}}$}

\put(12,-15){
\put(-4,9){\footnotesize $C_0$}
\put(-5,-3){{$\times$}}
\put(25,-20){\footnotesize $C_{1-}$}
\put(-46,-15){\footnotesize $C_{1+}$}
\put(20,-40){\footnotesize $C_{2-}$}
\put(-43,-40){\footnotesize $C_{2+}$}
\put(0,0){\circle{10}}
\put(-20,-20){\circle{10}}
\put(20,-20){\circle{10}}
\put(15,-40){\circle{10}}
\put(-15,-40){\circle{10}}
\qbezier(-3,-3)(-5,-5)(-17,-17)
\qbezier(3,-3)(5,-5)(17,-17)
\qbezier(-19,-26)(-18,-30.5)(-17,-35)
\qbezier(19,-26)(18,-30.5)(17,-35)
\qbezier(-10,-40)(-10,-40)(10,-40)
}

\put(110,-15){
\put(0,0){\circle{10}}
\put(-5,-3){{$\times$}}
\put(0,-20){\circle{10}}
\put(15,-40){\circle{10}}
\put(-15,-40){\circle{10}}
\put(-4,9){\footnotesize $C_0$}
\put(6,-22){\footnotesize $2 C_{1}$}
\put(20,-40){\footnotesize $C_{2-}$}
\put(-40,-40){\footnotesize $C_{2+}$}
\put(0,-25){\color{red}\circle*{2}}
\qbezier(0,-5)(0,-7)(0,-15)
\qbezier(0,-25)(-7,-31)(-14, -36)
\qbezier(0,-25)(7,-31)(14, -36)
}

}

\put(140,0){
\put(40,20){
\put(-6,14){$E_6$}
\put(0,0){\circle{10}}
\put(-5,-2){{$\times$}}
\put(-25,0){\footnotesize $C_0$}
\put(25,0){{or}}
\qbezier(0,-5)(0,-7)(0,-15)
\put(-25,-20){\footnotesize $2 C_{1}$}
\put(0,-20){\circle{10}}
\put(0,-20){
\put(0,-20){\circle{10}}
{\color{blue}\put(0,-20){\circle*{10}}
\put(0,-40){\circle*{10}}}
\put(0,-60){\circle{10}}
\multiput(0,0)(0,-20){3}{\qbezier(0,-5)(0,-10)(0,-15)}
\put(15,-20){\circle{10}}
\qbezier(5,-20)(7,-20)(10,-20)
\put(22,-20){\footnotesize $C_{2+}$}
{\color{blue} \put(-29,-20){\footnotesize $3C_{p+}$}
\put(-29,-40){\footnotesize $2C_{p-}$}}
\put(-25,-60){\footnotesize $C_{2-}$}
\put(-28,-80){{$\mathscr{E}_{tw} \ \text{and}\ \mathscr{E}_{xw}$}}
}
}

\put(120,20){
\put(-6,14){$E_6$}
\put(0,0){\circle{10}}
\put(-5,-2){{$\times$}}
\put(-20,0){\footnotesize $C_0$}
\put(25,0){{or}}
\qbezier(0,-5)(0,-7)(0,-15)
\put(-21,-20){\footnotesize $2 C_{1}$}
\put(0,-20){\circle{10}}
\put(0,-20){
\put(0,-20){\circle{10}}
{\color{blue}\put(0,-20){\circle*{10}}
\put(0,-40){\circle*{10}}}
\put(0,-60){\circle{10}}
\multiput(0,0)(0,-20){3}{\qbezier(0,-5)(0,-10)(0,-15)}
\put(15,-20){\circle{10}}
\qbezier(5,-20)(7,-20)(10,-20)
\put(22,-20){\footnotesize $C_{2-}$}
{\color{blue} \put(-29,-20){\footnotesize $3C_{p-}$}
\put(-29,-40){\footnotesize $2C_{p+}$}}
\put(-25,-60){\footnotesize $C_{2+}$}
\put(-28,-80){{$\mathscr{E}_{wt} \  \text{and}\  \mathscr{E}_{wx}$}}
}
}

\put(190,20){
\put(-6,14){$\tilde{E}^-_6$}
\put(0,0){\circle{10}}
\put(-5,-2){{$\times$}}
\put(-19,0){\footnotesize $C_0$}
\qbezier(0,-5)(0,-7)(0,-15)
\put(-25,-20){\footnotesize $2 C_{1}$}
\put(0,-20){\circle{10}}
\put(-2.5,-20){
{\color{red}
\color{blue}\put(0,-40){\circle*{10}}}
\put(0,-60){\circle{10}}
\put(0,0){\qbezier(0,-5)(0,-20)(0,-35)}
\put(0,-40){\qbezier(0,-5)(0,-10)(0,-15)}
\put(15,-20){{\color{blue}\circle*{10}} \put(-20,0){\qbezier(-5,0)(2,0)(5,0) }}
\put(30,-20){\circle{10} \put(-20,0){\qbezier(0,0)(2,0)(5,0) }}
\put(36,-21){\footnotesize $C_{2-}$}
{\color{blue} \put(7,-13){\footnotesize $2C_{p-}$}
\put(-29,-41){\footnotesize $2C_{p+}$}}
\put(-25,-61){\footnotesize $C_{2+}$}
\color{red}\put(0,-20){\circle*{2}}
\color{black}
\put(-25,-80){{$\mathscr{E}_{xt}\  \text{and}\  \mathscr{E}_{tx}$}}
}
}

}

\end{picture}
\caption{\label{codim3b4} \footnotesize{
Codimension-3 singularity in the base located at the points $\Pi_4:\beta_5=\beta_4=0$ of the $\SU(5)$-divisor.
The first two fibers on the left  have the structure of a projective $E_6$ Dynkin diagram. 
} }
\end{figure}

We will now consider the enhancement of the singular fiber at the points
$$\Pi_4: \beta_5=\beta_4=0,$$
 in the $\SU(5)$ divisor $D_{\su(5)}$. These  are  codimension-three points in the base of the elliptic fibration.  They are a sublocus of the  intersection of the curves $\Sigma_5$ and $\Sigma_{10}$.  
Above $\Pi_4$,  the nodes $C_{1+}$
 and $C_{1-}$ merge into a unique node $C_1$  of multiplicity 2. Moreover,    the nodes $C_{2+}$,  $C_{2-}$ and $C_1$ all intersect at the same point $p_4:  \beta_5=\beta_4=x=y=w=0$, which is the common intersection of the three lines $L_x$, $L_w$ and $L_t$. 
In the toric description of the resolution of the binomial variety, the point $p_4$ is the interior of the prism.
For any of our 6 small resolutions, its proper transform is the union of two ruling of  $\mathbb{F}_0$: 
\begin{equation}
\alpha_-\alpha_+=0. 
\end{equation}
This shows that $p_4$ is replaced   by  the  union of two $\mathbb{P}^1$s defined respectively by $\alpha_-=0$ and $\alpha_+=0$ inside $\mathbb{F}_0=\mathbb{P}^1\times \mathbb{P}^1$. We recall that we  parametrize the projective cone $\mathbb{F}_0$ by the projective coordinates $[\alpha_-:\sigma_-]$ and $[\alpha_+:\sigma_+]$.  We will denote the node $x=y=w=t=s=\alpha_-=0$  and $x=y=w=t=s=\alpha_+=0$ respectively as $C_{p-}$ and $C_{p+}$. 
  These two rational curves  intersect transversally at the point $\alpha_-=\alpha_+=0=x=y=w=\beta_5=\beta_4=0$.  
   In order to understand the structure of the new fiber above the point $\beta_5=\beta_4=0$ of the $\SU(5)$ divisor, we need to clarify  how the  two new nodes $C_{p\pm} $ connect to the nodes $C_1$, $C_{2+}$ and $C_{2-}$.  
   It is easy to see that in any of the six small resolutions the nodes  $C_{p+}$ (resp. $C_{p-}$)
    is connected to $C_{2+}$ (resp. $C_{2-}$) so that we have a chain $C_{2-}-C_{p-}-C_{p+}-C_{2+}$. In order to get the full structure of the fiber, we just need to determine how the node $C_1$ intersects that chain. 
    We have the following 3 behaviors according the small resolution we consider: 
   $C_1$ intersects $C_{p+}$ but not $C_{p-}$ (for ${\mathscr{E}}_{tw}$ and ${\mathscr{E}}_{xw}$),
 $C_1$ intersects $C_{p-}$ but not $C_{p+}$ (for ${\mathscr{E}}_{wt}$ and ${\mathscr{E}}_{wx}$),
  $C_1$ intersects both  $C_{p+}$ and $C_{p-}$ (for ${\mathscr{E}}_{tx}$ and ${\mathscr{E}}_{xt}$). The corresponding dual graphs are represented in figure \ref{codim3b4}. 

\subsection{Flop transitions and codimension-3 singular fibers}

\begin{figure}[ bht ]
\setlength{\unitlength}{.5 mm}
\begin{picture}(60,60)(-140,-40)

\put(0,-28){
\setlength{\unitlength}{.5 mm}
\color{black}
\put(0,44){$\mathscr{E}_{xt}$}
\put(32,0){$\mathscr{E}_{wx}$}
\put(-43,0){$\mathscr{E}_{tw}$}
\qbezier(-30,0)(0,0)(30,0)
\qbezier(-30,0)(-15,21.2132)(0,42.4264)
\qbezier(30,0)(15,21.2132)(0,42.4264)
\color{blue}
\qbezier(-30,0)(-30,15)(-30,28)
\qbezier(30,0)(30,15)(30,28)
\qbezier(-30,28)(-15,35.2132)(0,42.4264)
\qbezier(30,28)(15,35.2132)(0,42.4264)
}

\put(2,0){
\color{black}
\setlength{\unitlength}{.5 mm}
\put(5,-45){$\mathscr{E}_{tx}$}
\put(32,0){$\mathscr{E}_{wt}$}
\put(-43,0){$\mathscr{E}_{xw}$}
\qbezier(-30,0)(0,0)(30,0)
\qbezier(-30,0)(-15,-21.2132)(0,-42.4264)
\qbezier(30,0)(15,-21.2132)(0,-42.4264)
\color{blue}
\qbezier(-30,-28)(-15,-35.2132)(0,-42.4264)
\qbezier(30,-28)(15,-35.2132)(0,-42.4264)

}

\end{picture}
\caption{ 
\label{flop.transpo}
\footnotesize{The 6 small and flat resolutions of the $\SU(5)$ variety are related to each other by a network of flop transitions. Each triangles is an orbit of an element of $\mathscr{S}_3$ of order 3. These two orbits are exchanged by any transposition of $\mathscr{S}_3$.   The group $\mathscr{S}_3$ is isomorphic to the dihedral group $Dih_3$ of the triangle. The only permutation which preserves the structure of all fibers is $(xt)$.
The involution of the smooth fibers generated by the inverse of the group law of the elliptic fiber induces a birational transformation $\iota$  that acts as ${\mathscr{E}}_{ij}\mapsto {\mathscr{E}}_{ji}$ on the six resolutions. In string theory, it corresponds to $(-)^{F_L}\Omega$. On the hexagram, it corresponds to the central involution. It enlarges $\mathscr{S}_3$ to  $Dih_6$. 
 }
 }\end{figure}
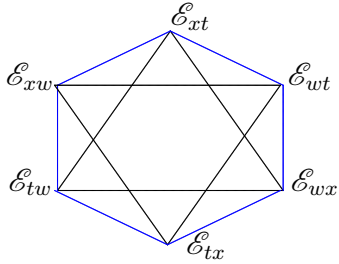

The different small resolutions of the  binomial variety $\mathscr{E}_{bin}$ lead to 6 different small resolutions of the  $\SU(5)$ geometry:
$$\{\mathscr{E}_{xt},\mathscr{E}_{wx},\mathscr{E}_{tw}, 
\mathscr{E}_{wt},\mathscr{E}_{tx},\mathscr{E}_{xw}
\}.$$  
Each of them corresponds to a  distinct resolution of the original  four-fold geometry. These six fourfolds have different topology and are related to each other by a network of flop transitions under which some rational curves inside codimension-two and codimension-three  singular fibers   shrink to a point and get replaced by other rational curves. These flop transitions do not modify the structure  of the fiber above codimension-2 loci. However in codimension-three in the base,  the dual graph is not necessary preserved over the points $\Pi_4$.   
At these points, the rational curves $C_{p\pm}$ can change their intersection with the curve $C_1$ leading in this way to three different  types of fibers. 
The fiber we get above  the point $\Pi_4$ are not Kodaira fibers. In the case of the resolutions 
$\mathscr{E}_{wx},\mathscr{E}_{tw}, \mathscr{E}_{wt}$  and $\mathscr{E}_{xw}$, the fiber is a projective $E_6$ Dynkin diagram that should not be confused with the affine $\tilde{E}_6$ which is usually conjectured to appear in the F-theory literature in  order to have the Yukawa couplings of the up-type quarks
\footnote{
The projective $E_6$ (resp. affine $\tilde{E}_6$) is composed of  6 (resp. 7) rational curves.  Once  the rational curve $C_0$ is removed, the remaining curves admit a dual graph of type projective $D_5$ (resp. projective $E_6$) Dynkin diagram.}. For the resolution $ \mathscr{E}_{xt}$  and $\mathscr{E}_{tx}$, the fiber is not even a Dynkin diagram but a bouquet composed of three 2-chains meeting at  a common point. 
We have seen that the  discrete group $\mathscr{S}_3=Dih_3$ acts transitively on these six fourfolds.  It can be described by the permutation of the  three three elements  $\{x,t,w\}$. This group  acts   transitively on the six small resolutions  $\mathscr{E}_{ab}$.  The group $\mathscr{S}_3$ contains 2 elements of order 3 and 3 elements of order 2 which are just transpositions. The elements of order 3 (that is $(xwt)$ or $(twx)$) organize the six small resolutions into two orbits: 
$$
\{ \mathscr{E}_{xt},\mathscr{E}_{wx},\mathscr{E}_{tw}\}, \quad \text{and}\quad 
\{ \mathscr{E}_{wt},\mathscr{E}_{tx},\mathscr{E}_{xw}\}.
$$
These two orbits are exchanged by the  transpositions $(xt)$, $(xw)$ or $(tw)$. 
Among all the permutations, only the  transposition $(xt)$  preserves the structure of all dual graphs. 
Using the action of the group $\mathscr{S}_3$,  it is convenient to  organize the six small resolutions into an hexagram as represented in figure \ref{flop.transpo}. The hexagram is composed of two triangles, each representing an orbit under the permutation of order three. The $\mathbb{Z}_2$ North-South symmetry is given by the permutation $(xt)$ which preserves all the dual graphs. The central involution which turns any $\mathscr{E}_{ij}$ into $\mathscr{E}_{ji}$ is induced by the inverse transformation on the smooth elliptic fibers. 
From the point of view of the modular group, that symmetry is given by minus the identity ( $-\mathrm{I}_2$) and corresponds to the perturbative string theory  operator $(-)^{F_L}\Omega$, where $\Omega$ is the worldsheet parity and $F_L$ is the left-moving sector spacetime fermion number   on the worldsheet. 
 On the hexagram, the map $\mathscr{E}_{ij}\leftrightarrow \mathscr{E}_{ji}$  corresponds to the central involution. It enlarges $\mathscr{S}_3\simeq Dih_3$ to  $Dih_6$.

\subsection{Some comments on the  exotic fibers}

In this section we will comments on the properties of the resolution we have obtained: the enhancement of fibers without increase of the rank, the orientifold picture and a no-go theorem using Batyrev theorem on the birational invariance of Betti numbers in the category of projective algebraic varieties.  
\begin{enumerate}
\item{\bf Strong coupling}. At the points $\Pi_4$, the  $j$-invariant vanishes. It follows that  the string couplings $g_s=(Im \tau)^{-1}$ is strongly coupled. 
Interestingly, the  string coupling constant is small ($j=\infty$) anywhere else in the 
$\SU(5)$ divisor. 
\item {\bf Conifold singularities in the orientifold picture}.
In the type IIB orientifold picture, we consider the double cover  over the base ramified at $b_2=0$ where $b_2=a_1^2+4 a_2$. Using the original Weierstrass equation of the fourfold before the blow-ups, we have  
$X:\xi^2=b_2$ with  $b_2=\beta_5^2+4 \beta_4 w$ and $\xi$ a section of $\mathscr{L}$:  
\begin{equation}
X: \xi^2=\beta_5^2+4 w \beta_4.
\end{equation}
The elliptic fourfold $Y$ is Calabi-Yau when $\mathscr{L}=K_{B}$ and it follows that when $Y$ is Calabi-Yau, the double cover $X$ is also Calabi-Yau. 
If we rewrite  $X$ as follows
\begin{equation}
X: (\beta_5+\xi)(\beta_5-\xi)-4 \beta_4 w =0. 
\end{equation}
It is clear that the points $\Pi_4:\beta_5=\beta_4=0$ in the divisor $D_{\su(5)}:w=0$ corresponds to conifold singularities of $X$ and these points are located on the  ramification locus (the orientifold). This has been noticed already by Donagi and Wijnholt  \cite{Donagi:2009ra}.
Conifold singularities of threefolds are very well understood and in the context of Calabi-Yau threefolds their physical interpretations  have been given by 
 Strominger \cite{Strominger:1995cz}. In the present case, there is a complication since we are in presence of an orientifold symmetry.  
The two small resolutions of the conifolds are
\begin{equation}
X_+
\begin{cases}
\lambda (\beta_5+\xi )-\mu w =0 \\
\mu (\beta_5-\xi)-4\lambda  \beta_4=0
\end{cases}\quad 
\overset{\xi \rightarrow -\xi}{\dashleftarrow \dashrightarrow}\quad 
X_-
\begin{cases}
\lambda (\beta_5-\xi )-\mu w =0 \\
\mu (\beta_5+\xi)-4\lambda  \beta_4=0
\end{cases}
\end{equation}
where   $[\lambda:\mu]$ denotes  projective coordinates of a   $\mathbb{CP}^1$.  
These two small resolutions $X_\pm$ are exchanged by the orientifold involution $\xi\mapsto -\xi$. This behavior is parallel to the  one we have discussed for the small resolutions of the fourfold: {\em a discrete symmetry of the singular space becomes a birational transformation between two different small resolutions}. 
We could also consider the blow-up of the conifold points where the singular points are replaced by ruled surfaces $\mathbb{F}_0=\mathbb{CP}^1\times\mathbb{CP}^1$ defined by a quadric equation in $\mathbb{CP}^3$: 
\begin{equation}
u_1 u_2 -u_3 u_4=0, [u_1, u_2, u_3, u_4]\in \mathbb{CP}^3.
\end{equation}
Here, the involution preserves the resolved space but exchanges the two rulings of $\mathbb{F}_0$ and the quotient is just a projective plane ( $\mathbb{F}_0/ \iota = \mathbb{P}^2$). 
It would be interesting to study the link between the exotic fibers in F-theory and the conifold points of  type IIB  orientifolds further. 

\item {\bf Degeneration without an increase of the rank.}
The fiber above the points $\Pi_4$ are examples of non-transverse collisions without an increase of the number of node of the fiber.   
  Collisions without increase of the rank of the singularity has been noticed before \cite{Codim3Sing}, but in a different context where exotic fibers had higher dimensional components.  It was deduced that in the heterotic dual,  the codimension-three singularities corresponded to a  point-like degeneration of certain  bundles.    The model of \cite{Codim3Sing} is studied further in appendix G of \cite{Beasley:2008dc}. 
The codimension-three singularity we have obtained here are different from the one studied in  \cite{Codim3Sing}   because we have a flat fibration( all the fibers have the same dimension). However, we could easily construct a non-flat resolution  $Z\rightarrow Y$   by  blowing-up the lines $L_x$, $L_w$ and $L_t$ after resolving the $I_5$ singularity. We will present the details somewhere else \cite{MboyoYau2}.  

\item {\bf Batyrev theorem and a no-go theorem.}
 Using $p$-adic analysis on algebraic varieties, Batyrev proved that birational Calabi-Yau varieties  have the same Betti numbers \cite{Batyrev}. This implies in our case that all the six small resolutions we have constructed have the same Betti numbers. Any other small resolution, will also have the same Betti numbers since it will be  birationally equivalent to the resolutions we have constructed here. This would implies the impossibility to obtain a small resolution where the fibers above the points $\Pi_4$ have a dual graph of type $\tilde{E}_6$. In other words, the conjectured fiber structure of $\SU(5)$ models is impossible since it will have a different Euler characteristic than the birational equivalent resolution obtain in this paper. The difference of Euler characteristic is due to the difference of fibers above $\Pi_4$. Obviously,  a no-go theorem is just as strong as its assumptions. Batyrev theorem is very strong since it only requires the two birational copies $Y_1$  and $Y_2$ to be smooth irreducible projective algebraic varieties related by a birational map that does not change the canonical class \cite{Batyrev}.

\end{enumerate}

\begin{figure}[!b h t ]
\setlength{\unitlength}{.5 mm}
\begin{picture}(25,305)(-40,-275)

\put(100,25){
\put(6,0){\footnotesize $C_0$}
\put(-3,-2){{$\times$}}
\put(25,-20){\footnotesize $C_{1-}$}
\put(-36,-20){\footnotesize $C_{1+}$}
\put(20,-40){\footnotesize $C_{2-}$}
\put(-31,-40){\footnotesize $C_{2+}$}
\put(0,0){\circle{10}}
\put(-20,-20){\circle{10}}
\put(20,-20){\circle{10}}
\put(15,-40){\circle{10}}
\put(-15,-40){\circle{10}}
\qbezier(-3,-3)(-5,-5)(-17,-17)
\qbezier(3,-3)(5,-5)(17,-17)
\qbezier(-19,-26)(-18,-30.5)(-17,-35)
\qbezier(19,-26)(18,-30.5)(17,-35)
\qbezier(-10,-40)(-10,-40)(10,-40)

\put(35,-70){
\linethickness{.3mm}
\qbezier(.5,-.5)(-7,7)(-15,15)
\qbezier(.5,-.5)(-1.1,2)(.2,4)
\qbezier(.5,-.5)(-2,1.1)(-4,0)
\put(-3,10){$\beta_5=0$}
}

\put(-35,-70){
\linethickness{.3mm}
\qbezier(-.5,-.5)(10,10)(15,15)
\qbezier(-.5,-.5)(1.1,2)(-.2,4)
\qbezier(-.5,-.5)(2,1.1)(4,0)
\put(-20,10){$P=0$}
}
}

\put(170,-90){
\put(-15,25){\color{blue}\circle*{10}}
\put(15,25){\circle{10}}
\put(12.5,23){{$\times$}}
\put(0,10){\circle{10}}
\put(0,-10){\color{blue}\circle*{10}}
\put(-15,-25){\circle{10}}
\put(15,-25){\circle{10}}
\put(6,7){\footnotesize $ 2C_{1}$}
\put(6,-10){\footnotesize \color{blue}$2C_t$}
\put(20,-25){\footnotesize $C_{2-}$}
\put(-34,-25){\footnotesize $C_{2+}$}
\put(20,25){\footnotesize $C_{0}$}
\put(-32,25){\footnotesize\color{blue} $C_x$}
\qbezier(0,5)(0,0)(0,-5.5)
\qbezier(5,12)(8,15)(14,21)
\qbezier(-5,12)(-8,15)(-14,21)
\qbezier(-5,-12)(-8,-15)(-14,-21)
\qbezier(5,-12)(8,-15)(14,-21)

%\put(-4,-55){
%\put(0,-5){\huge $\downarrow$}
%\put(10,0){$\beta_4=0$}
%}

}

\put(20,-75){
\put(6,20){\footnotesize $C_0$}
\put(-3,18){{$\times$}}
\put(25,0){\footnotesize $C_{1-}$}
\put(-36,0){\footnotesize $C_{1+}$}
\put(25,-20){\footnotesize $C_{2-}$}
\put(-36,-20){\footnotesize $C_{2+}$}
\qbezier(-20,-5)(-20,-10)(-20,-15)
\qbezier(20,-5)(20,-10)(20,-15)

\put(0,-40){\color{blue}\circle*{10}}
\put(-20,0){\circle{10}}
\put(20,0){\circle{10}}
\put(-20,-20){\circle{10}}
\put(20,-20){\circle{10}}
\put(0,20){\circle{10}}

\put(-3,-30){\footnotesize\color{blue} $C_{w}$}
\qbezier(-3,17)(-5,15)(-17,3)
\qbezier(3,17)(5,15)(17,3)
\qbezier(-3,-37)(-5,-35)(-17,-23)
\qbezier(3,-37)(5,-35)(17,-23)

}

\put(5,-210){
\put(-15,25){\color{blue}\circle*{10}}
\put(15,25){\circle{10}}
\put(12,23){{$\times$}}
\put(0,10){\circle{10}}
\put(0,-10){\color{blue}\circle*{10}}
\put(-15,-45){\color{blue}\circle*{10}}
\put(15,-45){\color{blue}\circle*{10}}
\put(0,-30){\circle{10}}
\put(6,-30){\footnotesize $ 2C_{2}$}
\put(6,7){\footnotesize $ 2C_{1}$}
\put(6,-10){\footnotesize \color{blue}$2C_{t}$}
\put(20,-45){\footnotesize \color{blue}$C_{w-}$}
\put(-31,-45){\footnotesize \color{blue}$C_{w+}$}
\put(20,25){\footnotesize $C_{0}$}
\put(-28,25){\footnotesize\color{blue} $C_x$}
\qbezier(0,5)(0,0)(0,-5.5)
\qbezier(5,12)(8,15)(14,21)
\qbezier(-5,12)(-8,15)(-14,21)
\qbezier(-5,-32)(-8,-35)(-14,-41)
\qbezier(5,-32)(8,-35)(14,-41)
\qbezier(0,-15)(0,-18)(0,-25.5)
}

\put(65,-210){
\setlength{\unitlength}{.40 mm}
\put(40,20){
\put(0,0){\circle{10}}
\put(-5,-2){{$\times$}}
\put(-25,0){\footnotesize $C_0$}
\put(20,0){or}
\qbezier(0,-5)(0,-7)(0,-15)
\put(-25,-20){\footnotesize $2 C_{1}$}
\put(0,-20){\circle{10}}
\put(0,-20){
\put(0,-20){\circle{10}}
{\color{blue}\put(0,-20){\circle*{10}}
\put(0,-40){\circle*{10}}}
\put(0,-60){\circle{10}}
\multiput(0,0)(0,-20){3}{\qbezier(0,-5)(0,-10)(0,-15)}
\put(15,-20){\circle{10}}
\qbezier(5,-20)(7,-20)(10,-20)
\put(22,-20){\footnotesize $C_{2+}$}
{\color{blue} \put(-25,-20){\footnotesize $3C_{p+}$}
\put(-25,-40){\footnotesize $2C_{p-}$}}
\put(-25,-60){\footnotesize $C_{2-}$}
\put(-28,-80){{$\mathscr{E}_{tw} \ \text{and}\ \mathscr{E}_{xw}$}}
}
}

\put(110,20){

\put(0,0){\circle{10}}
\put(-5,-2){{$\times$}
}
\put(-20,0){\footnotesize $C_0$}
\qbezier(0,-5)(0,-7)(0,-15)
\put(25,0){{or}}
\put(-21,-20){\footnotesize $2 C_{1}$}
\put(0,-20){\circle{10}}
\put(0,-20){
\put(0,-20){\circle{10}}
{\color{blue}\put(0,-20){\circle*{10}}
\put(0,-40){\circle*{10}}}
\put(0,-60){\circle{10}}
\multiput(0,0)(0,-20){3}{\qbezier(0,-5)(0,-10)(0,-15)}
\put(15,-20){\circle{10}}
\qbezier(5,-20)(7,-20)(10,-20)
\put(22,-20){\footnotesize $C_{2-}$}
{\color{blue} \put(-25,-20){\footnotesize $3C_{p-}$}
\put(-25,-40){\footnotesize $2C_{p+}$}}
\put(-25,-60){\footnotesize $C_{2+}$}
\put(-28,-80){{$\mathscr{E}_{wt} \  \text{and}\  \mathscr{E}_{wx}$}}
}

}

\put(180,20){
\put(0,0){\circle{10}}
\put(-5,-2){{$\times$}}
\put(-19,0){\footnotesize $C_0$}
\qbezier(0,-5)(0,-7)(0,-15)
\put(-25,-20){\footnotesize $2 C_{1}$}
\put(0,-20){\circle{10}}
\put(-2.5,-20){
{\color{red}
\color{blue}\put(0,-40){\circle*{10}}}
\put(0,-60){\circle{10}}
\put(0,0){\qbezier(0,-5)(0,-20)(0,-35)}
\put(0,-40){\qbezier(0,-5)(0,-10)(0,-15)}
\put(15,-20){{\color{blue}\circle*{10}} \put(-20,0){\qbezier(-5,0)(2,0)(5,0) }}
\put(30,-20){\circle{10} \put(-20,0){\qbezier(0,0)(2,0)(5,0) }}
\put(36,-21){\footnotesize $C_{2-}$}
{\color{blue} \put(9,-13){\footnotesize $2C_{p-}$}
\put(-25,-41){\footnotesize $2C_{p+}$}}
\put(-25,-61){\footnotesize $C_{2+}$}
\color{red}\put(0,-20){\circle*{2}}
\color{black}
\put(-25,-80){{$\mathscr{E}_{xt}\  \text{and}\  \mathscr{E}_{tx}$}}
}
}

\put(-10,0){
\linethickness{.2mm}
\put(0,30){\line(1,0){250}}
\put(250,30){\line(0,-1){120}}
\put(0,30){\line(0,-1){120}}
\put(0,-90){\line(1,0){250}}
}
}

\put(50,-170){
\linethickness{.2mm}
\qbezier(-.5,-.5)(17,17)(50,50)
\qbezier(-.5,-.5)(1.1,2)(-.2,4)
\qbezier(-.5,-.5)(2,1.1)(4,0)
\put(-28,15){$\beta_5=\beta_3=0$}

\put(50,50)
{\qbezier(0,0)(0,10)(0,20)}

\put(50,70){
\qbezier(0,0)(20,10)(40,20)
\qbezier(0,0)(-20,10)(-40,20)
}

\put(100,0){
\linethickness{.2mm}
\qbezier(.5,-.5)(-17,17)(-50,50)
\qbezier(.5,-.5)(-1.1,2)(.2,4)
\qbezier(.5,-.5)(-2,1.1)(-4,0)
\put(-6,15){$\beta_5=\beta_4=0$}
}

}

\end{picture}
\caption{ {Fiber degeneration of an  $\SU(5)$ GUTs with a small resolution. 
nodes $C_{x}$, $C_{w}$ and $C_t$ are coming from the resolution of the higher codimensional singularities. We have 6 possible resolutions $\mathscr{E}_{xw}$, $\mathscr{E}_{wx}$, $\mathscr{E}_{xt}$, $\mathscr{E}_{tw}$ and $\mathscr{E}_{wt}$. They are all related to each other by flop transitions related to permutation group of 3 elements. 
\label{enhancements}
  }
}
\end{figure}
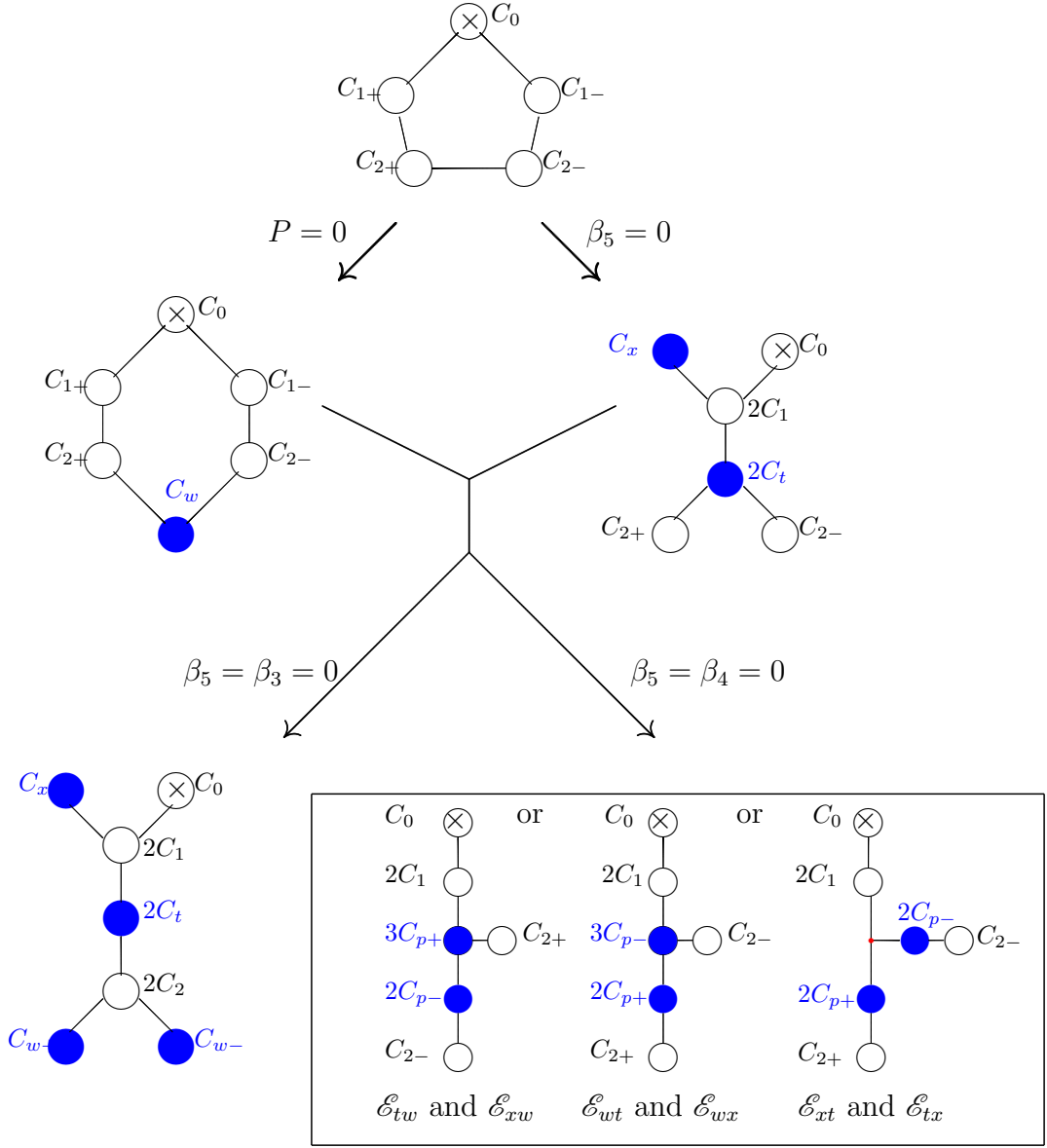

%\clearpage

\section{ Conclusions and discussions\label{conclusions}}

We have  presented an explicit desingularization of the elliptically fibered fourfold describing the $\SU(5)$ GUT geometry in F-theory. In codimension-one in the base, the resolution of the generic singularity generates a  fiber of Kodaira type $I_5$ over the $\SU(5)$  divisor. 
After resolving the singularity in codimension-one, there are  several left-over singularities in codimension-two  and codimension three. These higher codimension singularities are all visible in a unique patch where they can be described  elegantly by an affine binomial variety. The singular locus of the binomial variety consists of  a bouquet of  three lines  $L_w,L_t,L_x$ of  conifold singularities  all intersecting at a common point $p_3$ where the singularity worsen. The binomial geometry admits six small resolutions that we describe both torically and algebraically. 
In each of these small resolutions, the exceptional locus is a $\mathbb{CP}^1$ bundle over the bouquet of three lines $L_w, L_t, L_x$. At the origin of the bouquet, the $\mathbb{CP}^1$ fiber enhances to two intersecting $\mathbb{CP}^1$s. We then obtain a full desingularization of the $\SU(5)$ geometry by pulling  back the resolution obtained  for the binomial variety. We end up with six small resolutions related to each other by flop transitions. 

All the six small resolutions have the same fiber structure  in codimension-one and codimension-two in the  base and also in codimension-three for the fibers above the points $\beta_5=\beta_3=0$ in the $\SU(5)$ divisor. In codimension-one we have the $I_5$ fiber which described the $\SU(5)$ gauge group. In codimension-two,   we get enhancement to  $\tilde{A}_5$ and $\tilde{D}_5$ respectively along a curve $P=0$ and  $\beta_5=0$ in the $\SU(5)$ divisor. The dual graph $\tilde{A}_5$ corresponds to a Kodaira fiber of type $I_6$ while the dual graph $\tilde{D}_5$  corresponds to a Kodaira fiber $I^*_1$.  
Another interesting aspect of the resolution is that there are no fibers of type $\tilde{A}_6$ in codimension-three although above  the points $P=R=0$ in the $\SU(5)$ divisor, the  discriminant locus has an enhanced singularity.

The biggest surprise occurs in codimension-three for the fibers above the points $\Pi_4: \beta_4=\beta_5=0$ in the base. Above these points, the six small resolutions don't admit the same dual graph as we can see from table \ref{enhancements}. More precisely, over the  points $\Pi_4: \beta_5=\beta_4=0$ where an affine $\tilde{E}_6$ is usually expected in the physics literature, we get a fiber with dual graph $E_6$ for four of the six resolutions while the remaining two admits  singular fibers whose dual graph is not a Dynkin diagram  but a new exotic diagram that we call $\tilde{E}^-_6=T^-_{3,3,3}$:  that is a bouquet of three two-chains intersecting at the same point. 
Such fiber can be seen as a fiber of type $\tilde{E}_6$ with the central node contracted to a point. The dual graphs  $E_6$ and  $\tilde{E}^-_6$ have the same number of node as the graph $\tilde{A}_5$ or $\tilde{D}_5$. This is an example of enhancement of a  singular fiber with conservation of the rank of the fiber.  Using Batyrev theorem on the Betti numbers of birational equivalent Calabi-Yau varieties, once can deduce that the conjectured fiber structure of $\SU(5)$ models is impossible in the category of projective algebraic varieties. 
If one specializes the defining equation of the $\SU(5)$ models, the type of fibers will change, we will considering such models in a different work.

\begin{center}{\bf  Acknowledgments}\end{center}

M.E. would like to thank 
Murad Alim, 
Paolo Aluffi, 
Clay Cordova, 
Frederik Denef, Ron Donagi, Tamar Fridmann, 
Antonella Grassi, Jonathan Heckman,  Shamit Kachru, 
 Si Li,
David Morrison, 
James M$^{\text{C}}$Kernan, 
Keiji  Oguiso, Sakura Sch\"{a}fer-Nameki,   
Michael Szydlo, Washington Taylor,  
Li-Sheng Tseng  and Martin Wijnholt 
   for interesting discussions. The content of this paper has been presented in the workshop on 
Generalized Geometries and String Theory in March 2011 and at the String Math conference in June 2011 and in seminars at Harvard, Stanford, KITP and Caltech. 
We would like to thank the audience for their comments and suggestions. The authors would  like to thank Sheldon Katz for his very useful comments and for pointing out an arithmetic typo in the  multiplicity of nodes.

\appendix

\end{document}